# Nature of the 5f States in Actinide Metals

Kevin T. Moore*

*Chemistry and Materials Science Directorate, Lawrence Livermore National Laboratory, Livermore, California 94550, USA.*

Gerrit van der Laan[§]

*Diamond Light Source, Chilton, Didcot, Oxfordshire OX11 0DE, United Kingdom
and
STFC Daresbury Laboratory, Warrington WA4 4AD, United Kingdom*

## Abstract

Actinide elements produce a plethora of interesting physical behaviors due to the 5$f$ states. This review compiles and analyzes progress in understanding of the electronic and magnetic structure of the 5$f$ states in actinide metals. Particular interest is given to electron energy-loss spectroscopy and many-electron atomic spectral calculations, since there is now an appreciable library of core $d \rightarrow$ valence $f$ transitions for Th, U, Np, Pu, Am, and Cm. These results are interwoven and discussed against published experimental data, such as x-ray photoemission and absorption spectroscopy, transport measurements, and electron, x-ray, and neutron diffraction, as well as theoretical results, such as density-functional theory and dynamical mean-field theory.

*Contact Author
Tel:    925-422-9741
Fax:    925-422-6892
Email: moore78@llnl.gov

[§] Also at School of Earth, Atmospheric, and Environmental Sciences, University of Manchester, Manchester M13 9PL, United Kingdom.

**Contents**

# I. INTRODUCTION

## A. The Actinide Problem: Not enough Data!

The electronic and magnetic structure of most elemental metals in The Periodic Table is well understood. Some exceptions to this are manganese, which has a complicated magnetic structure that is still being investigated (Hobbs *et al.*, 2003; Hafner and Hobbs, 2003), cerium, which has an apparent isostructural fcc→fcc phase transformation with a volume change of ~17% that is due to a fundamental change in the behavior of the 4*f* states (Johansson, 1974; Koskenmaki and Gschneidner, 1978; Allen and Martin, 1982), and plutonium, which not only exhibits a wide variety of exotic behaviors due to the complex nature of the 5*f* states, but also sits at the nexus of an anomalous ~40% volume change that occurs in the actinide series (Zachariasen, 1973; Albers, 2001; Hecker, 2002).

Focusing on the actinides, it is apparent that only modest attention has been given to the light to middle metals in the series, excluding uranium (Lander *et al.*, 1994). There have been experimental investigations on the phase transformations (Ledbetter and Moment, 1976; Zocco *et al.*, 1990; Hecker *et al.*, 2004; Blobaum, 2006), electronic structure (Baer and Lang, 1980; Naegele *et al.*, 1984; Havela *et al.*, 2002; Gouder *et al.*, 2005), charge density waves (Smith *et al.*, 1980; Smith and Lander, 1984; Marmeggi *et al.*, 1990), phonon dispersion curves (Manley *et al.*, 2003a; Wong *et al.*, 2003), effects of self-induced irradiation (Schwartz *et al.*, 2005), and pressure-induced phase transformations (Roof *et al.*, 1980; Haire *et al.*, 2003; Heathman *et al.*, 2005). Yet even with these and other studies, there are still many unanswered questions, and arguments persist on topics such as the number of electrons in valence states, magnetism, angular momentum coupling, and the character of the bonding. The heaviest actinides have almost no experimental investigations, generating only a rudimentary level of understanding. Thus, the actinide series as a whole is modestly understood, with the level of comprehension decreasing with atomic number. The lack of experiments is due to the toxic and radioactive nature of the materials, which makes handling difficult and expensive. In addition, the cost of the materials themselves is exceedingly high, meaning experiments that need a large amount of materials further increase the expense of research.

Theoretical work on actinide metals is extensive, since theory allows one to circumvent the need to physically handle the materials[1]. The theoretical studies range from density-functional theory (DFT) with either the generalized gradient approximation (GGA) or the local density approximation (LDA) to multi-electron atomic spectral calculations to dynamical mean-field theory (DMFT). Regardless of this considerable body of work, the progress in understanding from these calculations has been hampered due to the extreme difficulty of the physics involved and the lack of a healthy body of experimental data from which to validate the theory.

In order to counter the lack of experimental data, we are progressively recording the various core *d* → valence *f* transitions of the actinides using electron energy-loss spectroscopy (EELS) in a transmission electron microscope (TEM) (Moore *et al.*, 2003; van der Laan *et al.*, 2004; Moore *et al.*, 2004a; 2004b; 2006b; 2006c; 2007a; 2007b; Moore and van der Laan, 2007; Butterfield *et al.*, 2008). Once acquired, the spectra are analyzed using multi-electron atomic spectral calculations (Thole and van der Laan, 1987; 1988a; 1988b; van der Laan and Thole, 1988a; 1996b) to discern fundamental aspects of the electronic structure of the 5*f* states in the actinide metals, such as angular momentum coupling mechanisms, electron filling, and limits on the number of valence electrons. We purposefully focus on the 5*f* states, since they are the main culprit for most of the odd behaviors observed in actinide metals, alloys, and compounds.

---

[1] Examples: Gupta and Loucks, 1969; Skriver *et al.*, 1978; Skriver, 1985; Solovyev *et al.*, 1991; Söderlind *et al.*, 1995; van der Laan and Thole, 1996; Penicaud, 1997; Fast *et al.*, 1998; Söderlind, 1998; Eriksson *et al.*, 1999; Savrasov and Kotliar, 2000; Savrasov *et al.*, 2001; Dai *et al.*, 2003; Kutepov and Kutepova, 2003; Söderlind and Sadigh, 2004; Wills *et al.*, 2004; Shick *et al.*, 2005; Pourovskii *et al.*, 2005; Moore *et al.*, 2006a; Shick *et al.*, 2006; Shim *et al.*, 2007; Svane *et al.*, 2007.

In this review, we cover progress in understanding the electronic structure of the 5*f* states in the actinide metal series. First, we will cover electron energy-loss spectroscopy (EELS) for the $O_{4,5}$ (5*d* → 5*f*) and the $N_{4,5}$ (4*d* → 5*f*) edges of thorium (Th), uranium (U), neptunium (Np), plutonium (Pu), americium (Am), and curium (Cm) metal.[2] Next, we will derive and examine the many-electron atomic spectral calculations for the 5*d* → 5*f* transitions, paying close attention to the *LS*, *jj*, and intermediate coupling mechanisms. Returning our attention to experiment, we will cover inverse, valence-band, and 4*f* core x-ray photoemission of the actinide metals up to Am. Finally, with all the EELS, many-electron atomic calculations, and photoemission (PE) spectra compiled, the electronic structure of each elemental metal will be discussed in turn. Further experimental results will be considered, such as x-ray absorption spectroscopy, transport measurements, and electron, x-ray, and neutron diffraction, as well as theoretical models, such as DFT and DMFT. The combination of experimental and theoretical results forms a cogent picture of the physics of the 5*f* states in the elemental actinide metals.

**B. Actinide Series Overview**

In the Hamiltonian used for electronic structure calculations, there are two standard ways to couple the angular momenta of multi-electronic systems: Russell-Saunders (*LS*) and *jj* coupling. In atoms where the spin-orbit coupling is weak compared to the Coulomb and exchange interaction, the orbital angular momenta $\ell$ of individual electrons are coupled to a total orbital angular momentum *L* and likewise the spin angular momenta *s* are coupled to a total *S*. Then *L* and *S* are coupled to form the total angular momentum *J*. This approach simplifies the calculation of the Coulomb and exchange interactions, which commute with *L*, *S*, and *J*, and hence are diagonal in these quantum numbers. For heavier elements with larger nuclear charge, relativistic effects give raise to a large spin-orbit interaction, which is diagonal in *j* and *J*, but not in *L* and *S*. Therefore, in the *jj* coupling scheme, the spin and orbital angular momenta, *s* and $\ell$, of each electron are coupled to form individual electron angular momenta *j* and then the different *j* are coupled to give the total angular momentum *J*. It is known that *LS* coupling holds quite well for transition metals (in the absence of crystal field) and for rare earth metals. Their atoms exhibit a Hund's rule ground state with maximum *S* and *L*, which are coupled by spin-orbit interaction antiparallel (parallel) to each other for less (more) than half filled shell, resulting in *J* = |*L*−*S*| (*J* = *L*+*S*). However, for the 5*f* states of the actinides the spin-orbit interaction is much stronger, giving a significant mixing of the Hund's rule ground state by other *LS* states with the same *J* value. Hence, the *LS* states are less pure and there is a tendency towards the *jj* coupling limit. The choice of the coupling limit has profound implications for the expectation value of the spin-orbit interaction, as well as for any other orbital-related interactions, such as the orbital magnetic moment. The safest way to address the 5f states is to use intermediate coupling, where the actual size of the spin-orbit and electrostatic interaction are accounted for. In this review we will show that this coupling can lead to a few surprises, such as the magnetism-driven state of a high-pressure phase of Cm metal.

---

[2] We discuss protactinium (Pa) and berkelium (Bk) in the later sections of the review even though we present no EELS, valence-band, or 4*f* photoemission spectroscopy.

In order to understand the bonding of the 5f metals across the actinide series it is instructive to consider the 4f and 5d metal series. Actinide metal bonding can be separated into two different behaviors, one where the 5f electrons strongly participate in bonding and one where they offer little or no cohesion. This is schematically illustrated in Fig. 1, where the Wigner-Seitz atomic radius (volume) is given for each element in the 5d, 4f, and 5f metal series. The 5d transition metals show a parabolic-like change in volume due to an increase in the number of d electrons. In traversing the series, the size of the atoms first decreases due to the filling of the 5d bonding states, then begins to increase as the antibonding states are filled. This parabolic-like behavior is indicative of a system with itinerant electrons that participate in the bonding. In the 4f rare-earth series the opposite case is observed, one where the volume changes little because the 4f electrons are localized and do not strongly participate in bonding. Rather, the $(spd)^3$ electrons act to bind the metals, and because the spd electrons do not vary in count from trivalent along the rare earth series, almost no change in packing density is observed. Exceptions in the rare earth series, which are omitted in Fig. 1, are Eu and Yb. Both these metals are divalent with $(spd)^2$ and thus have a larger volume than the rest of the trivalent rare earth metals. Finally, the 5f series shows both behaviors. First, a parabolic-like decrease in volume is observed with increasing f-electron count, similar to the 5d series. Then, a large jump in volume occurs in the vicinity of Pu, which is followed by little change for Am and beyond, similar to the rare earth series.

In the 5f states, there is a spin-orbit splitting of 1-2 eV between the $j = 5/2$ and 7/2 levels due to relativistic effects. This causes the 5f electrons to tend towards a jj coupling mechanism where the early actinide metals preferentially fill the $j = 5/2$ level (Söderlind, 1994; Moore et al., 2007a; 2007b). Thus, the first part of the actinide series in Fig. 1 shows a parabolic shape due to the filling of the bonding and then antibonding states in the $j = 5/2$ level. Here, the 5f states are delocalized, forming bands and are metallic. However, near the point where the $j = 5/2$ level is filled with the six electrons, the 5f electrons retract and localize, leaving the $(spd)^3$ electrons to perform the bulk of the bonding in the metal. The loss of 5f bonding causes the large volume increase in the actinide series shown in Fig. 1. Interestingly, the crystallographic volume change occurs over a span of six solid allotropic phases of Pu (α, β, γ, δ, δ′, ε; see Fig. 2), where α has the highest density and δ the lowest. After Pu, the actinide series appears similar to the rare earths due to the absence of appreciable f-electron bonding; accordingly, the 5f states behave in a more atomic fashion.

The changes that occur in the actinide series can be visualized in a different manner using the 'pseudo-binary' phase diagram shown in Fig. 3 (Smith and Kmetko, 1983). In this diagram, the binary phase diagrams for each two neighboring elemental metals are married together. The phase boundaries that are not exact matches between diagrams are extrapolated using logical guesses based on thermodynamic principles. Thus, while the 'pseudo-binary' phase diagram is not strictly correct, it does afford great insight into the general behavior of the actinide metal series. Examining Fig. 3 shows that as the early actinide metals are traversed several changes occur near Pu: the melting temperature reaches a minimum (Matthias et al., 1967; Kmetko and Hill, 1976), the number of phases increases to a maximum, and the crystal structures become exceedingly complex (Zachariasen and Ellinger, 1963). Whereas most metals are usually cubic or hexagonal, U, Np, and Pu exhibit tetragonal, orthorhombic, and even monoclinic crystal structures, the last being an atomic geometry usually found in minerals (Klein and Hurlbut, 1993).

Pressure also causes rapid and numerous changes in the actinides metals, particularly the elements beyond Pu. This is illustrated in the phase diagram in Fig. 4 (Lindbaum *et al.*, 2003; Heathman *et al.*, 2005; Heathman *et al.*, 2007), where the phase fields for Am, Cm, Bk, and Cf are numerous and complicated compared to the light actinides. This is due to the fact that as pressure is increased in these metals, the 5*f* states begin to actively bond, producing low-symmetry crystal structures. Thus, high-pressure research for Am, Cm, Bk, and Cf is of great interest and will be discussed in detail in the later sections of this review.

At early stages of actinide research, the low-symmetry crystal structures of the light actinide metals was ascribed to directional covalent bonds (Matthias *et al.*, 1967). However, over time it became evident that the 5*f* states in Th-Pu are delocalized and, to varying degrees, band-like. Arko *et al.* (1972) and Skriver *et al.* (1978) first showed that the 5*f* band is exceedingly narrow, on the order of 2 eV. In turn, it was shown that narrow bands prefer low-symmetry crystal structures as illustrated by the density-functional theory results in Fig. 5, where the calculated total energy of different crystal structures is plotted as a function of calculated bandwidth (Söderlind *et al.*, 1995a). Four separate metals are shown, each with different electron bonding states: Al (2*p* bonding), Fe (3*d* bonding), Nb (4*d* bonding), and U (5*f* bonding). For all four metals, low-symmetry crystal structures are observed when the band width is narrow, such as found in the 5*f* states of the actinides at ambient pressure. On the other hand, high-symmetry structures are found for wide bands, similar to 4*d* and 5*d* metals. This is even true for the 4*f* and 5*f* metals when they are pressurized to the point where the *f* states become broad enough to support high-symmetry cubic or hexagonal structures. The difference in band width between the *s*, *d*, and *f* states is schematically illustrated in the inset of the Fe panel of Fig. 5. Whereas the *s* band is on the order of 10 eV wide and the *d* band is on the order of 6 eV wide, the *f* band is only about 2 eV wide. Of course, pressure varies these bandwidths, where positive pressure widens the band while negative pressure narrows the bands, reducing the crystal structure symmetry (Söderlind *et al.*, 1995a).

At ambient pressure, U, Np, and Pu exhibit a large number of crystal structures as the temperature is raised to melting, due to small energy differences between allotropic phases. The small energy differences between crystals is a result of a narrow 5*f* band with a high density of states at the Fermi energy and a slightly broader *d* band, each of which are incompletely filled and close in energy. The effect on the actinides can be seen in the rearranged Periodic Table shown in Fig. 6(a), which contains the 4*f*, 5*f*, 3*d*, 4*d*, and 5*d* metals (Smith and Kmetko, 1983). At ground state, the metals in the blue area exhibit superconductivity and the metals in the red area exhibit a magnetic moment. The white band is a transition region where metals are on the borderline between localized (magnetic) to itinerant (conductive) valence electron behavior. The metals at the transition between magnetic and superconducting behavior exhibit numerous crystallographic phases with small energy difference between crystal structures. This increase in the number of phases near the transition is clearly shown in Fig. 6(b), where gray scale indicates the number of solid allotropic phases observed. The diagonal of lighter shades (more phases) in Fig. 6(b) matches the white band in Fig. 6(a). Thus, each of the metals that lie on the localized-itinerant band has frustrated valence electrons and exhibits numerous solid allotropic crystal structures. U, Np, Pu, and Am all exhibit numerous phases with Pu showing an unsurpassed six different crystal structures that are almost energetically degenerate.

Returning our attention to Figs. 3 and 4, we now see more clearly why the crystal structure of the actinide metals near Pu is so sensitive to temperature and pressure. The close energy levels of bonding states and subsequent large degree of hybridization in the actinides near the itinerant-localized transition allows the metal's behavior to be easily changed or 'tuned' via pressure, temperature, and chemistry. This again is due to the small energy differences between crystal structures of the metals on, or near, the transition. Examples where pressure is used to change the 5$f$ states of actinides from localized to delocalized via a diamond anvil cell are Am (Lindbaum *et al.*, 2001), Cm (Heathman *et al.*, 2005), Bk (Haire *et al.*, 1984), Cf (Peterson *et al.*, 1983), and Bk-Cf alloys (Itie *et al.*, 1985). In each case, as the pressure is increased, the structure changes from a high-symmetry, high-volume cubic or hexagonal structures to low-symmetry, low-volume orthorhombic or monoclinic structures that are indicative of active 5$f$ bonding. Examining Am, Cm, and Bk in Fig. 4 clearly illustrates this loss of symmetry as pressure is increased. In an opposite but similar manner, Pu metal can be transformed from the low-symmetry monoclinic $\alpha$ phase to the high-symmetry face-centered cubic $\delta$ phase by raising the temperature from ambient conditions to ~600 K, as shown in Fig. 2. The high-temperature $\delta$ phase of Pu can also be retained to room temperature by the addition of a few atomic percent of Al, Ce, Ga, or Am (Hanson and Anderko, 1988; Hecker, 2002). Indeed, the crystal structure of U, Np, Pu, and Am can all be changed by alloying with small amounts of dopants due to the sensitivity of the 5$f$ states. The fact that the crystal structure of most of the middle actinides can be easily altered by pressure, temperature, and chemistry makes for rather interesting and unique physics.

Given the small energy differences between multiple structures in actinide metals near the itinerant-localized transition in Fig. 6(a), subtle changes can have dramatic effects on the magnetic behavior of U, Np, Pu, and Am. An example of this is shown in Fig. 7, where the superconducting and magnetic transition temperatures for a number of U metals, alloys, and compounds are plotted against the U-U interatomic spacing. The original theory by Hill (1970) is that the degree of overlap of the $f$-electron wave functions between neighboring $f$-electron atoms dictates whether an actinide compound is magnetic or superconducting, independent of crystal structure or other atomic species present in the compound or alloy. Superconducting compounds tend to have short distances between $f$-electron atoms, while magnetic compounds to have large distances between $f$-electron atoms. Most materials follow this behavior and for U the superconducting-magnetic transition is found to be near 3.5 Å. Exceptions to this are $U_2PtC_2$, $UGe_3$, $UPt_3$, $UB_{13}$, and UN. $UGe_3$ has a U-U distance of 0.42 nm, yet is non-magnetic. Following the Hill criteria, the $f$ states in $UGe_3$ should be localized with an atomic magnetic moment. The lack of magnetism is due the 5$f$ electrons hybridizing into bands with the Ge $p$ states, thus breaking down the Hill criteria. Thus, while this type of plot is useful, being able to successfully predict the transition between magnetism and superconductivity in Ce, Np, and Pu (Smith, 1980), it fails for some cases.

While magnetism in actinide compounds is widely studied and accepted (Santini *et al.*, 1999), magnetism in pure Pu is often debated, even though there is no convincing experimental evidence of moments in any of the six allotropic phases of the metal (Lashley *et al.*, 2005; Heffner *et al.*, 2006). EELS and x-ray absorption (XAS) clearly show that Pu is at or near a 5$f^5$ configuration with at least one hole in the $j$ = 5/2 level (Moore *et al.*, 2003; 2006b; 2007a; 2007b). With the hole in the $j$ = 5/2 level, there should be an incomplete cancellation of electron spin and, accordingly a measurable magnetic moment. How then is there no magnetic moment in the metal? Several explanations have been proposed, including Kondo shielding (Shim *et al.*, 2007) and electron pairing correlations (Chapline *et al.*, 2007). Kondo shielding, which is schematically shown in Fig. 8(a), occurs when $s$, $p$, and $d$ conduction electrons cloak the local magnetic moment that should be present in Pu due to the 5$f^5$ configuration. Electron pairing correlations is another theory to explain the absence of magnetism in Pu, and is shown schematically in Fig. 8(b). When lattice distortions are present that lead to internal electric fields, spin-orbit effects can cause the spontaneous appearance of spin currents and pairing of itinerant $f$ electrons with opposite spin. Simple symmetry considerations imply that an electric field can lead to spin pairing only if spin-orbit interactions are important, which is indeed known to be true for Pu. Recent magnetic susceptibility measurements have in fact shown that magnetic moments on the order of 0.05 $\mu_B$/atom form in Pu as damage accumulates due to self-irradiation (McCall *et al.*, 2006). This suggests that small perturbations to the gentle balance of electronic and magnetic structure of Pu metal may destroy or degrade what ever mechanism is responsible for the lack of magnetism in Pu. Finally, spin fluctuations have been proposed as the reason for the anomalous low-temperature resistivity behavior of Pu that shows no magnetism in the metal (Nellis *et al.*, 1970; Arko *et al.*, 1972). Spin fluctuations can be thought of as spin alignments that have lifetimes less than ~$10^{-14}$ seconds and are therefore too short to see via specific heat, susceptibility, or nuclear magnetic resonance (Brodsky 1978).

The next logical question is what exactly is the underlying physics that causes the large volume change at the localized-delocalized transition in the actinide series? To address this, we should first look towards a metal with similar issues that also falls on the localized-itinerant transition in Fig. 6, the 4$f$ metal Ce. At ambient pressure, Ce metal exhibits four allotropic crystal structures between absolute zero and its melting temperature at 1071°K: α, β, γ, and δ. There are large hystereses between the transformations of α, β, and γ, causing phase boundaries to be kinetic approximations and mixtures of two or even three phases to metastably persist in the thermodynamically single-phase fields (McHargue and Yakel, 1960; Rashid and Altstetter, 1966). When fcc γ-Ce transforms to fcc α-Ce upon cooling, it undergoes an isostructural volume collapse of 17%. Several interpretations as to why this collapse occurs are available, such as the promotion of the single 4$f$ electron from localized and non-bonding to delocalized and bonding (Lawson and Tang, 1949), a metal-to-insulator Mott transition (Johansson, 1974), and a Kondo volume-collapse (Allen and Martin, 1982). The promotional model was challenged when Gustafson *et al.* (1969) showed there was no significant change in the number of $f$ electrons between α- and γ-Ce via positron lifetime and angular correlation measurements. The promotion model was further questioned by Compton scattering data (Kornstädt *et al.*, 1980) and x-ray absorption measurements of the *L* edges (Lengeler *et al.*, 1983) that showed no substantial valence change between α- and γ-Ce. In disagreement with a metal-to-insulator Mott transition, photoemission experiments (Allen *et al.*, 1981) show the $f$ level is located between 2 and 3 eV below the Fermi energy in both phases, never crossing the Fermi level. Magnetic form factor (Murani *et al.*, 2005) and phonon densities of states (Manley *et al.*, 2003b) measurements also disagree with a metal-to-insulator Mott transition by showing that the magnetic moments remain localized in both phases. To date, the Kondo volume collapse, where the 4$f$ level is always below the Fermi energy and results in a localized 4$f$ magnetic moment, seems the most plausible scenario. Indeed, DMFT calculations of the optical properties of α− and γ-Ce (Haule *et al.*, 2005) are in agreement with the optical data of van der Eb *et al.* (2001), supporting the Kondo picture. Nonetheless, the Ce issue is still not yet resolved and there is theoretical evidence supporting a combination of all three effects (Held *et al.*, 2001)!

Similar arguments are made for Pu, which also sits directly on the localized-delocalized transition, but has five 5$f$ electrons rather than one 4$f$ electron as in the case of Ce. The mixed-level model (Eriksson *et al.*, 1999; Joyce *et al.*, 2003; Wills *et al.*, 2004) postulates that while all five 5$f$ electrons are actively bonding in α-Pu, there is only one actively bonding in δ-Pu with the other four electrons effectively localized. This argument has some similarity to the promotion model for Ce. DMFT provides a description of the electronic structure of strongly-correlated materials by treating both the Hubbard bands and quasiparticle bands on equal footing (Kotliar *et al.*, 2006). In strongly-correlated materials, there is a competition between the tendency towards delocalization of the bonding states, which leads to band formation, and the tendency towards localization, which leads to atomic-like behavior. This frustrated behavior is exactly what is shown for Ce and Pu in Fig. 6. The DMFT approach can discern between a metal-to-insulator Mott transition and a Kondo collapse scenario. Much like Ce, the exact physics driving the volume anomaly near Pu remains unanswered. Since Pu metal has necessitated the greatest advances in computational techniques for the actinide metals due its difficult and interesting physics, a detailed and historical perspective of DFT and DMFT will be presented in the Pu section in the later half of this review.

If the complex physics of the 5*f* states and the toxic nature of the materials are not enough, add the fact that most actinide metals accumulate damage over time due to self-induced radiation. This comes in the form of α, β, and γ decay that occur in different amounts, depending on element and isotope (Poenaru *et al.*, 1996). For example, α decay occurs in Pu, as shown in Fig. 9. In this process, a He atom is ejected with an energy of ~5 MeV and a uranium atom recoils with an energy of ~86 keV (Wolfer, 2002). The He atom creates little damage to the lattice; however, the U atom dislodges thousands of plutonium atoms from their normal positions in the crystal lattice, producing vacancies and interstitials know as Frenkel pairs. Within ~200 nanoseconds most of the Frenkel pairs annihilate, leaving a small amount of damage in the lattice. Over time, this damage accumulates in the form of defects, such as vacancies, interstitials, dislocations, and He bubbles (Hecker, 2004; Schwartz *et al.*, 2005). What this means is that not only are actinide metals intrinsically complicated, but lattice damage accumulates over time due to self irradiation, further complicating the physics of the materials.

This section is meant to serve as a general overview of the actinide series, where the physics is presented in a manner that is approachable for members of numerous scientific and engineering communities. More precise and detailed discussion of the electronic and magnetic structure of each elemental 5*f* metal, where various experimental and theoretical data are discussed, will be given in the later half of this review. Before this, we will compile all the $O_{4,5}$ and $N_{4,5}$ EELS edges, derivate and summarize the multi-electron atomic spectral calculations used to analyze the EELS data, and assemble the published inverse, valence-band, and 4*f* photoemission data. While there are several books on actinide physics and chemistry, such as Freeman and Darby (1974), Freeman and Lander (1985), and Morss *et al.*, (2006), we will strive throughout this review to reference the original literature publications when possible. Lastly of note, unpublished research that is in progress will be discussed and referenced at points throughout the review, since we feel presenting such data and ideas, albeit preliminary, does service for the community.

## II. ELECTRON ENERGY-LOSS SPECTROSCOPY

Before progressing to the EELS spectra, the question of why a TEM is used for spectral acquisition should be addressed. This has several answers. First, the TEM utilizes small samples, allowing one to avoid the handling of appreciable amounts of toxic and radioactive materials. The alternative is XAS performed at a multi-user synchrotron radiation facility, which is less well-adapted for the delicate and secure handling of radioactive materials. Second, the technique is bulk sensitive due to the fact that the electrons traverse ~40 nm of metal, this being the appropriate thickness for quality EELS spectra of actinide materials. A few nanometers of oxide do form on the surfaces of the TEM samples, but this is insignificant in comparison to the amount of metal sampled through transmission of the electron beam. Third, metals at or near the localized-itinerant transition in Fig. 3(a) exhibit numerous crystal structures that can coexist in metastable equilibrium due to close energy level between phases. Therefore, acquiring single-phase samples of metals at or near this transition, such as Mn (four phases), Ce (four phases), and Pu (six phases), is uncertain, making spectroscopic techniques with low spatial resolution questionable. Finally, actinide metals readily react with hydrogen and oxygen, producing many unwanted phases in the material during storage or preparation for experiments. The TEM has the spatial resolution to image and identify secondary phases (Hirsch *et al.*, 1977; Reimer, 1997; Fultz and Howe, 2001), ensuring examination of only the phase(s) of interest. An example of this is shown in Fig. 10, where (a) is a bright-field TEM image of an fcc $CmO_2$ particle in a dhcp α-Cm metal matrix, (b) is an [0001] diffraction pattern of the metal, and (c) is an [001] diffraction pattern of $CmO_2$. A field-emission-gun TEM, such as the one used in these experiments, can produce an electron probe of ~5 Å, meaning recording spectra from a single phase when performing experiments is easily achieved. Quantitatively measuring the reflections in the electron diffraction pattern in Fig. 10(b) and (c), as well as other crystallographic orientations, proves that the correct phase is examined (Zuo and Spence, 1992; Moore *et al.*, 2002a).

EELS spectra collected in the TEM can be compared to XAS spectra and many-electron atomic spectral calculations with complete confidence. At first this may not seem a reasonable comparison, since in XAS the transitions are electric-dipole whereas in EELS there is also an electric-quadrupole transition due to momentum transfer (Reimer, 1995; Egerton, 1996). However, the work by Moser and Wendin (1988; 1991) on the $O_{4,5}$ ($5d \rightarrow 5f$) and $N_{4,5}$ ($4d \rightarrow 5f$) edge of Th shows that as the energy of the incident electron is increased, the EELS spectral shape became more similar to XAS, and at around 2 keV they are close to identical. The incident energy of the TEM electron source used in these studies is 297 keV, ensuring the electron transitions are close to the electric-dipole limit. The use of apertures that remove high-angle Bragg and plural scattering (Moore *et al.*, 1999a; 1999b; 2002b) further refine the quality of EELS, providing spectra that are practically identical to XAS for transition metals (Blanche *et al.*, 1993), rare earths (Moore *et al.*, 2004b), and actinides (Moore *et al.*, 2006c). Looking specifically at actinides, direct comparison of Th and U $O_{4,5}$ edge can be made between EELS (Moore *et al.*, 2003; 2004a) and XAS (For Th: Cukier *et al.*, 1978; Aono *et al.*, 1981; for U: Cukier *et al.*, 1978; Iwan *et al.*, 1981). Similar comparisons can be made for the $N_{4,5}$ edge of U for EELS (van der Laan *et al.*, 2004) and XAS (Kalkowski *et al.*, 1987). In all cases, the EELS and XAS spectra are identical within error.

This equivalence between EELS and XAS is becoming stronger now that monochromated TEMs are available that have an energy resolution comparable to monochromatized synchrotron radiation (Walther and Stegmann, 2006; Lazar *et al.*, 2006). In fact, some TEMs can even resolve better than 50 meV (Bink *et al.*, 2003), which is close to the energy of phonon excitations. An example of monochromated EELS with 100 meV energy resolution as compared to synchrotron XAS is shown in Fig. 11. The $N_{4,5}$ ($4d \rightarrow 4f$) and $M_{4,5}$ ($3d \rightarrow 4f$) transitions of γ-Ce are shown for EELS spectra acquired in a monochromated TEM, XAS from the Advanced Light Source at the Lawrence Berkeley National Laboratory, and many-electron atomic spectral calculations (Moore *et al.*, 2004b). Note the similarity between spectra, particularly the two experimental spectra, which clearly illustrates that EELS in a TEM is practically identical to synchrotron-radiation-based XAS. Ironically, a monochromated source is not needed for EELS measurements on actinides, since the intrinsic core-hole lifetime broadening for the actinide $d \rightarrow f$ transitions is ~2 eV (Kalkowski *et al.*, 1987). Nonetheless, the comparison of EELS in a monochromated TEM and synchrotron-radiation-based XAS in Fig. 11 firmly demonstrates the equivalence of techniques when a high primary electron energy is utilized for EELS.

### A. The $O_{4,5}$ ($5d \rightarrow 5f$) edge

Having proven beyond reasonable doubt that a TEM can acquire EELS spectra that are directly comparable to XAS and many-electron atomic spectral calculations, we may proceed to compile and analyze the spectra. The $O_{4,5}$ edge of the α phase of Th, U, Np, Pu, Am, and Cm metal are shown in Fig. 12. Immediately noticeable is that the spectra for each elemental metal contain a broad edge, which is often referred to as the giant resonance (Wendin, 1984; 1987; Allen, 1987; 1992). There is also a smaller structure in Th, U, and Np that is normally referred to as a pre-peak. The giant resonance is ill-defined because the core $5d$ spin-orbit interaction is smaller than the core-valence electrostatic interactions in the actinide $O_{4,5}$ ($5d \rightarrow 5f$) transition (Ogasawara et al., 1991; Ogasawara and Kotani, 1995, Ogasawara and Kotani, 2001; Moore and van der Laan, 2007; Butterfield et al., 2008).  This effectively smears out the transitions, encapsulating both the $O_4$ ($5d_{3/2}$) and $O_5$ ($5d_{5/2}$) peaks within the giant resonance, thus making differentiation between them difficult if not impossible. In other words, the pre-peak is not a dipole-allowed transition, since these are contained within the giant resonance (Moore and van der Laan, 2007; Butterfield et al., 2008).

The shape of the rare earth $N_{4,5}$ ($4d \rightarrow 4f$) edges (Ogasawara and Kotani, 1995; Starke et al., 1997) and the $3d$ transition metal $M_{2,3}$ ($3p \rightarrow 3d$) edges (van der Laan, 1991b) are similar to the actinide $O_{4,5}$, due to the fact that each exhibit a giant resonance. In all three of these cases, the core-valence electrostatic interactions dominate the core spin-orbit interaction. The $4f$ metals show a pre-peak structure that is similar to the light actinides and is rather insensitive to the local environment (Dehmer et al., 1971; Starace, 1972; Sugar, 1972). The $3d$ metals show a pre-peak structure that is strongly dependent on the crystal field and hybridization (van der Laan, 1991a). Since the $5f$ localization is between those of $4f$ and $3d$, the $O_{4,5}$ pre-peak behavior for the actinides is expected to show only a mild dependence on the environment. The degree of dependence should change across the actinide series, since the $5f$ states of U are more delocalized than the those of Am and Cm. However, there are no pre-peaks in Pu, Am, and Cm, so there is no way to validate this assertion. There is a slight change in the pre-peak structure of the $O_{4,5}$ edge between α-U and $UO_2$, where a small shoulder appears on the high energy side of the peak at about 98 eV in $UO_2$ (Kalkowski et al., 1987, Moore and van der Laan, 2007).

The presence of the pre-peak in the actinide $O_{4,5}$ edge up to but not including Pu is interpreted by Moore et al. (2003) as being due to a failure of LS coupling in the $5f$ states of α- and δ-Pu. They argue that the loss of the pre-peak at Pu is due to filling of the $j = 5/2$ level during the process of EELS or XAS. Assuming that Pu is operating in $jj$ coupling with 5 electrons in the $j = 5/2$ (which can hold only 6 electrons), the level is filled when the $5f$ occupation goes from 5 to 6 during the $d^{10}5f^5_{5/2} \rightarrow d^9 5f^6_{5/2}$ transition. The filling of the $j = 5/2$ level shuts off the angular momentum coupling between partially occupied $5d$ and $5f$ states, removing the possibility of pre-peak(s).

The idea is based on the $N_{4,5}$ edge of the rare earth metals, which also consist of pre-peaks and a giant resonance.  The $N_{4,5}$ edge structure is explained by the Coulomb and exchange interactions between the partially occupied $4d$ and $4f$ final state levels, which drive the splitting on the scale of 20 eV between the angular-momentum-coupled states (Dehmer et al., 1971; Starace, 1972; Sugar, 1972). The pre-peaks in the $N_{4,5}$ edge persist across the rare earth series until $4f^{13}$, where only a single line is observed (Johansson et al., 1980). Finally, Yb ($4f^{14}$) shows no edge due to a filled $4f$ state. Because the pre-peaks persist across the rare earth series, the only filling effects that shut off the angular momentum coupling between the partially occupied $d$ and $f$ states are observed at the end of the series. However, the $O_{4,5}$ pre-peak in the actinide series disappears at Pu because the large $5f$ spin-orbit interaction splits the $j = 5/2$ and $7/2$ levels, causing a preferential filling of the $5/2$ level (intermediate or $jj$ coupling).

The argument of Moore *et al.* (2003) concerning the angular-momentum coupling of the Pu $5f$ states appears to be correct, since it has been subsequently supported by the $4d \rightarrow 5f$ transition for EELS and XAS (van der Laan *et al.*, 2004; Moore *et al.*, 2007a; 2007b). However, the specific interpretation of the pre-peak structure has become seemingly more complicated than first thought. For instance, it should be expected that pre-peak structure will return in the $O_{4,5}$ EELS and XAS spectra of Am and beyond due to angular momentum coupling between the partially occupied $5d$ and $5f_{7/2}$ states (Moore *et al.*, 2004a). However, none of the $O_{4,5}$ EELS edges after Np in Fig. 12 show pre-peak structures. Could it be that the $O_{4,5}$ edges for Am and Cm have numerous and small pre-peaks, similar to the early rare-earth $N_{4,5}$ edges, and that these are effectively lost by the 2 eV core-hole lifetime broadening of the main peak in the actinide $d \rightarrow f$ transitions? The prepeaks have a narrow linewidth when the corresponding final states have a long life time, i.e., have no decay channels. However, if these states can interact with those of the main peak the lifetime broadening becomes large.

Examining the electric-dipole transitions $5d\ ^{10}5f^{n} \rightarrow 5d\ ^{9}5f^{n+1}$ with and without $5d$ core spin-orbit interaction by means of atomic multiplet calculations further complicate the situation (Moore and van der Laan, 2007; Butterfield *et al.*, 2008). Results of the behavior of the giant resonance and pre-peak of the $O_{4,5}$ EELS edges for Th, U, and Pu show that when the $5d$ spin-orbit interaction is switched off, the pre-peak structure vanishes, meaning the pre-peak(s) are a consequence of first-order perturbation by the $5d$ spin-orbit interaction. This result is most clear for $5f$ counts of 0 and 1, but becomes rapidly more complicated for higher values of $5f$ count. Nonetheless, examining the calculated $O_{4,5}$ edges for $5f$ counts of 0, 1, 2, and 5 show that in all cases the pre-peak intensity increases with the size of the $5d$ spin-orbit interaction relative to the electrostatic interactions, while the angular quantum number for the $5f$ states ($j = 7/2$ or $5/2$) strongly influences the precise spectral shape of the pre-peak structure and the position of the giant resonance. Thus, the $O_{4,5}$ pre-peak size and structure is dependent on the spin-orbit interaction of *both* the $5d$ and $5f$ states.

## B. The $N_{4,5}$ ($4d \rightarrow 5f$) edge

For the actinide $N_{4,5}$ ($4d \rightarrow 5f$) edge, the situation is completely opposite of the $O_{4,5}$ transition: the core spin-orbit interaction is dominant over the electrostatic interaction (Moore and van der Laan, 2007). Because of this, the spin-orbit split white-lines of the $N_4(d_{3/2})$ and $N_5(d_{5/2})$ states are clearly resolved. The $N_{4,5}$ edge for the α phase of Th, U, Np, Pu, Am, and Cm metal is shown in Fig. 13, where each spectrum is normalized to the $N_5$ ($4d_{5/2}$) peak height. Immediately noticeable is the gradually growing separation between the $N_4$ and $N_5$ peaks from Th to Cm, in pace with the increase in $4d$ spin-orbit splitting with atomic number. Also noticeable is that the $N_4$ ($4d_{3/2}$) peak reduces in intensity going from Th to Am, but then the trend reverses, giving a larger intensity for Cm. The behavior of the $N_4$ peak in the EELS spectra in Fig. 13 directly reflects the filling of the angular momentum levels in the $5f$ state. Selection rules govern that a $d_{3/2}$ electron can only be excited into an empty $f_{5/2}$ level, which means that the ratio of the $N_4$ ($d_{3/2}$) and $N_5$ ($d_{5/2}$) peak intensities serves as a measure for the relative occupation of the $5f_{5/2}$ and $5f_{7/2}$ levels. The $N_4$ peak reduces rapidly as the atomic number increases because the majority of the 5$f$ electrons are occupying the $f_{5/2}$ level. By the time Am is reached, the $N_4$ peak is almost extinct because the $f_{5/2}$ level is close to full with six 5$f$ electrons (there is only a minor amount of electrons in the $f_{7/2}$ level). Thus, there is little room for an electron from $d_{3/2}$ to be excited into the $5f_{5/2}$ level. For Cm, the $N_4$ peak then increases relative to the $N_5$ peak because the 5$f$ electron occupation is spread out in the $j = 5/2$ and 7/2 levels, becoming more *LS*-like.

The branching ratio, $B = I(N_5)/[I(N_5) + I(N_4)]$, where $I(N_4)$ and $I(N_5)$ are the integrated intensities in the $N_4$ ($4d_{3/2}$) and $N_5$ ($4d_{5/2}$) edges, respectively, is extracted by calculating the second derivative of the EELS spectra and integrating the area beneath the peaks above zero (Fortner and Buck, 1996; Wu *et al.*, 2004; Yang e*t al.*, 2006). This technique is beneficial because it reduces the signal-to-noise ratio in the spectrum and circumvents the need to remove the background intensity with an inverse power-law extrapolation (Egerton, 1996; Williams and Carter, 1996). The branching ratio for the α phase of Th, U, Np, Pu, Am, and Cm metal are shown in Table I. Full analysis of the $N_{4,5}$ ($4d \rightarrow 5f$) EELS edges in Fig. 13 will later be performed in conjunction with many-electron atomic spectral calculations and the spin-orbit sum rule. For this reason, the detailed discussion of the spectra will be handled in the subsequent theory section.

Before turning our attention to theory, we would like to consider a few relevant topics. First, the $M_{4,5}$ ($3d \rightarrow 5f$) EELS are known for several Th, U, and Pu materials, meaning the branching ratio of the $M_4$ ($3d_{3/2}$) and $M_5$ ($3d_{5/2}$) white-line peaks has been extracted and analyzed (Fortner and Buck, 1996; Buck and Fortner, 1997; Fortner *et al.*, 1997; Buck *et al.*, 2004; Colella *et al.*, 2005). Their data, which is also acquired using a TEM, show that the $M_{4,5}$ edge is sensitive to changes in the environment of the actinide element, exhibiting changes in the branching ratio for various *f*-electron materials. This is in agreement with the sensitivity of the $N_{4,5}$ EELS edge for U and Pu materials (Moore *et al.*, 2006b). What is more, the branching ratio for both the $N_{4,5}$ and $M_{4,5}$ edges of U (Kalkowski *et al.*, 1987) is in accordance with the EELS results. In particular, the branching ratio from the $N_{4,5}$ XAS edge is 0.676, while for EELS it is 0.686. Interestingly, the branching ratio in EELS is systematically higher than in XAS, usually by about 0.01 (van der Laan *et al.*, 2004). This is a small difference, but repeatedly appears when the $N_{4,5}$ branching ratio is extracted from both EELS and XAS. The branching ratio of the $M_{4,5}$ edge of U is similar to the $N_{4,5}$ edge, since in both cases the influence of the core-hole interaction on the 5$f$ states is small compared to the core hole spin-orbit splitting. Thus, the overall picture of EELS and XAS using both the $N_{4,5}$ and $M_{4,5}$ edges is in accordance for the actinide metals.

It is possible to measure the element-specific local magnetic moment by using polarized x-rays. The difference between the absorption spectra measured using x-rays of opposite polarization with the beam along the sample magnetization direction gives the magnetic x-ray dichroism (Thole *et al.*, 1985a; van der Laan *et al.*, 1986a). For magnetic materials this effect is very large at the transition metal $L_{2,3}$ edges (van der Laan and Thole, 1991a) and rare earth $M_{4,5}$ edges (van der Laan *et al.*, 1986a; Goedkoop *et al.*, 1988). Sum rules allow one to extract the expectation values of the spin and orbital magnetic moments in the ground state (Thole *et al.*, 1992, van der Laan, 1998). This effect has also been used to study uranium compounds, where the strong 5$f$ spin-orbit interaction gives rise to a large contribution of the magnetic dipole term to the effective spin magnetic moment (Collins *et al.*, 1995; Yaouanc *et al.*, 1998). Recently, the potential to do magnetic circular dichroism experiments in a TEM, *without* a spin-polarized electron source was illustrated by Schattschneider *et al.* (2006; 2007). Such experiments could be done in a standard TEM with a field-emission electron source using correct scattering conditions with appropriate geometry and apertures. The actinide $N_{4,5}$ edge would be the most applicable EELS edge for such analysis given the reasonable intensity compared to the $M_{4,5}$ edge and the

~40 eV spin-orbit splitting of the 4*d* states. At present, dichroic experiments in a TEM can achieve a 40 nm spatial resolution, with the prospect of reaching 10 nm (Schattschneider *et al.*, 2007), meaning actinide magnetism could be investigated at the nanoscale. Circular dichroism experiments in a TEM are still in the "proof of principle" phase and need more work to become a viable and robust technique. In addition, a TEM is unlikely to have the strong dichroism signal that is available using synchrotron radiation. However, the possibility of circular dichroism experiments in a TEM opens many avenues for future lab-based experiments with actinides.

### III. MANY-ELECTRON ATOMIC SPECTRAL CALCULATIONS

In Sec. III A we treat the influence of the spin-orbit and electrostatic interaction on the electronic configuration in the *jj*, *LS*, and intermediate coupling schemes. In this context it is important to be aware of the difference between "coupled eigenstates" and "coupled basis states", e.g., the *LS*-coupled Hund's ground state can be written in *jj* coupled basis states, in which case it will have off-diagonal matrix elements. Both the spin-orbit interaction and electrostatic interaction are important in the case of the 5*f* electrons, in which case we have intermediate coupled eigenstates that can be written in either *LS* or *jj* coupled basis states. In the former the electrostatic interaction is diagonal and in the latter the spin-orbit interaction is diagonal. Recoupling coefficients can be used to switch between *LS*-coupled and *jj*-coupled basis states. We present examples for the two-particle state: $f^2$ and $f^{12}$. Furthermore, general relations are given for the expectation values of the spin-orbit operator and occupation numbers of the *j* levels. Calculated ground state expectation values for all actinide elements are given in the case of each coupling scheme.

In Sec. III B we give a general derivation of the spin-orbit sum rule, which relates the angular dependent part of the spin-orbit interaction to the EELS or XAS branching ratio, i.e., the intensity ratio of the core *d* spin-orbit split *j*-manifolds in the $f^n \rightarrow d^9 f^{n+1}$ transition. The applicability of the sum rule to the actinides is critically discussed.

In Sec. III C we show how the multiplet calculations are performed. As an example, we present the $5f^0 \rightarrow d^9 5f^1$ transition for the $O_{4,5}$, $N_{4,5}$, and $M_{4,5}$ edges, corresponding to the 5*d*, 4*d* and 3*d* core levels, respectively. Numerical results for other $f^n$ configurations are given using tables and figures.

### A. Ground-state Hamiltonian

For *n* electrons moving about a point nucleus of charge the Hamiltonian can be written in the *central field approximation* as

$$H = H_{el} + H_{so} , \qquad (1)$$

where $H_{el}$ and $H_{so}$ are the electrostatic and spin-orbit interaction, respectively (Cowan, 1968; 1981). Other interactions, such as crystal field, are usually much smaller, leading only to small perturbations. The interaction can be separated in an *angular and radial part*. The angular part depends on the angular quantum numbers of the basis states of the configuration and are independent of the radial wave functions. General analytical methods for calculating these coefficients have been developed by Racah (1942; 1943; 1949) and computerized by Cowan (1968). The basis wave functions are assumed to be an antisymmetrized product of one-electron functions. These wave functions are eigenfunctions of the total angular momentum *J* and its component $M_J$. The states are characterized by quantum numbers $\alpha LS$, where $\alpha$ is a suitable quantity for distinguishing between terms having the same values of the orbital and spin angular momenta, $L$ and $S$.

As said, Racah algebra offers powerful tools to analyze the angular part, and in this respect coupled tensor relations are particularly useful for both the spin-orbit and electrostatic interactions. In the treatment of the Hamiltonian, we use the fact that a scalar product $[\mathbf{T}^{(k)} \cdot \mathbf{U}^{(k)}]^0$ of the multipole tensor operators $\mathbf{T}^{(k)}$ and $\mathbf{U}^{(k)}$ with rank $k$ that act separately on parts $a$ and $b$ of the system, such as spin and orbital space, or on different particles like in the case of Coulomb interaction, can be written as

$$\langle \alpha j_a j_b JM | \mathbf{T}^{(k)} \cdot \mathbf{U}^{(k)} | \alpha' j'_a j'_b J'M' \rangle = \delta_{J,J'} \delta_{M,M'} (-1)^{j'_a+j_b+J} \begin{Bmatrix} j'_a & j'_b & J \\ j_b & j_a & k \end{Bmatrix}$$
$$\times \sum_{\alpha''} \langle \alpha j_a \| T^{(k)} \| \alpha'' j'_a \rangle \langle \alpha'' j_b \| U^{(k)} \| \alpha' j'_b \rangle \quad (2)$$

Coming back to the Hamiltonian in Eq. (1), two different basis sets, $LS$ and $jj$ coupled wave functions, are of particular interest. The electrostatic interaction is diagonal in $LS$ coupling whereas the spin-orbit interaction is diagonal in $jj$ coupling. In the $LS$-coupling scheme the various one-electron orbital momenta $\ell$ are coupled together successively to give a total orbital momentum, and the various one-electron spin momenta $s$ are coupled to give a total spin,

$$\{[((\ell_a s_a) L_a S_a, \ell_b s_b) L_b S_b, \cdots, \ell_n s_n] L_n S_n\} J_n \,, \quad (3)$$

with triangulation rules such as $L_b = |L_a - \ell_b|, \ldots L_a + \ell_b$. In the other scheme of $jj$-coupling, each $\ell$ and $s$ are coupled to give a total angular momentum $j$, and the various $j$ are then coupled to give successive values of $J$,

$$\{[(\ell_a s_a j_a) J_a, (\ell_b s_b j_b)] J_b, \cdots, (\ell_n s_n j_n)\} J_n \,. \quad (4)$$

For a given electronic configuration we arrive in both coupling schemes at the same set of allowed values of the total angular momentum, $J_n = |L_n - S_n|, \ldots, L_n + S_n$. This means that the Hamiltonian of Eq. (1) is block-diagonal in $J$. For each $J$ block the states can be transformed between $LS$ and $jj$-coupling using recoupling coefficients that can be expressed in terms of 9-$j$ symbols,

$$\langle [(\ell_a \ell_b) L, (s_a s_b) S] J | [(\ell_a s_a) j_a, (\ell_b s_b) j_b] J \rangle = [L, S, j_a, j_b]^{1/2} \begin{Bmatrix} \ell_a & \ell_b & L \\ s_a & s_b & S \\ j_a & j_b & J \end{Bmatrix} \,, \quad (5)$$

where $[x, y, \ldots] \equiv (2x+1)(2y+1)\ldots$. These coefficients form the transformation matrix $T^J_{LS, j_a j_b}$.

## 1. Spin-orbit interaction

The spin-orbit interaction for the $\ell$-shell is given by a one-electron operator

$$H_{so} = \zeta_\ell(r) \sum_{i=1}^{n} \mathbf{l}_i \cdot \mathbf{s}_i, \tag{6}$$

where $\mathbf{l}_i$ and $\mathbf{s}_i$ are the orbital and spin angular momentum operators of the $i$-th electron of the $\ell^n$ configuration. In the following we will write for brevity the angular part as

$$\mathbf{l} \cdot \mathbf{s} \equiv \sum_{i=1}^{n} \mathbf{l}_i \cdot \mathbf{s}_i. \tag{7}$$

The Hamiltonian $H_{so}$ commutes with $\mathbf{J}^2$ and $J_z$ and is hence diagonal in $J$ and independent of the magnetic quantum number $M_J$. It does not commute with $\mathbf{L}^2$ or $\mathbf{S}^2$ and can thus couple states of different $LS$ quantum numbers. The spin-orbit coupling constant $\zeta_\ell$ is defined as the radial integral

$$\zeta_\ell = \tfrac{1}{2}\alpha^2 \int_0^\infty R^2(r) \frac{dV}{dr} r\, dr, \tag{8}$$

where $\alpha \approx 1/137$ is the fine-structure constant. Hartree-Fock values of $\zeta_\ell$ for representative elements of the various transition metal series, Cm $5f^7$, Gd $4f^7$, and Mn $3d^5$, are given in Table II. It is seen that the spin-orbit parameter for rare earths is about five times larger than for $3d$ metals. For actinides the spin-orbit parameters are about twice as large as for rare earths metals. The spin-orbit coupling constant $\zeta_\ell$ usually requires hardly any scaling for comparison with experimental results; the values calculated using the method by Watson and Blume (1965) are quite accurate for both core and valence electrons.

For the spin-orbit interaction of a single electron, application of Eq. (2) gives

$$\langle \ell s j | \mathbf{l} \cdot \mathbf{s} | \ell s j \rangle = (-1)^{j+\ell+s} \begin{Bmatrix} \ell & \ell & 1 \\ s & s & j \end{Bmatrix} \langle s \| s^{(1)} \| s \rangle \langle \ell \| \ell^{(1)} \| \ell \rangle, \tag{9}$$

where the reduced-matrix elements are

$$\langle s \| s^{(1)} \| s \rangle = [s(s+1)(2s+1)]^{1/2},$$

$$\langle \ell \| \ell^{(1)} \| \ell \rangle = [\ell(\ell+1)(2\ell+1)]^{1/2}. \tag{10}$$

Explicit evaluation of the 6$j$-symbol in Eq. (9) combined with Eq. (10) results in

$$\langle \ell s j | \mathbf{l} \cdot \mathbf{s} | \ell s j \rangle = \frac{1}{2}[j(j+1) - \ell(\ell+1) - s(s+1)]. \tag{11}$$

Thus, the spin-orbit interaction splits the $\ell$ state ($\ell \neq 0$) into a doublet $\ell \pm s$ with spin-orbit expectation values

$$\langle \ell s j | \mathbf{l} \cdot \mathbf{s} | \ell s j \rangle = \begin{cases} -\frac{1}{2}(\ell+1) & \text{for } j_1 = \ell - s \\ \frac{1}{2}\ell & \text{for } j_2 = \ell + s \end{cases}, \quad (12)$$

with energies

$$E_j = \langle \ell s j | \mathbf{l} \cdot \mathbf{s} | \ell s j \rangle \zeta_\ell. \quad (13)$$

The energy separation between these two $j$ levels is ($\ell \neq 0$)

$$E_{j_2} - E_{j_1} = \frac{1}{2}(2\ell+1)\zeta_\ell, \quad (14)$$

and the weighted average energy is

$$\sum_{i=1,2}(2j_i+1)E_{j_i} = 0, \quad (15)$$

where $(2j_i+1)$ is the degeneracy of the $j_i$ level, which is equal to the number of components $m_j = -j, \ldots, j$, so that $2j_1+1 = 2\ell$ and $2j_2+1 = 2\ell+2$.

Equation (12) leads to a useful general expression for $\langle \mathbf{l} \cdot \mathbf{s} \rangle$, which is valid in intermediate coupling, including the *LS* and *jj* coupling limits. Since the spin-orbit operator is always block diagonal in *J*, there are no cross terms between different *J* values, nor between different *j* values. For the configuration $\ell^n$ with $n = n_{j_1} + n_{j_2}$, where $n_{j_1}$ and $n_{j_2}$ are the number of electrons in the levels $j_1$ and $j_2$, respectively, application of Eq. (12) gives

$$\langle \ell^n J | \mathbf{l} \cdot \mathbf{s} | \ell^n J \rangle = \sum_{j=j_1, j_2} \langle j | \mathbf{l} \cdot \mathbf{s} | j \rangle n_j = -\frac{1}{2}(\ell+1)n_{j_1} + \frac{1}{2}\ell n_{j_2}. \quad (16)$$

This result is independent of the specific value of *L*, *S*, and *J*.

To give an example of Eq. (16) for the *f* shell, consider when the $f_{5/2}$ and $f_{7/2}$ level have $\langle \mathbf{l} \cdot \mathbf{s} \rangle$ = -2 and 3/2, respectively. For $f^2$ the *jj* coupled basis functions (5/2,5/2), (5/2,7/2), and (7/2,7/2) have $\langle \mathbf{l} \cdot \mathbf{s} \rangle$ = -4, -1/2, and 3, respectively, independent of the value of *J* which ranges from $j_2 - j_1$ to $j_2 + j_1$. In the case of *LS* and intermediate coupling, $n_{j_1}$ and $n_{j_2}$ are no longer restricted to half-integer values. An arbitrary state of the configuration $\ell^2$ is given by

$$\psi(\ell^2) = c_{11}\psi(j_1, j_1) + c_{12}\psi(j_1, j_2) + c_{22}\psi(j_2, j_2), \quad (17)$$

where the *c*'s are wavefunction coefficients, with $n_{j_1} = 2c_{11}^2 + c_{12}^2$ and $n_{j_2} = c_{12}^2 + 2c_{22}^2$, so that

$$\langle\psi|\mathbf{l}\cdot\mathbf{s}|\psi\rangle = -\tfrac{1}{2}(\ell+1)n_{j_1} + \tfrac{1}{2}\ell n_{j_2} = -(\ell+1)c_{11}^2 - \tfrac{1}{2}c_{12}^2 + \ell c_{22}^2. \tag{18}$$

The expectation value $\langle\mathbf{l}\cdot\mathbf{s}\rangle$ for the ground state of $\ell^n$ is always negative because the spin-orbit interaction couples $\ell$ and $s$ antiparallel. Generally, $\langle\mathbf{l}\cdot\mathbf{s}\rangle < 0$ if $n_{j_1} > 2j_1 + 1$, i.e., if the number of electrons with $j_1 = \ell - s$ exceeds the statistical value.

## 2. Electrostatic interactions

Returning to the second term in Eq. (1), we see that for the electrostatic interactions of $n$ electrons in an atom with nuclear charge $Ze$ the non-relativistic Hamiltonian is [Condon and Shortley, 1963]

$$H_{el} = -\frac{\hbar^2}{2m}\sum_{i=1}^{n}\nabla_i^2 - \sum_{i=1}^{n}\frac{Ze^2}{r_i} + \sum_{i<j}^{n}\frac{e^2}{r_{ij}}. \tag{19}$$

The first term describes the kinetic energy of all electrons, the second gives the potential energy of all electrons in the potential of the nucleus, and the third describes the repulsive Coulomb potential of the electron-electron interaction.

Since the Schrödinger equation for the Hamiltonian with $n > 1$ is not exactly solvable, one makes the approximation that each electron moves independently in a central field build up from the nuclear potential and the average potential of all other electrons. The electron-electron interaction is taken as a perturbation potential. The matrix elements of this potential,

$$\left\langle \alpha S L J M_J \left| \sum_{i<j}^{n}\frac{e^2}{r_{ij}} \right| \alpha' S' L' J' M_J' \right\rangle = E_C + E_X, \tag{20}$$

are independent of the quantum numbers $J$ and $M_J$, and diagonal in $L$ and $S$, but not diagonal in $\alpha$, which represents additional quantum numbers required to fully specify the states. This means there are non-zero matrix elements between different states with the same $L$ and $S$ quantum numbers. The symbol $r_{ij}$ stands for the distance $|r_i - r_j|$ between electrons $i$ and $j$. In Eq. (20), $E_C$ and $E_X$ give the Coulomb and exchange energy, respectively. The Coulomb potential can be expanded in Legendre polynomials $P_k$ which can be written as,

$$\frac{1}{r_{ij}} = \sum_{k=0}^{\infty}\frac{r_<^k}{r_>^{k+1}}P_k(\cos\omega_{ij}) = \sum_{k=0}^{\infty}\frac{r_<^k}{r_>^{k+1}}\left[C_i^{(k)*}(\theta_1,\phi_1)\cdot C_j^{(k)}(\theta_2,\phi_2)\right], \tag{21}$$

where

$$C_q^{(k)}(\theta,\phi) \equiv \sqrt{\frac{4\pi}{2k+1}}Y_{kq}(\theta,\phi), \tag{22}$$

are reduced spherical harmonics, and $r_<$ and $r_>$ are the lesser and greater of the distance of the electrons $i$ and $j$ to the nucleus and $\omega$ gives the angle between these vectors. The two-electron integral can be expressed as

$$\left\langle n_a \ell_a, n_b \ell_b; SL \left| \frac{e^2}{r_{12}} \right| n_a \ell_a, n_b \ell_b; SL \right\rangle$$

$$= \sum_k \left[ f_k(\ell_a, \ell_b) F^k(n_a \ell_a, n_b \ell_b) + g_k(\ell_a, \ell_b) G^k(n_a \ell_a, n_b \ell_b) \right]$$  (23)

where $f_k$ and $g_k$ are the angular coefficients and $F^k$ and $G^k$ are the radial integrals of the matrix elements. Since the operators $C_1^{(k)}$ and $C_2^{(k)}$ act on different particles, we can use the coupled tensor relation in Eq. (2) to obtain

$$f_k(\ell_a, \ell_b) = (-1)^{\ell_a + \ell_b + L} \left\langle \ell_a \| C_1^{(k)} \| \ell_a \right\rangle \left\langle \ell_b \| C_2^{(k)} \| \ell_b \right\rangle \begin{Bmatrix} \ell_a & \ell_a & k \\ \ell_b & \ell_b & L \end{Bmatrix},$$  (24)

$$g_k(\ell_a, \ell_b) = (-1)^S \left\langle \ell_a \| C_1^{(k)} \| \ell_b \right\rangle \left\langle \ell_b \| C_2^{(k)} \| \ell_a \right\rangle \begin{Bmatrix} \ell_a & \ell_b & k \\ \ell_b & \ell_a & L \end{Bmatrix},$$  (25)

with reduced matrix elements

$$\left\langle \alpha \ell \| C^{(k)} \| \alpha' \ell' \right\rangle = \delta(\alpha, \alpha')(-1)^\ell [\ell, \ell']^{1/2} \begin{pmatrix} \ell & k & \ell' \\ 0 & 0 & 0 \end{pmatrix}.$$  (26)

The radial integrals, $F^k$ and $G^k$, of the electrostatic interaction can be treated as empirically adjustable quantities to fit the observed energy levels and their intensities, but are theoretically defined as the Slater integrals

$$F^k(n_a \ell_a, n_b \ell_b) = e^2 \int_0^\infty \frac{2 r_<^k}{r_>^{k+1}} R^2_{n_a \ell_a}(r_1) R^2_{n_b \ell_b}(r_2) \, dr_1 dr_2 ,$$  (27)

$$G^k(n_a \ell_a, n_b \ell_b) = e^2 \int_0^\infty \frac{2 r_<^k}{r_>^{k+1}} R_{n_a \ell_a}(r_1) R_{n_b \ell_b}(r_2) R_{n_a \ell_a}(r_2) R_{n_b \ell_b}(r_1) \, dr_1 dr_2 .$$  (28)

The direct integrals $F^k$ represent the actual electrostatic interaction between the two electronic densities of electrons $n_a \ell_a$ and $n_b \ell_b$. The exchange integrals $G^k$ arise due to the quantum mechanical principle that fermions are indistinguishable, so that the wave function is totally antisymmetric with respect to permutation of the particles. Consequently, the $G^k$ are zero for equivalent electrons and, hence, absent in the expression for the configuration $\ell^n$. The symmetry properties of the 3$j$-symbol in Eq. (26) give that $F^k(\ell, \ell)$ has non-zero values for $k = 0, 2, \cdots, 2\ell$, and $G^k(\ell, \ell')$ has non-zero values for $k = |\ell - \ell'|, \cdots, \ell + \ell'$.

Table II gives the calculated atomic Hartree-Fock values of the atomic radial parameters of the Slater integrals for representative elements of the various transition metal series. The values of the Slater integrals $F^k$ for the different metals are comparable in size. However, in the metal their value depends on the degree of delocalization of the valence electrons. In localized atomic systems the electrostatic and exchange parameters require a typical scaling to 80% of the Hartree-Fock value to account for interactions with configurations omitted in the calculation (Thole *et al.*, 1985b); however, in fully itinerant systems this can be drastically smaller (van der Laan, 1995).

Table III shows the parameter values of the Slater integrals (reduced to 80%) and the 5$f$ spin-orbit parameter across the actinide elements. As a rule, $F^2 > F^4 > F^6$ (Cowan, 1981). It is seen that while the Slater integrals increase by 45% from $5f^2$ to $5f^{12}$, the 5$f$ spin-orbit interaction increases by 200%. Hence, in intermediate coupling the relative importance of the spin-orbit interaction increases along the series.

### 3. *LS*-coupling scheme

We will evaluate here the expectation value of the spin-orbit interaction in *LS* coupling, which is not zero when $L$ and $S$ are coupled to a given $J$ within the allowed range. For a state $|\alpha LSJM_J\rangle$, the spin-orbit interaction is diagonal in $J$ and independent of $M_J$, but is not diagonal in $\alpha$, $L$, and $S$, so that states with different $L$ and $S$ are coupled. The matrix elements can be written as

$$\langle \alpha LSJM_J | \mathbf{l} \cdot \mathbf{s} | \alpha' L'S'J'M_{J'}\rangle = \delta_{JJ'}\delta_{M_J M_{J'}}(-1)^{L'+S+J}[\ell(\ell+1)(2\ell+1)]^{1/2} \\ \times \begin{Bmatrix} S & L & J \\ L' & S' & 1 \end{Bmatrix} (\alpha LS\|V^{11}\|\alpha'L'S') \qquad (29)$$

Thus, the dependence of the interaction on $J$ is given by the 6-$j$ symbol, while the dependence on the other quantum numbers is given by Racah's double-tensor operator $V^{11}$ with reduced matrix elements

$$(\alpha LS\|V^{11}\|\alpha'L'S') = \sum_{i=1}^{2} \langle s_1 s_2 S\|s_i^{(1)}\|s_1 s_2 S'\rangle \langle \ell_1 \ell_2 L\|l_i^{(1)}\|\ell_1 \ell_2 L'\rangle \quad , \qquad (30)$$

where

$$\langle s_1 s_2 S\|s_i^{(1)}\|s_1 s_2 S'\rangle = [s(s+1)(2s+1)(2S+1)(2S'+1)]^{1/2} \begin{Bmatrix} S & 1 & S' \\ s & s & s \end{Bmatrix} \qquad (31)$$

and a similar expression for $\langle \ell_1 \ell_2 L\|l_i^{(1)}\|\ell_1 \ell_2 L'\rangle$.

The lowest energy corresponds to the so-called Hund's rule ground state, which has maximum values of $L$ and $S$. For a more than half-filled shell, the ground state has $J = L+S$. For a less than half-filled shell, $J = |L–S|$. The value of $\langle \mathbf{l} \cdot \mathbf{s} \rangle$ for the ground state is always negative, because the spin-orbit interaction couples $\ell$ and $s$ antiparallel.

Useful expressions can be given for the ground state of a free atom. In *LS* coupling, $L$, $S$, and $J$ are good quantum numbers, and explicit evaluation of the 6-$j$ symbol in Eq. (29) results in

$$\left\langle \alpha LSJ \middle| \mathbf{l}\cdot\mathbf{s} \middle| \alpha LSJ \right\rangle = \frac{E_{\alpha LSJ} - E_{\alpha LS}}{\zeta} = \frac{1}{2}\left[J(J+1) - L(L+1) - S(S+1)\right]\frac{\zeta(\alpha LS)}{\zeta}, \qquad (32)$$

where $\zeta(\alpha LS)$ is the effective spin-orbit splitting factor and $E_{\alpha LSJ} - E_{\alpha LS}$ is the energy dependence in a spin-orbit-split $LS$ term. For $LS$ terms of maximum spin, Eq. (32) is reduced with

$$\frac{\zeta(\alpha LS)}{\zeta} = \begin{cases} n^{-1} & \text{for } n < 2\ell + 1 \\ 0 & \text{for } n = 2\ell + 1 \\ -n_h^{-1} & \text{for } n > 2\ell + 1 \end{cases}, \qquad (33)$$

where $n_h = 4\ell + 2 - n$ is the number of holes in the $\ell$ shell. For the Hund's rule ground state we have

$$\left\langle \alpha LSJ \middle| \mathbf{l}\cdot\mathbf{s} \middle| \alpha LSJ \right\rangle_{\text{Hund}} = \begin{cases} -(L+1)S/n & \text{for } J = |L-S| \text{ if } L \geq S \\ -L(S+1)/n & \text{for } J = |L-S| \text{ if } S \geq L \\ -LS/n_h & \text{for } J = L + S \end{cases}. \qquad (34)$$

Note that $\mathbf{l}\cdot\mathbf{s} \equiv \sum_{i=1}^{n} \mathbf{l}_i \cdot \mathbf{s}_i$ is an $n$-particle operator, thus $\langle \mathbf{l}\cdot\mathbf{s} \rangle$ is not per electron/hole, despite the deceiving appearance in Eq. (34). The expectation values for the Hund's state of all actinide elements are listed in Table IV.

**4. *jj*-coupling scheme**

The *jj* coupling model is appropriate when the electrostatic interactions are weak compared to the spin-orbit interaction. In *jj*-coupling, first all $j = \ell - s$ levels are filled before the $j = \ell + s$ levels get their fair share. The total angular momentum $J$ is a good quantum number and for the ground state its value is the same as in $LS$ coupling, namely $J = |L - S|$ and $L + S$ for less and more than half-filled shell, respectively.

For the one-particle state, the *jj* and $LS$ coupled state is obviously one and the same state (i.e., $^2F_{5/2}$ and $^2F_{7/2}$ for $f^1$ and $f^{13}$, respectively, using the spectroscopic notation $^{2S+1}L_J$). However, they are different for a many-particle state. To make this explicit, let us examine the two-particle case in more detail. The allowed $LS$ states for $f^2$ are $^1S, ^1D, ^1G, ^1I, ^3P, ^3F,$ and $^3H$. In *jj*-coupling, there are two electrons in the $j = 5/2$ level, which in the ground state couple to a total angular momentum $J = 4$. This level is contained in $^1G_4, ^3F_4,$ and $^3H_4$. The *jj* coupled states can be transformed to $LS$ coupled states using Eq. (5), which gives a transformation matrix ($^3F_4, ^1G_4, ^3H_4$) → [(7/2,7/2)$_4$, (5/2,5/2)$_4$, (7/2,5/2)$_4$] equal to

$$T = \frac{1}{7} \begin{pmatrix} 2\sqrt{\frac{22}{3}} & 3\sqrt{2} & -\sqrt{\frac{5}{3}} \\ -\frac{2}{\sqrt{3}} & \sqrt{11} & \sqrt{\frac{110}{3}} \\ \sqrt{\frac{55}{3}} & -2\sqrt{5} & 4\sqrt{\frac{2}{3}} \end{pmatrix}, \quad (35)$$

which gives a $jj$ coupled ground state

$$f^2 : \psi(5/2,5/2)_4 = -\frac{2}{7\sqrt{3}} \psi(^3F_4) + \frac{1}{7}\sqrt{11}\,\psi(^1G_4) + \frac{1}{7}\sqrt{\frac{110}{3}}\psi(^3H_4). \quad (36)$$

The character is obtained from the square of the wave function coefficient, which gives 2.7% $^3F_4$, 22.5% $^1G_4$, and 74.8% $^3H_4$.

For $f^{12}$ with two holes in the $f$ shell the allowed $LS$ states are the same as for $f^2$. In the ground state, the two holes with $j = 7/2$ couple to $J = 6$, which is contained in $^3H_6$ and $^1I_6$. The transformation matrix $(^3H_6, {}^1I_6) \rightarrow [(7/2,7/2)_6, (7/2,5/2)_6]$ is given by

$$T = \begin{pmatrix} \sqrt{\frac{6}{7}} & \sqrt{\frac{1}{7}} \\ \sqrt{\frac{1}{7}} & \sqrt{\frac{6}{7}} \end{pmatrix}, \quad (37)$$

which results in

$$f^{12} : \psi(7/2,7/2)_6 = \sqrt{\frac{6}{7}}\,\psi(^3H_6) + \sqrt{\frac{1}{7}}\,\psi(^1I_6). \quad (38)$$

The character of this state is 85.7% $^3H_6$ and 14.3% $^1I_6$.

The two examples above show that when going from the $LS$ to the $jj$ coupled ground state, other $LS$ states are mixing in. This leads to a ground state that contains a significant singlet spin, i.e., this amount of low-spin character is needed to produce the $jj$-coupled ground state.

The transformation matrix in Eq. (35) can also be used to express the $LS$ coupled ground state in $jj$ coupled basis states,

$$f^2 : \psi(^3H_4) = -\frac{1}{7}\sqrt{\frac{5}{3}}\,\psi(7/2,7/2) + \frac{1}{7}\sqrt{\frac{110}{3}}\,\psi(5/2,5/2) + \frac{4}{7}\sqrt{\frac{2}{3}}\,\psi(5/2,7/2), \quad (39)$$

$$f^{12}(^3H_6) = \sqrt{\frac{6}{7}}\,\psi(7/2,7/2) + \sqrt{\frac{1}{7}}\,\psi(7/2,5/2). \quad (40)$$

Using Eq. (18), we find for $f^2 : \psi(^3H_4)$ that $n_{5/2} = 1.714$, $n_{7/2} = 0.286$, and $\langle \mathbf{l} \cdot \mathbf{s} \rangle = -3$. The latter value is the same as obtained from $-L(S+1)/n$ [Eq. (34)]. For $f^{12} : \psi(^3H_6)$ we find that $n_{5/2}^h = 0.143$, $n_{7/2}^h = 1.857$, and $\langle \mathbf{l} \cdot \mathbf{s} \rangle$ = -2.5, which is the same value as obtained from $-LS/n_h$ in Eq. 34. For all $f^n$ configurations the electron occupation numbers are given in Table IV. Thus, when going from the $jj$ coupled ground state to the $LS$ coupled ground state, other $j$ states are mixed in. The ground state of intermediate coupling, which is discussed next, has occupation numbers that are between these two extreme cases.

## 5. Intermediate-coupling scheme

We are now ready to confront both terms of the Hamiltonian in Eq. (1) simultaneously. In intermediate coupling, both spin-orbit and electrostatic interactions are taken into account by choosing appropriate values of the radial parameters for the configuration at hand. Hence, it can be expected that this coupling will provide excellent results for realistic situations, especially in the case of actinides where both interactions are equally important. However, an analytical separation of the total Hamiltonian into angular and radial part, although straightforward, becomes rather tedious, since the new basis states are linear combinations of sets of $LS$ or $jj$ coupled states. Instead, the matrix diagonalization approach, combined with the use symmetry arguments, is most efficient. To automate this, Robert Cowan (1968; 1981) developed code for modern computers.

As a manageable example, we will express the intermediate coupled ground state of the $f^2$ and $f^{12}$ configurations in both $LS$ and $jj$ coupled states. Two particles in the same shell $\ell$ coupled to $L$ have an electrostatic energy of $E = \sum_k f_k F^k$, where

$$f_k = (-1)^L [\ell]^2 \begin{pmatrix} \ell & k & \ell \\ 0 & 0 & 0 \end{pmatrix}^2 \begin{Bmatrix} \ell & \ell & L \\ \ell & \ell & k \end{Bmatrix}, \tag{41}$$

with $k = 0, 2, \cdots, 2\ell$. The first step is to simplify the problem by using symmetry restrictions. For equivalent electrons, the Pauli principle requires that $L+S$ must be even, and together with the triangulation rule, $0 \leq L \leq 2\ell$, the possible states for $f^2$ (or $f^{12}$) will be $^1S_0, ^1D_2, ^1G_4, ^1I_6, ^3P_{0,1,2}$, $^3F_{2,3,4}$, and $^3H_{4,5,6}$. Adding up all these states with a degeneracy of $(2J+1)$, gives a total of 91 $M_J$-sublevels. This number is equal to the binomial $(4\ell + 2, 2)$, as it should be.

The Hund's rule ground state $f^2$ $^3H_4$ is mixed by spin-orbit interaction with $^3F_4$ and $^1G_4$. The Hamiltonian for $f^2$ $J = 4$ in matrix form using the $LS$-coupled basis states ($^3F_4, ^1G_4, ^3H_4$) is

$$H^{(LS)}_{J=4} = \begin{pmatrix} E(^3F) + \frac{3}{2}\zeta & \sqrt{\frac{11}{3}}\zeta & 0 \\ \sqrt{\frac{11}{3}}\zeta & E(^1G) & -\sqrt{\frac{10}{3}}\zeta \\ 0 & -\sqrt{\frac{10}{3}}\zeta & E(^3H) - 3\zeta \end{pmatrix}, \tag{42}$$

with the electrostatic energies obtained by Eq. (41) as

$$E(^3F) = F_0 - \frac{2}{45}F^2 - \frac{1}{33}F^4 - \frac{50}{1287}F^6,$$

$$E(^1G) = F_0 - \frac{2}{15}F^2 + \frac{97}{1089}F^4 + \frac{50}{4719}F^6,$$

$$E(^3H) = F_0 - \frac{1}{9}F^2 - \frac{17}{363}F^4 - \frac{25}{14157}F^6. \tag{43}$$

The diagonal matrix elements of the spin-orbit interaction, which depend on *L*, *S*, and *J*, are obtained using in Eq. (32). The Hamiltonian in *LS*-coupled basis states is diagonal in the electrostatic interaction, but not in the spin-orbit interaction. The off-diagonal matrix elements of the spin-orbit interaction mix the singlet state into the triplet ground state. This mixing is largest for the *jj* coupled state.

The Hund's rule ground state of $f^{12}$ is $^3H_6$, which is mixed by spin-orbit interaction with the $^1I_6$ state. The Hamiltonian for $f^{12}$ $J=6$ in the *LS*-coupled basis states ($^3H_6$, $^1I_6$) is

$$H_{J=6}^{(LS)} = \begin{pmatrix} E(^3H) + \frac{5}{2}\zeta & \sqrt{\frac{3}{2}}\zeta \\ \sqrt{\frac{3}{2}}\zeta & E(^1I) \end{pmatrix}, \tag{44}$$

with the electrostatic energies obtained by Eq. (41) as

$$E(^3H) = F_0 - \frac{16}{45}F^2 - \frac{8}{33}F^4 - \frac{400}{1287}F^6,$$

$$E(^1I) = \frac{1}{9}F^2 + \frac{1}{121}F^4 + \frac{25}{184041}F^6. \tag{45}$$

Equations (42) and (44) reveal that the intermediate coupled state gains significant singlet character. The spin character of the atomic ground state of all actinides in intermediate coupling is given in Table V.

The matrix elements in *jj*-coupled basis states can be obtained using the transformation $H_J^{(jj)} = T \cdot H_J^{(LS)} \cdot T^{-1}$, where *T* is the transformation matrix from Eq. (5). For $f^2$ and $f^{12}$ (equivalent electrons!), the Pauli principle allows the $(5/2,5/2)_{0,2,4}$, $(5/2,7/2)_{1,2,3,4,5,6}$, and $(7/2,7/2)_{0,2,4,6}$ states, again with 91 $M_J$-sublevels in total. The $J = 6$ level is contained in the (5/2,7/2) and (7/2,7/2) states. Using the transformation matrix from Eq. (37), we obtain

$$H_6^{(jj)} = \frac{1}{7}\begin{pmatrix} 6E(^3H) + E(^1I) + 21\zeta & \sqrt{6}[E(^3H) - E(^1I)] \\ \sqrt{6}[E(^3H) - E(^1I)] & E(^3H) + 6E(^1I) - \frac{7}{2}\zeta \end{pmatrix}. \tag{46}$$

The Hamiltonian in *jj*-coupled basis states is diagonal in the spin-orbit interaction, but not in the electrostatic interaction, which is thus the opposite situation as for the *LS*-coupled states. Eq. (46) confirms that $\langle \mathbf{l} \cdot \mathbf{s} \rangle$ is 3 and -1/2 for the (7/2,7/2) and (5/2,7/2) states, respectively, in agreement with Eq. (16).

Using Eq. (16) it can be shown that intermediate-coupled eigenstates have a value of $\langle \mathbf{l} \cdot \mathbf{s} \rangle$ that is in between the *LS* and *jj* coupled limits. The negative value of $\langle \mathbf{l} \cdot \mathbf{s} \rangle$ becomes more negative going from the Hund's rule state to the intermediate coupled state due to the increasing spin-orbit interaction. This change is strong if the spin-orbit coupling can mix in other *LSJ* states. A first-order change in the expectation value occurs when there are excited states that can mix with the ground state. The spin-orbit coupling only mixes states with $\Delta L = 0, \pm 1$, $\Delta S = 0, \pm 1$, and $\Delta J = 0$. For instance, the $^3F_4$ and $^3H_4$ both mix with the $^1G_4$ level, but not with each other [c.f., Eq. (42)].

## B. Spin-orbit sum rule

### 1. *w*-tensors

For the benefit of a general approach, we introduce first the *w*-tensors. The electronic and magnetic state of an electron configuration $\ell^n$ can be characterized by *LS*-coupled multipole moments $\langle w^{xyz} \rangle$, where the orbital moment *x* and spin moment *y* are coupled to a total moment *z* (van der Laan, 1997b; 1998). These tensor operators allow a systematic classification; moments with even *x* describe the shape of the charge distribution and moments with odd *x* describe an orbital motion, e.g., $w^{000}$ is the number operator, $w^{011}$ is the spin-orbit operator, $w^{011}$ is the spin magnetic moment operator, and $w^{101}$ is the orbital magnetic moment operator, etc. The relation between these *LS*-coupled operators $w^{xyz}$ and the standard operators is given in Table VI. The *w*-tensors have a shell-independent normalization $\langle w^{xyz} \rangle = (-1)^{z+1}$. Consequently, the conversion to the standard operators depends on $\ell$, i.e., for the spin-orbit coupling operator we have $w^{110}_{\ell=2} = \sum_i \mathbf{l}_i \cdot \mathbf{s}_i$ and $w^{110}_{\ell=3} = \frac{2}{3} \sum_i \mathbf{l}_i \cdot \mathbf{s}_i$ for the *d* and *f* shell, respectively.

Use of the expressions for *n* and $\langle \mathbf{l} \cdot \mathbf{s} \rangle$ from Eq. (16) immediately gives the expectation values as

$$\langle w^{000} \rangle = n_{j_1} + n_{j_2} , \tag{47}$$

$$\langle w^{110} \rangle = -\frac{\ell+1}{\ell} n_{j_1} + n_{j_2} . \tag{48}$$

We can also define operators for the hole state, which are denoted by an underscore, $\langle \underline{w}^{000} \rangle \equiv 4\ell + 2 - \langle w^{000} \rangle$ and $\langle \underline{w}^{110} \rangle \equiv -\langle w^{110} \rangle$. Using $n_{j_i} + n^h_{j_i} = 2j_i + 1$, it follows that

$$\langle \underline{w}^{000} \rangle = n_h = n^h_{j_1} + n^h_{j_2}, \tag{49}$$

$$\left\langle \underline{w}^{110} \right\rangle = -\frac{\ell+1}{\ell} n^h_{j_1} + n^h_{j_2}. \tag{50}$$

Comparison between Eqs. (47)-(50) shows that the expression for the *w*-tensors retains its form if all electron operators are replaced by hole operators.

Since in the *jj* coupling limit the $j_1 = \ell-s$ levels are filled first, and only when these are completely full do the $j_2 = \ell+s$ levels fill, we obtain

$$\left\langle w^{110} \right\rangle = -\frac{\ell+1}{\ell} n \quad \text{for } n \leq 2\ell, \tag{51}$$

$$\left\langle w^{110} \right\rangle = -n_h \quad \text{for } n > 2\ell, \tag{52}$$

The fact that $\left\langle \underline{w}^{110} \right\rangle / n_h = 1$ for $j_2 = \ell + s$ is a consequence of the normalization of the *w*-tensors.

For use in many-particle calculations, the one-electron *w*-operators can be written in terms of creation and annihilation operators, $a^\dagger$ and $a$, as

$$w^{000} = \sum_{\lambda\lambda'\sigma\sigma'} a^\dagger_{\ell\lambda s\sigma} a_{\ell\lambda's\sigma'} = \sum_{j_i j'_i m_i m'_i} a^\dagger_{j_i m_i} a_{j'_i m'_i} = \sum_{j_i} \bar{n}_{j_i} = \bar{n}, \tag{53}$$

$$\begin{aligned}
w^{110} &= (\ell s)^{-1} \sum_{\lambda\lambda'\sigma\sigma'} \left\langle \ell\lambda s\sigma \middle| \mathbf{l}\cdot\mathbf{s} \middle| \ell\lambda's\sigma' \right\rangle a^\dagger_{\ell\lambda s\sigma} a_{\ell\lambda's\sigma'} \\
&= (\ell s)^{-1} \sum_{j_i j'_i m_i m'_i} \left\langle j_i m_i \middle| \mathbf{l}\cdot\mathbf{s} \middle| j_i m'_i \right\rangle a^\dagger_{j_i m_i} a_{j_i m'_i} = -\frac{\ell+1}{\ell} \bar{n}_{j_1} + \bar{n}_{j_2},
\end{aligned} \tag{54}$$

where $\lambda$, $\sigma$, and $m_i$ are the magnetic components of $\ell$, $s$ and $j_i$, respectively, and $\bar{n}$ are the number operators. To derive the right hand-side of the above expressions, we used the fact that $\mathbf{l}\cdot\mathbf{s}$ is diagonal in $j_i m_i$.

The creation and annihilation operators remove the need to employ coefficients of fractional parentage (Judd, 1967). The anticommutation relations of these operators correctly handle the wave functions constructed for an *n*-electron atom from linear combinations of *n* one-electron spin-orbitals. If the angular momenta are not coupled together, antisymmetrization is accomplished by forming determinantal functions, e.g., for a two-particle state

$$\psi_{ij}(q_1, q_2) = \frac{1}{\sqrt{2}} \begin{vmatrix} \psi_i(q_1) & \psi_j(q_1) \\ \psi_i(q_2) & \psi_j(q_2) \end{vmatrix}, \tag{55}$$

with

$$\left\langle \psi_{ij}(q_1,q_2) \middle| \psi_{kl}(q_1,q_2) \right\rangle = \delta_{ik}\delta_{jl} - \delta_{il}\delta_{jk} = \left\langle 0 \middle| a_j a_i a^\dagger_k a^\dagger_l \right\rangle. \tag{56}$$

## 2. Derivation of the sum rule

The branching ratio of a spin-orbit-split core-valence transition in EELS or XAS is linearly related to the expectation value of the spin-orbit operator of the valence states. This spin-orbit sum rule was first derived by Thole and van der Laan (1988b) using *LSJ* coupled states and coefficients of fractional parentage. The application to 3*d* transition metals was presented in Thole and van der Laan (1988a) and van der and Thole (1988a). Here, we give a simple derivation using creation and annihilation operators in a *jj* coupled basis (van der Laan and Thole, 1996b; van der Laan, 1998).

The branching ratio is determined by the angular part of the spin-orbit interaction and not by its magnitude, which is given by its radial part $\zeta_\ell$. Therefore, it is complementary to measurements of the energy splittings that include the spin-orbit parameter $\zeta_\ell$. Strictly speaking, the sum rule applies to the line strength. The x-ray absorption intensity is obtained by multiplying the line strength with the photon energy $h\nu$ (Thole and van der Laan, 1988a).

We consider the case where a sufficiently large spin-orbit interaction splits the core level $c$ into two manifolds $j$, i.e., $j_- = c - s$ and $j_+ = c + s$, and assume that the splitting due to the core-valence interaction is much smaller. The valence state $\ell$ contains the spin-orbit levels $j_i$, i.e., $j_1 = \ell - s$ and $j_2 = \ell + s$. The structure of the valence state is not important, it can be mixed, e.g. by electrostatic interaction, hybridization, and/or band effects.

We consider an excitation $csj \to \ell s j_i$ in a many-electron system. The transition probability for electric $2^Q$-pole radiation with polarization $q$ is given by a one-electron operator

$$T_q = \sum_{\gamma\lambda\sigma\sigma' m m_i} \left\langle csjm \middle| C_q^{(Q)} \middle| \ell s j_i m_i \right\rangle R_{c\ell}\, a^\dagger_{j_i m_i} a_{jm}$$

$$\propto \sum_{\gamma,\lambda,\sigma,m,m_i} \begin{pmatrix} j & s & c \\ m & \sigma & \gamma \end{pmatrix} \begin{pmatrix} c & Q & \ell \\ \gamma & q & \lambda \end{pmatrix} \begin{pmatrix} \ell & s & j_i \\ \lambda & \sigma' & m_i \end{pmatrix} P_{c\ell}\, a^\dagger_{j_i m_i} a_{jm}$$

(57)

where $\gamma$, $\lambda$, $\sigma$, $m$, and $m_i$ are the components (i.e., magnetic sublevels) of $c$, $\ell$, $s$, $j$ and $j_i$, respectively, $a_{jm}$ is the annihilation operator for a core electron with quantum numbers $jm$ and $a^\dagger_{j_i m_i}$ is the creation operator for a valence electron with quantum numbers $j_i m_i$ and $C_q^{(Q)}$ is a normalized spherical harmonic, $R_{c\ell}$ stands for the radial matrix element of the $c \to \ell$ electric-dipole transition, and $P_{c\ell} = \langle c \| C^{(1)} \| \ell \rangle R_{c\ell}$ (Thole *et al.*, 1994). The right hand side of this expression contains the 3-*j* symbols, corresponding to the coupling of orbital and spin momenta in the core shell $(c, s, j)$, light and orbital momenta $(c, Q, \ell)$, and orbital and spin momenta in the valence shell $(\ell, s, j_i)$, summed over the intermediate components $\gamma$ and $\lambda$. Equation (57) can be recoupled to

$$T_q = (-1)^{j_i - j} [jc\ell]^{1/2} \begin{Bmatrix} j & Q & j_i \\ \ell & s & c \end{Bmatrix} \sum_{\sigma m m_i} \begin{pmatrix} j & Q & j_i \\ m & q & m_i \end{pmatrix} P_{c\ell}\, \delta_{\sigma\sigma'}\, a^\dagger_{j_i m_i} a_{jm}.$$

(58)

From a many-electron ground state $|g\rangle$, the intensity summed over the final states $|f\rangle$ of the core $j$ level is

$$I_q = \sum_f \langle g|T_q^\dagger|f\rangle\langle f|T_q|g\rangle |P_{c\ell}|^2 \quad . \tag{59}$$

Assuming that there is no overlap between the two manifolds belonging to the core $j$ levels, the final states can be removed by extending the set of functions to the whole Hilbert space and using the closure relation, $\sum_f |f\rangle\langle f| = 1$. The angular dependent part of the intensity over the $j$ core edge can then be written

$$\sum_f \sum_{j_i j_i' mm' m_i m_i'} \langle g|a_{jm'}^\dagger a_{j_i' m_i'}|f\rangle\langle f|a_{j_i m_i}^\dagger a_{jm}|g\rangle = \sum_{j_i j_i' mm' m_i m_i'} \langle g|a_{jm'}^\dagger a_{j_i' m_i'} a_{j_i m_i}^\dagger a_{jm}|g\rangle$$

$$= \delta_{mm'} \sum_{j_i j_i' m_i m_i'} \langle g|a_{j_i' m_i'} a_{j_i m_i}^\dagger|g\rangle \tag{60}$$

where we moved $a_{jm'}^\dagger$ to the right by applying the anticommutator rules, and removed the core shell operators by using $a_{j'm'}^\dagger|g\rangle = 0$, since $|g\rangle$ does not contain holes in the core level.

For the isotropic spectrum (i.e., averaged over polarization $q$ and magnetic state $m_i$) there are no cross terms between different $j_i m_i$ states and the diagonal elements of effective operator $a_{j_i m_i} a_{j_i m_i}^\dagger$ acting on the ground state counts the number of holes $n_{j_i}^h$ in the valence spin-orbit level $j_i$, i.e.,

$$\sum_{j_i j_i' m_i m_i'} \langle g|a_{j_i' m_i'} a_{j_i m_i}^\dagger|g\rangle = \delta_{j_i' j_i} \sum_{m_i} \langle g|\delta_{m_i' m_i}|g\rangle = \delta_{j_i' j_i} \delta_{m_i' m_i} n_{j_i}^h . \tag{61}$$

Using this in Eqs. (58) and (59), the integrated intensities for the transition $j \to j_i$ become

$$I(j \to j_i) = \langle g\|T^\dagger T\|g\rangle n_{j_i}^h = [jc\ell]\begin{Bmatrix} j & Q & j_i \\ \ell & s & c \end{Bmatrix}^2 |P_{c\ell}|^2 n_{j_i}^h . \tag{62}$$

One might recognize the right-hand side of this equation as the one-electron multipole transition probability $csj \to \ell s j_i$ times the number of holes in $j_i$. We could have used this as the starting equation, however, a main purpose of the above derivation is to demonstrate explicitly that this equation, and hence the emerging sum rule, is valid for an $n$-electron state. The creation and annihilation operators are acting on a many-electron state (see Sec. III 6 A). The simplicity arises because the transition probability is given by a one-electron operator.

Using Eq. (62) as the key result, we obtain the angular part of integrated intensities (i.e., omitting the radial factor $|P_{c\ell}|^2$) for the transitions with $Q = 1$, $\ell = c + 1$ as

$$I(j_- \to j_1) = \ell^{-1}(2\ell+1)(\ell-1) n_{j_1}^h ,$$

$$I(j_- \to j_2) = 0 ,$$
$$I(j_+ \to j_1) = \ell^{-1} n^h_{j_1} ,$$
$$I(j_+ \to j_2) = (2\ell - 1) n^h_{j_2} . \tag{63}$$

Thus, the $j_-$ edge only probes the $j_1$ level. The $j_+$ edge probes both the $j_1$ and $j_2$ level, but is $\ell(2\ell-1)$ times more sensitive to the latter. Under the assumption that $P_{c\ell}$ is constant, the intensity over each edge is

$$I_{j_-} = \ell^{-1}(2\ell+1)(\ell-1) n^h_{j_1} , \tag{64}$$

$$I_{j_+} = (2\ell-1) n^h_{j_2} + \ell^{-1} n^h_{j_1} , \tag{65}$$

which gives the total intensity and branching ratio as

$$I_{\text{total}} \equiv I_{j_+} + I_{j_-} = (2\ell-1)(n^h_{j_2} + n^h_{j_1}) , \tag{66}$$

$$B \equiv \frac{I_{j_+}}{I_{j_+} + I_{j_-}} = \frac{n^h_{j_2} + [\ell(2\ell-1)]^{-1} n^h_{j_1}}{n^h_{j_2} + n^h_{j_1}} , \tag{67}$$

Substitution of the definitions $n_h \equiv n^h_{j_2} + n^h_{j_1}$ and $\langle w^{110} \rangle \equiv \frac{\ell+1}{\ell} n^h_{j_1} - n^h_{j_2}$ gives

$$I_{\text{total}} = (2\ell-1) n_h = (2c+1) n_h , \tag{68}$$

$$B = -\frac{\ell-1}{2\ell-1} \frac{\langle w^{110} \rangle}{n_h} + \frac{\ell}{2\ell-1} = -\frac{c}{2c+1} \frac{\langle w^{110} \rangle}{n_h} + \frac{c+1}{2c+1} , \tag{69}$$

and rearrangement gives the spin-orbit expectation value per hole

$$\frac{\langle w^{110} \rangle}{n_h} = -\frac{2c+1}{c}(B - B_0) , \tag{70}$$

where

$$B_0 = \frac{c+1}{2c+1} = \frac{2j_+ + 1}{2(2c+1)} , \tag{71}$$

is the statistical value. q.e.d.

The branching ratio $B \equiv I_{j_+}/(I_{j_+} + I_{j_-})$ and the intensity ratio $R \equiv I_{j_+}/I_{j_-}$ are easily converted into each other by using $B = R/(R+1)$ or $R = B/(1-B)$. However, only $B$ has the advantage of being directly proportional to $\langle w^{110} \rangle / n_h$, as well as having the convenient mathematical limits of $0 \leq B \leq 1$. The lower limit, which would mean all intensity is in the $j_-$ level, is however physically not achievable. We will investigate the allowed range of $B$ using the above sum rule results, and apply this to the $f$ shell. When all the valence holes are in the $j_2 = 7/2$ level, we have $n^h_{j_2} = n_h$ and $n^h_{j_1} = 0$, so that $\langle w^{110} \rangle / n_h = -1$, and $B = 1$ is the upper limit. In this limit, all the intensity is in the $j_+ = c + 1/2 = 5/2$ core level, which is due to the fact that transitions from core $j_- = 3/2$ to valence $j_2 = 7/2$ are forbidden. For the other extreme case, where all valence holes are in the $j_1 = 5/2$ level, $n^h_{j_1} = n_h$ and $n^h_{j_2} = 0$, so that $\langle w^{110} \rangle / n_h = \ell^{-1}(\ell+1) = 4/3$, and $B = [\ell(2\ell-1)]^{-1} = 1/15$ is the lower limit. In this limit most, but not all, intensity is in the $j_- = c - 1/2 = 3/2$ core level. The reason is that transitions to the $j_1 = 5/2$ valence level are allowed from both the $j_+ = 5/2$ and $j_- = 3/2$ core level. Thus the minimum value of the branching ratio is set by the dipole selection rules. Note furthermore that since $\ell$ and $s$ prefer to be coupled antiparallel, $\langle w^{110} \rangle$ is negative so that the branching ratio is larger than the statistical ratio (in the absence of core-valence interactions). The calculated values of $\langle w^{110} \rangle$ for the three coupling schemes are given in Table VII.

### 3. Limitations of the sum rule

There are a few theoretical and experimental considerations that one should keep in mind when using the sum rule. The derivation of the sum rule requires a separation of the transition probability into a radial and angular part (sometimes called dynamic and geometric part), whereby it is assumed that the radial part $P_{c\ell}$ is constant for each transition in the spectrum. Not only does the relativistic radial matrix element depend on the core and valence $j$ values, but there can also be variations within each $j \rightarrow j_1$ transition array. Although each $j$ edge extends only over a narrow range of a few eV, transitions to a different part of the valence band can have a different cross-section in the case of hybridization and/or band structure. Such effects are expected to be small for the strongly localized $f$ shell of rare earths but could be more pronounced for actinides, particularly the lighter and more delocalized $5f$ metals Th, Pa, U, Np, and $\alpha$-Pu.

The sum rule is based on the assumption that it is possible to integrate the signal of a core level over regions assigned by a good quantum number, in this case the total angular momentum $j$. However, core-valence interactions between the two $j$ edges, the so-called $jj$ mixing, can induce a transfer of spectral weight, thereby invalidating the spin-orbit sum rule (Thole and van der Laan, 1988a; van der Laan and Thole 1988a). The importance of this effect will be discussed in the next subsection. Alternatively, the sum rules can be used indirectly, by calculating the absorption spectrum and comparing the branching ratio to that of the measured spectrum. Multiplet calculations in intermediate coupling fully take the $jj$ mixing fully into account. Moreover, if one switches off the core-valence interaction in the calculation, the branching ratio gives exactly the ground-state spin-orbit interaction. Band structure calculations, on the other hand, have difficulty to properly include the core-valence interaction, resulting in a different line shape and branching ratio. Band theory can also have difficulty calculating the correct number of valence holes, but is nevertheless often consulted to assess the number of holes needed to convert to $\langle w^{110} \rangle$.

For the analysis of XAS and EELS spectra it is often sufficient to consider only electric-dipole transitions, certainly at energies below a few keV. At higher energies electric-quadrupole transitions will start to play a minor role. Furthermore, dipole transitions are not only allowed to $\ell = c + 1$ states but also to $\ell = c - 1$ states. In the case of actinides this would mean transitions to unoccupied $p$ states; however, these are much smaller than to $f$ states.

In the case of XAS, there are some experimental complications due to saturation effects which occur in both electron yield and fluorescence. Saturation effects happen because the electron escape depth cannot be neglected with respect to the x-ray attenuation length (van der Laan and Thole 1988b), especially at grazing incidence. The electron escape depth in XAS is of the order of several nm, which makes the electron yield signal surface sensitive, necessitating more sophisticated surface science preparation. Near the surface the spin-orbit interaction can be different from the bulk due to symmetry breaking from loose bonds and chemical contamination of the surface, such as oxidization or hydriding. This, of course, is very important for the actinides, which readily react with oxygen and hydrogen.

Besides an isotropic part $\langle w^{110} \rangle$, the spin-orbit interaction also contains an anisotropic part $\langle w^{112} \rangle$, which relates to the difference in probability for $\ell$ and $s$ parallel and perpendicular to an arbitrary axis. Like $\langle w^{110} \rangle$ is related via the sum rule to the branching ratio of the isotropic spectrum, $\langle w^{112} \rangle$ is related via another sum rule to the branching ratio of the linear dichroism (van der Laan, 1999). Therefore, to determine purely $\langle w^{110} \rangle$ one needs to make sure to take an isotropic average over the measurements, and to keep in mind that synchrotron radiation beam is naturally polarized. Measurements with linearly polarized soft x-rays have evidenced a small but significant contribution from $\langle w^{112} \rangle$ in 3$d$ transition metal thin films (Dhesi *et al.*, 2001; 2002). In more localized materials the anisotropy could be much larger, and measurements of $\langle w^{112} \rangle$ for actinides have so far not been done. The anisotropic spin-orbit interaction plays a major role in the magnetocrystalline anisotropy of a material, which makes it an important quantity to measure (van der Laan, 1999).

It is sometimes forgotten that the intensity is given as the product of line strength and photon energy, while the sum rule applies to the line strength. This requires a small correction when the two core $j$ levels are far apart. Furthermore, a reliable $I_0$ monitor is essential, because the photon flux is not constant as a function of energy.

Another source of errors can arise from the choice of the integration limits. The *background* at energies below and above the edge has usually not the same height, making separation of discrete and continuum states difficult. Furthermore, the application of the sum rule requires the choice of a specific energy point to separate the intensities of the two edges. Such a choice becomes ambiguous when the signal is not entirely zero between the two edges.

**4. *jj* mixing**

As for all XAS sum rules, the expectation value is obtained per hole, since the core electron is excited into the unoccupied valence states (van der Laan, 1996a; 1997a). In the same way as the spin magnetic moment sum rule in x-ray magnetic circular dichroism (XMCD), the sum rule in Eq. (70) is strictly valid for the case where core-valence electrostatic interactions are absent. Equation (70) shows that in order to extract the value of $\langle w^{110} \rangle / n_h$ from the branching ratio, $B$, we need to know the value of $B_0$, which, in the case that the sum rule is exact, is equal to the statistical value. Empirically, we can define $B_0$ as the value of the branching ratio when the valence spin-orbit is zero, or – what amounts to the same – when we take the average over all the spin-orbit split sublevels. The value of $B_0$ will then only depend on the core-valence interaction (Thole and van der Laan, 1988a). If the spin-orbit-split core levels are mixed due to core-valence interactions, a correction term $\Delta$ is needed, which is proportional to the difference between $B_0$ and the statistical value, such that

$$\frac{\langle w^{110} \rangle}{n_h} = -\frac{2c+1}{c}\left(B - \frac{c+1}{2c+1}\right) + \Delta, \tag{72}$$

$$\Delta \equiv \frac{2c+1}{c}\left(B_0 - \frac{c+1}{2c+1}\right). \tag{73}$$

It can be shown using first-order perturbation theory (van der Laan *et al.*, 2004) that $\Delta$ is proportional to the ratio between the core-valence exchange interaction $G^1(c,\ell)$ and the core spin-orbit interaction $\zeta_c$. Table VIII shows that there is a remarkable linear relationship between $G^1(c,\ell)/\zeta_c$ and $\Delta$ over a wide range of different edges in $3d$, $4d$, $4f$, $5d$, and $5f$ metals. These values were obtained from relativistic atomic Hartree-Fock calculations using Cowan's code (Cowan, 1981), where $B_0$ was calculated using the weighted average over the different $J$ levels in the ground state.

From Table VIII we can make the following observations. For $3d$ transition metals the application of the spin-orbit sum rule for the $L_{2,3}$ branching ratio is severely hampered by the large $(2p,3d)$ exchange interaction, which is of similar size as the $2p$ spin-orbit interaction (Thole and van der Laan, 1988a; van der Laan and Kirkman, 1992). The same is true for the $M_{4,5}$ edges of the lanthanides, where the $(3d,4f)$ exchange interaction is strong compared to the $3d$ spin-orbit interaction (Thole *et al.*, 1985b). However, even in the case of the rare earth elements the trend in the branching ratio can be used the obtain the relative population of spin-orbit split states, as was demonstrated for Ce systems (van der Laan *et al.*, 1986b).

On the other hand, the sum rule holds quite well for the $L_{2,3}$ edges of $4d$ and $5d$ transition metals, as might be expected for a deep $2p$ core level that has small $G^1(c,\ell)$ and large $\zeta_c$. The situation is also favourable for the $M_{4,5}$ and $N_{4,5}$ edges of the actinides, giving a small $G^1(c,\ell)/\zeta_c$ for the $3d$ and $4d$ core levels. The latter result is quite surprising. In spite of the fact that the Th $4d$ core level is shallower than the Zr $2p$ or Hf $2p$ and lies in between the Ti $2p$ and La $3d$, the core-valence interactions do not spoil the sum rule. Calculations show that $B_0$ for the actinide $M_{4,5}$ and $N_{4,5}$ edges varies between only 0.59 and 0.60 for the light actinides and, therefore, is very close to the statistical ratio of 0.60 (van der Laan *et al.*, 2004). This means that the EELS and XAS branching ratios depend almost exclusively on the $5f$ spin-orbit expectation value per hole, thus affording an unambiguous probe for the $5f$ spin-orbit interaction in actinide materials.

Table I gives the experimental branching ratio, $B$, of the $N_{4,5}$ EELS spectra shown in Fig. 13 and the expectation value of the $5f$ spin-orbit interaction per hole, $\langle w^{110} \rangle/(14 - n_f) - \Delta$, for the $\alpha$ phase of Th, U, Np, Pu, Am, and Cm metal. The experimental electron occupation numbers $n_{5/2}$ and $n_{7/2}$ of the $f_{5/2}$ and $f_{7/2}$ levels are obtained by solving Eqs. (47) and (48).

**C. Many-electron spectral calculations**

Multiplet theory provides the most accurate method for calculating the atomic core level spectra at the $M_{4,5}$, $N_{4,5}$, and $O_{4,5}$ edges, each of which is associated with the transitions $f^n \to d^9 f^{n+1}$. It has become the preferred method to calculate core level spectra of localized rare-earth and 3$d$ transition-metal systems (van der Laan and Thole, 1991a). Contrary to band structure calculations, multiplet theory treats spin-orbit, Coulomb, and exchange interactions on an equal footing, which is essential in the treatment of the localized character of the valence states. This was first highlighted by the excellent agreement obtained by multiplet calculations in the case of the rare earths $M_{4,5}$ edges (Thole *et al.*, 1985b). The calculations for the actinides are done in similar way as for the rare earths, only the values of the radial parameters are different. Moreover, it is straightforward to include crystal field and hybridization in the calculation (van der Laan and Thole, 1991a; van der Laan and Kirkman, 1992), although this does put a heavy penalty on the computing time.

It turns out, however, that the crystal field interaction is not strong for the actinides and that other mechanisms based on hybridization can be more significant. With the exception of some uranium compounds (Yaouanc *et al.*, 1998), there is no evidence of appreciable crystal field interaction. Instead, the 5$f$ electrons hybridize with either the 6$d$ conduction states or neighbouring atom $p$ states. The influence of the crystal field on the branching ratio is only important when the crystal field becomes in the order of the spin-orbit coupling and so mixes in other $\alpha J$ levels (van der Laan and Thole, 1996b). This occurs readily for the $d$ transition metals, which typically have crystal fields of a few eV and a spin-orbit coupling of a few hundredths of an eV. In the actinides, where the 5$f$ spin-orbit splitting is of the order of an eV (c.f., Table III), the importance of the crystal field is negligible.

In a nutshell, the atomic multiplet calculation of the spectrum is done as follows. First, the initial- and final-state wave functions are calculated in intermediate coupling using the atomic Hartree-Fock method with relativistic correction (Cowan, 1968; 1981), much as has been shown in Sec. III A 5 for the initial state. After empirical scaling of the output parameters for the Slater integrals and spin-orbit constants, the electric-multipole transition matrix elements are calculated from the initial state to the final state levels of the specified configurations. At low-energies only the electric-dipole transitions are relevant. The electric-dipole selection rules from the ground state strongly limit the number of accessible final states, with the consequence that compared to the total manifold of final states, the XAS lines are only located in a narrow energy region (Thole *et al.*, 1985b). The multiplet structure usually involves many states, which necessitates the use of large computer codes. The number of levels for each configuration $\ell^n$ is equal to the binomial$(4\ell + 2, n)$, which can become quite large. For instance, in the transition $f^6 \to d^9 f^7$ there are 3,003 and 48,048 levels in the initial- and final-state, respectively, resulting in matrices with large dimensions that need diagonalization. Fortunately, selection rules and symmetry restrictions will strongly reduce the size of the calculation.

We can only present here examples of manageable size, and therefore show the calculation of the spectra for the $f^0 \to d^9 f^1$ transition, which gives already some flavor of the method. The $\underline{d} f^1$ final state, where the underscore denotes a hole in a core $d$ shell, can have $L = 0, 1, 2, 3, 4, 5$ and $S = 0, 1$, with 140 levels in total. Selection rules restrict the final states that are assessable from the initial state. Dipole transitions from the ground state $f^0$ ($^1S_0$) are allowed only to the final state $^1P_1$, which is mixed by spin-orbit interaction with the triplet final states $^3D_1$ and $^3P_1$. The final-state Hamiltonian is $H = H_{\text{el}} + \langle \mathbf{l} \cdot \mathbf{s} \rangle \zeta_d + \langle \mathbf{l} \cdot \mathbf{s} \rangle \zeta_f$. The spin-orbit eigenvalues in *jj* coupling are obtained from Eq. (16) as $\langle \mathbf{l} \cdot \mathbf{s} \rangle$ = -1 and 3/2 for the $d_{5/2}$ and $d_{3/2}$ hole, respectively, and $\langle \mathbf{l} \cdot \mathbf{s} \rangle$ = 3/2 and -2 for the $f_{7/2}$ and $f_{5/2}$ electron, respectively, so that

$$E(\underline{d}_{3/2}f_{5/2}) = \frac{3}{2}\zeta_d - 2\zeta_f,$$
$$E(\underline{d}_{5/2}f_{5/2}) = -\zeta_d - 2\zeta_f,$$
$$E(\underline{d}_{5/2}f_{7/2}) = -\zeta_d + \frac{3}{2}\zeta_f. \tag{74}$$

In order to add the electrostatic interaction, we can write the Hamiltonian in *LS* coupled basis states. The transformation matrix $(^3D_1, {}^3P_1, {}^1P_1) \rightarrow (d_{3/2}f_{5/2}, d_{5/2}f_{5/2}, d_{5/2}f_{7/2})$ is obtained from Eq. (5) as

$$T = \begin{pmatrix} \sqrt{\frac{2}{5}} & \sqrt{\frac{1}{5}} & \sqrt{\frac{2}{5}} \\ -\sqrt{\frac{16}{35}} & \sqrt{\frac{18}{35}} & \sqrt{\frac{1}{35}} \\ -\sqrt{\frac{1}{7}} & -\sqrt{\frac{2}{7}} & \sqrt{\frac{4}{7}} \end{pmatrix}. \tag{75}$$

Using $H_J^{(LS)} = T^{-1} \cdot H_J^{(jj)} \cdot T$ and adding the diagonal electrostatic interactions, gives the final-state Hamiltonian in matrix form as

$$H_1^{(LS)} = \begin{pmatrix} E(^3D) - \frac{3}{2}\zeta_f & \frac{1}{2}\sqrt{2}(\zeta_d + \zeta_f) & \zeta_d - \zeta_f \\ \frac{1}{2}\sqrt{2}(\zeta_d + \zeta_f) & E(^3P) - \frac{1}{2}\zeta_d - \zeta_f & \frac{1}{2}\sqrt{2}\zeta_d + \sqrt{2}\zeta_f \\ \zeta_d - \zeta_f & \frac{1}{2}\sqrt{2}\zeta_d + \sqrt{2}\zeta_f & E(^1P) \end{pmatrix}, \tag{76}$$

where the energies $E(^3D)$, $E(^3P)$, and $E(^1P)$ are given by linear combinations of the Slater integrals $F^0$, $F^2$, $F^4$, $F^6$, $G^1$, $G^3$, and $G^5$. Here, we will refrain from writing these energies in linear combinations of Slater integrals, since this yields seven Slater parameters instead of three energy values. However, in the calculation of more complicated final states the Slater parameters largely reduce the parameter space. After substituting the values of parameters (e.g., obtained by Hartree-Fock calculations), the matrix in Eq. (76) is diagonalized to obtain the final-state eigenstates and eigenvalues. Since the normalized dipole-selection rules from the ground state lead to $\langle \psi(^1S) | \hat{r} | \psi(^1P) \rangle = 1$ and $\langle \psi(^1S) | \hat{r} | \psi(^3D) \rangle = \langle \psi(^1S) | \hat{r} | \psi(^3P) \rangle = 0$, the intensity of each eigenstate is equal to the amount of $^1P$ character. (The character is equal to the square of the wave function coefficient.) In the remaining of this section, we will present the spectra for the $O_{4,5}$, $N_{4,5}$, and $M_{4,5}$ edges by substituting the calculated Hartree-Fock values for the spin-orbit and electrostatic parameters.

## 1. The $O_{4,5}$ ($5d \rightarrow 5f$) edge

For the Th $O_{4,5}$ transition ($5d \rightarrow 5f$) the spin-orbit parameters are $\zeta_{5f} = 0.21$ and $\zeta_{5d} = 2.70$ eV, and the electrostatic energies are $E(^3D) = -0.315$, $E(^3P) = -2.823$, and $E(^1P) = 15.187$ eV, which are taken relative to the average energy (i.e., we take $F^0 = 0$, which leads to a rigid shift of the total spectrum). Diagonalization of the matrix in Eq. (76) gives the following eigenvalues and corresponding eigenstates

$$E_1 = -5.303; \quad \psi_1 = 0.392\,\psi(^3D) - 0.920\,\psi(^3P) + 0.025\,\psi(^1P),$$
$$E_2 = -0.274; \quad \psi_2 = -0.906\,\psi(^3D) - 0.381\,\psi(^3P) + 0.186\,\psi(^1P),$$
$$E_3 = 15.753; \quad \psi_3 = 0.161\,\psi(^3D) - 0.095\,\psi(^3P) + 0.982\,\psi(^1P). \tag{77}$$

From the square of the $\psi(^1P)$ coefficients we obtain the intensities as $I_1 = 0.06\%$, $I_2 = 3.45\%$, and $I_3 = 96.5\%$. Thus the spectrum consists of two small peaks at low energy and one intense peak at high energy. These correspond to the weak pre-peaks at low energy and the giant resonance at high energy in the experimental spectrum. The pre-peaks are switched on by the spin-orbit interaction. In the absence of this interaction all the intensity is in the high energy peak, which is the dipole-allowed transition in $LS$ coupling. With increasing spin-orbit interaction, $L$ and $S$ cease to be good quantum numbers and only $J$ remains a good quantum number, so that levels with the same $J$ will mix. Equation (77) shows that prepeaks have mainly triplet character, whereas the giant resonant has mainly singlet character. While all states are rather pure in $LS$ character, they are strongly mixed in core $j$ character, namely $\psi_1$ has 80% $d_{5/2}$ and 20% $d_{3/2}$, while $\psi_2$ and $\psi_3$ both have 60% $d_{5/2}$ and 40% $d_{3/2}$ character.

The above spectral analysis can also be considered from the viewpoint of perturbation theory. For the $O_{4,5}$ edge, the electrostatic interaction is much larger than the spin-orbit interaction, which can be considered as a perturbation. First-order perturbation theory gives an energy separation between the triplet and singlet spin states of $\Delta E = \sqrt{(\Delta E_{el})^2 + (\Delta E_{so})^2}$ and a relative intensity for the "forbidden" triplet states of $I_{triplet}/I_{singlet} = (\Delta E_{so})^2 / 2(\Delta E_{el})^2$, where $\Delta E_{el}$ and $\Delta E_{so}$ are the effective splitting due to electrostatic and spin-orbit interaction, respectively. Comparing this to the values obtained from exact matrix diagonalization shows that this simple perturbation model holds reasonably well. Therefore, the relative intensity of the pre-peak structure is a sensitive measure for the strength of the $5d$ core spin-orbit interaction relative to the $5d,5f$ electrostatic interaction.

The picture for the other light elements $f^n$ is similar, but becomes rapidly more complicated with increasing $n$. The main peaks in the spectrum are due to the allowed transitions $\Delta S = 0$, $\Delta L = -1, 0, 1$. For a ground state $f^1(^2F_{5/2})$ the allowed transitions to a final state in $LS$ coupled basis are $^2D$, $^2F$, and $^2G$, with $J = 3/2$, 5/2, and 7/2. The small $5d$ spin–orbit interaction allows "forbidden" transitions with $\Delta S = 1$ to final states with quartet spin ($S = 3/2$), which are at lower energy due to the $5d,5f$ exchange energy. The splitting within the main peak is due to both Coulomb interaction and spin-orbit interaction and these cannot be separated. For $f^2$ ($^3H_4$), the dipole-allowed transition are to $^3G$, $^3H$, and $^3I$ states with $J = 3$, 4, and 5. In intermediate coupling, the ground state is a mixture of different $LS$ states, namely 88% $^3H_4$, 1% $^3F_4$, and 11% $^1G_4$. Analysis of the pre-peak structure shows it contains a mixture of mainly triplet and quintet spin states.

For less than a half-filled shell, there are always "forbidden" states with high spin at lower energies. The reason for this is that for a ground state $5f^n$ with maximum spin $S$, the maximum spin for the final state $5d^9 5f^{n+1}$ is $S+1$ for $n \leq 6$, and $S$ for $n \geq 7$ (c.f., Fig. 3 in Thole and van der Laan, 1988a). The energy separation between the states with different spin is determined by the exchange interaction. A necessary requirement of a sharp pre-peak is that its decay lifetime is long: A high spin state has the advantage that there are no, or only a few, states with the same $S$ into which it can decay. The excited state is then called "double forbidden". Complications in the LS picture arise from the fact that the ground state is strongly mixed, e.g. $5f^3$ has 84% $S = 4$ and 16% $S = 2$, and $5f^5$ has 67% $S = 6$ and 27% $S = 4$ (see Table V). This increased mixing of the spin states causes a decrease in energy separation between the pre-peak and giant resonance with increasing atomic number.

Figure 15 shows the calculated actinide $O_{4,5}$ spectra in the presence (black thick line) and absence (red thin line) of 5d core spin-orbit interaction for the ground state configurations $f^0$ to $f^9$. The decay channels that give rise to the broadening were not taken into account, instead all spectral lines were broadened with the same Lorentzian line shape of $\Gamma = 0.5$ eV. The pre-peak region and giant resonance are expected to be below and above ~5 eV, respectively. In all cases, it is clearly seen that when the spin-orbit interaction is switched on addition structure appears at low energy. This corresponds to the high-spin states that become allowed. This picture holds up quite well up to $n = 6$. For higher value of $n$ the original main peaks disappear at the cost of the low energy peaks. As mentioned above, for $n \geq 7$ the final state has the same spin multiplicity as the ground state and there are no forbidden spin transitions. States of the same spin are mixed by the 5d spin-orbit interaction, which increases in strength over the series (from $\zeta_{5d} = 2.70$ eV for Th to 4.31 eV for Cm).

The calculated actinide $O_{4,5}$ absorption spectra from Fig. 15 were convoluted using a Fano line shape broadening for the giant resonance and shown in Fig. 14. In general, the agreement between the experimental EELS $O_{4,5}$ edges in Fig. 12 and the calculated $O_{4,5}$ edges in Fig. 14 is quite satisfying. First, the pre-peak and giant resonance in the calculated $O_{4,5}$ edge for $n = 0$ and $n = 1$ are similar in form and intensity to the Th $O_{4,5}$ EELS edge and the calculated $O_{4,5}$ edge for $n = 3$ is similar to U (Moore *et al.*, 2003). Further, note that the width of the calculated $O_{4,5}$ edge in Fig. 14 reduces by about half when going from $n = 5$ to $n = 6$, which is exactly what is observed between Pu and Am in the $O_{4,5}$ EELS in Fig. 12. This is due to the fact that the $j = 5/2$ level is almost entirely full in Am, meaning the $d_{3/2} \rightarrow f_{5/2}$ transition is almost completely removed when going from Pu to Am.

Summarizing, the final states are close to the LS coupling limit for the $O_{4,5}$ edge. The 5d core spin-orbit interaction, which is much smaller than the electrostatic interaction, switches on the intensity of the pre-peaks with high-spin that are located at lower energy.

## 2. The $N_{4,5}$ ($4d \rightarrow 5f$) edge

For the Th $N_{4,5}$ transition ($4d \rightarrow 5f$) the spin-orbit parameters are $\zeta_{5f} = 0.23$ and $\zeta_{4d} = 15.38$ eV and the $4d^9 5f^1$ final-state electrostatic energies are $E(^3D) = -0.055$, $E(^3P) = -0.993$, and $E(^1P) = 0.267$ eV, which are taken relative to the average energy of the configuration (i.e., $F^0 = 0$). Solving the final-state Hamiltonian in Eq. (76) gives the following eigenvalues with eigenstates

$$E_1 = -16.489; \quad \psi_1 = 0.529\,\psi(^3D) + 0.565\,\psi(^3P) - 0.633\,\psi(^1P),$$
$$E_2 = -15.058; \quad \psi_2 = -0.847\,\psi(^3D) + 0.302\,\psi(^3P) - 0.437\,\psi(^1P),$$

$$E_3 = 22.508; \quad \psi_3 = 0.056\,\psi(^3D) - 0.767\,\psi(^3P) - 0.639\,\psi(^1P), \tag{78}$$

where the $^1P$ character gives the intensities 40.1%, 19.1%, and 40.8%, respectively. Hence, we obtain a double peak at −16.5 and −15 eV and a single peak at 22.5 eV. These energy positions are close to those expected for the pure $j = 5/2$ and $j = 3/2$ levels, which are at $-\frac{1}{2}c\zeta = -15.38$ eV and $\frac{1}{2}(c+1)\zeta = 23.7$ eV, respectively. Thus, we can truly assign these peaks to the $N_5$ and $N_4$ edges. The branching ratio is obtained as the intensity ratio $I(N_5)/[I(N_4) + I(N_5)] = 0.592$, which is close to the statistical ratio of 0.6. This is the value expected from the sum rule because for $f^0$ the spin-orbit interaction is zero. While the states in Eq. (78) are strongly mixed in $LS$ quantum numbers, they are rather pure in the core hole $j$ quantum number: $\psi_1$ and $\psi_2$ have 99.99% $d_{5/2}$ character and $\psi_3$ has 99.99% $d_{3/2}$ character. This high purity is of course due to the fact that the core spin-orbit interaction is much larger than the core-valence interaction. Hence the $N_{4,5}$ edge is ideally suited for the sum-rule analysis, which requires a negligible $jj$ mixing.

The $N_{4,5}$ spectra calculated in intermediate coupling for $^{92}$U $5f^1$ to $f^5$ and $^{100}$Fm $5f^7$ to $f^{13}$ are shown in Fig. 16, convoluted by 2 eV, which corresponds to the intrinsic lifetime broadening (Kalkowski, 1987). Taking this a step further, Fig. 17 unites the theoretical and experiment EELS results from Figs. 13 and 16, respectively. It shows the ground-state spin-orbit interaction per hole, $\langle w^{110}\rangle/(14-n) - \Delta$, as a function of the number of 5f electrons ($n_f$). The three theoretical coupling schemes are shown: $LS$, $jj$, and intermediate. The dots indicate the results of the spin-orbit analysis using the experimentally measured $N_{4,5}$ branching ratio of each metal in Fig. 13. The lower panel shows the electron occupation numbers $n_{5/2}$ and $n_{7/2}$ calculated in the three coupling schemes as a function of $n_f$. Again, the dots indicate the experimental results: the ground-state $n_{5/2}$ and $n_{7/2}$ occupation numbers of the 5f shell from the spin-orbit analysis of the EELS spectra in Fig. 13.

Using the theoretical framework discussed above, the magnetic moments of each element can be addressed. Utilizing this opportunity, the atomic spin magnetic moment $m_s = -2\langle S_z\rangle = -2\sum_i s_{z,i}$ and orbital magnetic moment $m_l = -\langle L_z\rangle = -\sum_i l_{z,i}$ (in $\mu_B$) for all 14 actinide elements is plotted against $n_f$ in Fig. 18(a) and (b), respectively. In each frame the calculated values for each of the three theoretical coupling schemes are shown: $LS$, $jj$, and intermediate. Immediately noticeable is that for the $f^1$ to $f^5$ the spin and orbital moments are aligned antiparallel, meaning there is partial cancellation between the spin and orbital moments. For $f^6$, there is no spin or magnetic moment for all three coupling mechanisms. For $f^7$ to $f^{13}$, the spin and orbital moments are parallel, meaning they sum additive, creating strong magnetic moments. Note that the spin magnetic moments become exceedingly large for either $jj$ or intermediate coupling for $f^7$, meaning that if Cm exhibits either of these coupling mechanisms it will produce a large spin polarization and subsequent magnetic moment. This will be the basis for one of the large changes in 5f behavior across the actinide series, and will be discussed in great detail in the Cm section using Fig. 18(a)-(c).

Finally, we mention that a similar behavior as for the $N_{4,5}$ edge is found for the $M_{4,5}$ edge. This edge has also been measured using resonant magnetic x-ray scattering (Tang *et al.*, 1992). For the Th $M_{4,5}$ transition ($3d \rightarrow 5f$) the spin-orbit parameters are $\zeta_{5f} = 0.23$ and $\zeta_{4d} = 66.00$ eV and the $4d^9 5f^1$ final-state electrostatic energies are $E(^3D) = -0.071$, $E(^3P) = -0.564$, and $E(^1P) = 2.147$ eV. Diagonalization of the matrix in Eq. (76) gives spectral peaks with energies of −66.83, −64.51, and 99.27 eV with relative intensities 39.6%, 19.7%, and 40.7%, respectively.

Summarizing, for the $M_{4,5}$ and $N_{4,5}$ edges the final states are close to $jj$ coupling scheme, because the 3d and 4d spin-orbit interaction is much larger than the core-valence electrostatic interactions, so that the spectrum is split into a $j = 5/2$ and $j = 3/2$ structure. The branching ratio of these edges is related to the ground state spin-orbit interaction, and, therefore, can be used to understand the angular momentum coupling mechanisms for the actinide elements.

## IV. PHOTOEMISSION SPECTROSCOPY

### A. Basics

Photoelectron emission (PE) spectroscopy is a well-known tool to study the composition and electronic structure of materials. The small electron elastic escape depths render the technique rather surface sensitive, thus good vacuum conditions are required to conduct measurements of a prepared surface of a sample. The photoelectron inelatic mean free path varies as a function of kinetic energy with a minimum around 40 eV. Bulk sensitivity is gained using high photon energy, but at the cost of reduced cross-section and often reduced photon energy resolution. One distinguishes between XPS and UPS, when using soft x-rays and ultraviolet radiation, respectively. XPS is primarily performed with Al or Mg $K\alpha$ radiation from a lab x-ray source or monochromatized radiation from a synchrotron; UPS is mainly performed using He I or He II gas discharge lamp in the laboratory. A benefit of the different photon energies, e.g., using tunable synchrotron radiation, is that it gives different relative cross-sections of the transitions involved, thereby providing a way to distinguish between them. PE can also be performed in angle-resolved fashion using a single-crystal sample, or at resonance using x-ray energies that coincide with a core-valence excitation (Terry *et al.*, 2002).

In the PE process, a photon $h\nu$ is absorbed under emission of an electron with kinetic energy, $E_{\text{kin}}$. Energy conservation requires that $E_{\text{kin}} = h\nu + E_N - E_{N-1}$, where $E_N$ and $E_{N-1}$ are the energies of the $N$-electron initial state and the ($N$–1)-electron final state. The energy $E_B = h\nu - E_{\text{kin}} = E_{N-1} - E_N$ is abusively called the electron binding energy; however, it is really the particle removal energy. Only when the electrons do not see each other (i.e. if there is no correlation) does the PE give the one-electron density-of-states (DOS). However, in correlated materials, such as many of the actinide metals, PE should first and foremost be regarded as a probe of the many-electron state.

Inverse photoelectron spectroscopy (IPES) gives information about the *unoccupied* states above the Fermi level. In this technique, the sample is irradiated with a monochromatized beam of electrons, and the photons emitted during the decay process are measured as a function of energy. IPES is inherently surface-sensitive due to the low energy of the incident electrons, while Bremsstrahlung isochromat spectroscopy (BIS) is more bulk sensitivity due to higher energy of the electrons. One drawback to the technique is that the signal levels are approximately five orders of magnitude weaker than in regular PE (Smith, 1988), meaning the total signal is rather weak.

### B. Theory in a nutshell

From calculational point of view, PE is similar to XAS, even though in PE the excited electron goes into a continuum state, and ultimately into the detector, instead of to an unoccupied valence state. As we will see, this has large implications for the screening of the photo-excited hole. Thus, in the actinide atom, the valence and core PE can be represented by electric-dipole transitions $5f^n \rightarrow 5f^{n-1}\varepsilon$ and $5f^n \rightarrow \underline{c}5f^n\varepsilon$, respectively, where $\varepsilon$ is a continuum state far above the Fermi level and $\underline{c}$ denotes a core hole.

In the so-called sudden approximation, one assumes that the excited photoelectron has no interaction with the state left behind, so that in the calculation the photoelectron state can be decoupled from the atomic state. The PE spectrum, as a function of binding energy, $E_B$, is expressed as

$$I(E_B) \propto \sum_{nmm'\sigma\sigma'} \left|\langle f_n | \varepsilon^\dagger_{m'\sigma'} c_{m\sigma} | g \rangle\right|^2 \delta(E_B + E_g - E_{f_n}), \quad (79)$$

where $|g\rangle$ and $|f_n\rangle$ are the ground and final state with energy $E_g$ and $E_{f_n}$, respectively, and $c_{m\sigma}$ is the annihilation operator of an electron $c$ with quantum numbers $m$ and $\sigma$, and $\varepsilon^\dagger_{m'\sigma'}$ is the creation operator of a continuum electron $\varepsilon$ with quantum numbers $m'$ and $\sigma'$.

The calculated PE of rare earth metals shows intense multiplet structure (van der Laan and Thole, 1993; Lademan *et al.*, 1996) and satellite peaks in mixed valence Ce compounds (Fuggle *et al.*, 1980). For the actinides the multiplet structure in intermediate coupling of the 5$f$ PE has been calculated by Gerken and Schmidt-May (1983).

The calculation of BIS and IPES is similar to that of PE, by replacing Eq. (79) by its time reversed counterpart, where an electron is annihilated in a state above the Fermi level and an additional $f$ electron is created. It was shown that the calculated BIS spectra of rare earths, arising from the electric-dipole transition $4f^n \varepsilon_d \to 4f^{n+1}$, show a detailed multiplet structure (van der Laan and Thole, 1993), evidencing that the belief that BIS simply measures the unoccupied DOS is an oversimplification.

**1. Screening of the photoinduced hole**

Much insight can be gained from a simple two-level model (van der Laan *et al.*, 1981). Consider an initial state $g$ composed of two basis states $\psi_a$ and $\psi_b$ with an energy difference, $\Delta = E_b - E_a$, and mixed by an interaction or hybridization with matrix elements $V = \langle \psi_a | H | \psi_b \rangle$. Introducing the mixing parameter $\theta$, defined by $\tan 2\theta = 2V/\Delta$, the ground state can be written as

$$\psi_g = \psi_a \sin\theta - \psi_b \cos\theta. \quad (80)$$

After electron emission, the final-state basis functions are $\psi'_a = \varepsilon^\dagger c \psi_a$ and $\psi'_b = \varepsilon^\dagger c \psi_b$ with an energy difference $\Delta' = E'_b - E'_a$ and mixing $V' = \langle \psi'_a | H | \psi'_b \rangle$, giving a mixing parameter $\theta'$ defined by $\tan 2\theta' = 2V'/\Delta'$. This gives the 'bonding' and 'antibonding' final states as

$$\begin{aligned}\psi_M &= \psi'_a \sin\theta' - \psi'_b \cos\theta', \\ \psi_S &= \psi'_a \cos\theta' + \psi'_b \sin\theta',\end{aligned} \quad (81)$$

with an energy separation of $E_S - E_M = (\Delta'^2 + 4V'^2)^{1/2}$. Using Eq. (81) we obtain the relative intensity ratio of the satellite to main peak, given as

$$\frac{I_S}{I_M} = \frac{\left|\left\langle \psi_S \left| \varepsilon^\dagger c \right| \psi_g \right\rangle\right|^2}{\left|\left\langle \psi_M \left| \varepsilon^\dagger c \right| \psi_g \right\rangle\right|^2} = \left(\frac{\sin\theta'\cos\theta - \cos\theta'\sin\theta}{\cos\theta'\cos\theta + \sin\theta'\sin\theta}\right)^2 = \tan^2(\theta' - \theta), \tag{82}$$

where the indices $M$ and $S$ relate to the main and satellite peak, respectively. This demonstrates the important fact that the satellite intensity depends only on the difference in hybridization between the initial and final state. Thus, if the PE process induces no change in the hybridization, i.e., $\theta' = \theta$, then all the intensity is in the main peak and the satellite intensity is zero. Since PE creates a hole in either the valence band or core level, there will be will in general a change in the energy difference between the two basis states due to screening, so that $\Delta' \neq \Delta$ and a satellite peak will be present.

For instance, consider the ground state as a mixture of $|5f^n\rangle$ and $|5f^{n+1}\underline{k}\rangle$ and the final state a mixture of $|\underline{c}5f^n\varepsilon\rangle$ and $|\underline{c}5f^{n+1}\underline{k}\varepsilon\rangle$, where $\underline{k}$ denotes a reservoir of hole states near the Fermi level. The underlying physical picture is one in which the $f$ electrons fluctuate among the two different atomic configurations by exchanging electrons with a reservoir. Quantum mechanically, the electrons can, for short periods of time, preserve their atomic character in a superposition of two atomic valence states with different number of 5$f$ electrons, while at the same time also maintaining their metallic, delocalized hopping between neighboring sites (Shim *et al.*, 2007). Correlations are strongest when the electrons are on the same atom. If the energy difference between the initial states is taken as $\Delta$, then the energy difference between the final states is $\Delta' = \Delta - Q$, where $Q$ is the $c$-5$f$ Coulomb interaction. For $Q = 0$ we obtain only the main peak, which will contain multiplet structure. When the Coulomb interaction is switched on, the satellite peak appears, also showing multiplet structure. If $Q>\Delta$, then the satellite peak is at a lower intensity than the main peak. This gives a well-screened peak that can have a higher intensity than the main peak, depending on the precise values of $Q$, $\Delta$, and $T$.

It should be noted that mixing of different configurations also occurs in the case of XAS and EELS, but in this case the photo-excited core electron goes into an $f$ state, which provides a very effective way to screen the core hole. Here, the final state energy is $\Delta' = \Delta - Q + U$, where $U$ is the $f$-$f$ Coulomb interaction, and as a rough guide $U/Q \approx 0.8$. Thus, the core hole is well screened, resulting in a low satellite intensity.

In localized systems the core hole potential gives rise to a poorly screened photoemission peak. In metallic systems, on the other hand, the core hole can be screened by valence electrons from surrounding atoms, giving rise to a well screened peak, which is at lower binding energy compared to the unscreened peak. The relative peak intensities will be a function of the mixing integral and the Coulomb interaction. Looking at the 3$d$ transition metals, nickel metal is known as a correlated-electron system and shows a very clear satellite structure in PE (van der Laan *et al.*, 2000). Other 3$d$ metals, such as cobalt, have satellites that are substantially weaker (Panaccione *et al.*, 2001), providing clear evidence that the satellite structure is due to mixing of localized $d^n$ configurations. When delocalization sets in, an asymmetric line shape starts to appear and the satellite peaks diminish. The sudden creation of the core hole in PE gives rise to the creation of low-energy electron-hole pair excitations that show up as a peak asymmetry that can be fitted by a Doniach-Sunjic (DS) line shape (Doniach and Sunjic, 1970). The peak asymmetry becomes larger with increasing density of states at the Fermi level.

**2. Kondo resonance**

The hopping between localized and itinerant states gives rise to a Kondo peak near the Fermi level (Allen, 1985). Gunnarsson and Schönhammer (1983) evaluated the single-impurity degenerate Anderson Hamiltonian, where the symmetrized projection of conduction band states hybridize to the local state (Allen, 1985). Nominally trivalent Ce $f^1$ provides a useful example to demonstrate emergent Kondo and quasi-particle properties (Allen, 2002). First assume $V = 0$ and $U \neq 0$. The spectral function, made up by joining the PE and BIS spectra at $E_F$, will contain an $f^1 \rightarrow f^0$ ionization peak (in PE) and $f^1 \rightarrow f^2$ affinity peak (in BIS), with a separation equal to $U$. This system has a localized magnetic moment and a Curie law magnetic susceptibility. When $V$ is switched on, the mixed ground state $\left| f^1 \underline{k} \right\rangle + \left| f^0 \right\rangle$ is a singlet (due to its $f^0$ character) and as temperature decreases the susceptibility changes from Curie-like to Pauli-like around a Kondo temperature $T_K$. The $V = 0$ spectrum now has an additional peak: the Kondo resonance at $E_F$. The appearance of a Kondo resonance is thus the spectral manifestation of a change in ground state with a disappearing magnetic moment. Consider now the opposite case, the so-called Fermi liquid, with $U = 0$ and $V \neq 0$. Since electrons can be added and removed at no extra energy, the spectral function is a single peak at the Fermi level. With equal occupancy of all orbitals the magnetic moment is zero. Switching on $U$, the ionization and affinity peaks appear while the mixing requires the presence of a quasi-particle peak, i.e. Kondo resonance at $E_F$.

Pu is an example of a strongly correlated material, in which the valence electrons interact with each other. The mixing of the 5$f$ electrons and $s$, $p$, $d$ electrons determine many of the key properties of Pu, such as its lack of magnetism and poor conductivity. It is suggested that Kondo shielding is responsible for the lack of a magnetic moment in Pu (Savrasov *et al.*, 2001; Shim *et al.*, 2007). The Kondo regime describes the dynamical screening effects in situation where magnetic moments are present on the mean field level in Anderson model. Thus, δ-Pu is in a paramagnetic state with its moment entirely screened by the Kondo effect. In a Kondo model, a resonance is observed at the Fermi level in the form of a quasiparticle peak, and this is found in the Pu valence-band PE spectrum (Gouder *et al.*, 2001).

Multiplet effects are clearly visible in the Pu 5$f$ PE and provide widths to the Hubbard bands, as pointed out previously in studies of americium (Savrasov *et al.*, 2006). In fact, multiplet structure in combination with band-like spectra have employed in a variety of PE studies on actinides (Okada, 1999; Gouder *et al.*, 2005; Svane, 2006; Shick *et al.*, 2007). What is more, this mixture of band-like and atomic-like behavior forces the question when doing calculations whether to start from a band or atomic limit. Many progressive theoretical approaches of δ-Pu start from the atomic limit, since it exhibits a more atomic nature, as will be shown in subsequent sections.

## C. Experimental results

### 1. Inverse photoemission

IPES and BIS have only been performed on α-Th and α-U. The BIS spectra for each metal are shown in Fig. 19 (Baer and Lang, 1980), where the intensity is plotted as function of energy above the Fermi level. It is seen that both spectra have a double-lobe structure, where the peaks in the Th spectrum are well above the Fermi level while the peaks in U are close enough to the Fermi level that one is cut off at $E_F$. To date, no other IPES or BIS has been published for the other actinide metals. Of great importance would be Pu, specifically the α and δ phases. While there is available valence-band photoemission to compare with, there is no inverse photoemission, leaving the states above the Fermi level unmeasured except through EELS and XAS.

## 2. Valence-band photoemission

The valence-band XPS or UPS spectra for α-Th (Baer and Lang, 1980), α-U (Baer and Lang, 1980), α-Np (Naegele *et al.*, 1987), α- and δ-Pu (Gouder *et al.*, 2001), and α-Am (Naegele *et al.*, 1984) are collected in Fig. 20. The intensity for α-Th is scaled up compared to the other actinides, because its spectrum is much lower in intensity due to the small *f* density of states at the Fermi level. This is clear when looking at the BIS of α-Th in Fig. 19, where the intensity near the Fermi level is low in relation to that of α-U. In fact, the double peak structure in the α-Th valence-band spectrum is primarily of *d* character with only a small *f* contribution. A saw-tooth shape is then observed in the spectrum of α-U, α-Np, α-Pu, and δ-Pu; the overall asymmetric DS line shape of these metals is indicative of a relatively delocalized system. The amount of fine structure in the spectra grows when moving along the actinide series from U to Pu, indicating increased localization of the 5*f* states. Fine structure is entirely absent in the α-U spectra, slightly present in the α-Np spectra, even more present in the α-Pu spectra, then clearly visible in the δ-Pu spectra. Finally, α-Am shows a valence-band spectrum that is well removed from the Fermi level, evidence that the *f* states are mostly localized.

## 3. 4*f* core photoemission

The 4*f* XPS spectra for α-Th (Moser *et al.*, 1984), α-U (Moser *et al.*, 1984), α-Np (Naegele *et al.*, 1987), α- and δ-Pu (Arko *et al.*, 2000b), and α-Am (Naegele *et al.*, 1984) are collected in Fig. 21. Due to the large 4*f* spin-orbit interaction there are two distinct manifolds, i.e., $4f_{5/2}$ and $4f_{7/2}$. The lighter metals exhibit an asymmetric DS line shape for the $4f_{5/2}$ and $4f_{7/2}$ peaks due to the delocalized nature of Th, U, and Np; however, Pu and Am start to show broad peaks due to unresolved multiplet structure. The degree of 5*f* delocalization is reflected through the satellite peaks at the high-energy side of the $4f_{5/2}$ and $4f_{7/2}$ structure. These are indicative of poor screening of the photo-induced core hole. In a delocalized system with a high density of 5*f* states at the Fermi level, the $4f_{5/2}$ and $4f_{7/2}$ peaks are asymmetry due to a considerable number of low-energy electron-hole pairs generated during the photoemission process (Doniach and Sunjic, 1970). The bulk of screening in the light actinides is performed by the delocalized 5*f* electrons (Johansson *et al.*, 1980), but this occurs to varying degrees. Looking along the actinide metal series, the satellite peaks are present in Th, entirely absent in U, weak in Np where they appear merely as a change in slope where the satellite peak should be, stronger in α-Pu, then even stronger in δ-Pu. The Am spectrum consists almost entirely of the poorly-screened peak, with a very small amount of weight at the position of the well-screened peak.

## D. Photoemission as a probe for 5*f* localization in Pu

The PE experiments by Gouder *et al.* (2001) and Havela *et al.* (2002) represent what is arguably the most clean and informative photoemission data available for Pu. What is more, they clearly show that the 4*f* PE is sensitive to, and can track the progress of, 5*f* localization in Pu. This is shown in Fig. 22 where the valence-band photoemission spectrum of Pu is tracked as single monolayers (ML) of Pu are deposited on a Mg substrate. At 1 ML the spectrum shows a large peak at about 0.8 eV, strongly resembling the spectrum of δ-Pu in Fig 20. As the deposited thickness increases, the peak at 0.8 eV diminishes and the main peak at the Fermi level become prominent. The shift of intensity from the peak at 0.8 eV to the peak at the Fermi level, gives direct evidence of a change towards more delocalized 5*f* states. Thus, the fine structure of valence-band photoemission of Pu is a sensitive probe for the degree of localization of the 5*f* states. One drawback to the Gouder *et al.* (2001) and Havela *et al.* (2002) data is there is no structural information, such as low-energy electron diffraction, to determine the phase(s) being examined. Nonetheless, the data clearly illustrate changes in localization of 5*f* states can be tracked using valence-band photoemission.

The 4*f* core photoemission spectra can also be used to track the degree of 5*f* localization, as shown in the inset of Fig. 22. In this case, there is a peak on the high-energy side of the $4f_{5/2}$ and $4f_{7/2}$ peaks due to poor screening of the photo-induced core hole, as described above. The strength of this poorly screened peak is large enough in the 1 ML spectrum that it appears similar to Am, where the 5*f* states are almost completely localized. However, by the time 9 ML is achieved the two peaks are sharper with a more asymmetric DS line shape, indicating a relative delocalization of the 5*f* states. 4*f* core photoemission can also give insight into the 5f valence. Using an Anderson impurity model, Cox *et al.* (1999) find from fitting the 4*f* spectra 5.03 and 4.95 5*f*-electrons for the ground states of the δ and α phases, respectively, corresponding to valence state distributions of ~0 and 0.06 for $f^4$, 0.97 and 0.88 for $f^5$, and 0.03 and 0.06 for $f^6$.

The above data clearly show that valence-band and 4*f* photoemission can be used to track the degree of 5*f* localization in Pu, as well as other 5*f*-electron compounds and alloys (Havela *et al.* 2003; Gouder *et al.* 2005). In fact, XPS is considerably more sensitive to 5*f* localization than EELS and XAS measurements. EELS measurements on α- and δ-Pu show only small changes in both the $O_{4,5}$ and $N_{4,5}$ edges (Moore *et al.*, 2003; 2006b). As mentioned, this is because in EELS the excited core electron goes into an unoccupied *f* state that efficiently screens the core hole, whereas in PE the core electron is excited into a continuum state leaving the core hole largely unscreened. Regardless, all techniques have their strengths and so EELS, XAS, many-electron atomic calculations, and inverse, valence-band, and 4*f* photoemission will all be used to examine the electronic and magnetic structure of each elemental actinide metal in detail in the following sections.

## V. ELECTRONIC STRUCTURE OF ACTINIDE METALS

### A. Thorium

A brief historical perspective of Th is in order, given that it was one of the first actinides rigorously examined for electronic and magnetic structure. When experiments first began on actinides it was generally assumed that the 5*f* states would behave similar to the 4*f* states of the rare earths, which are localized and atomic-like. Initial calculations on Th by Gupta and Loucks (1969) artificially removed the 5*f* states in order to obtain good agreement with de Haas-van Alphen experiments by Thorsen *et al.* (1967) and Boyle and Gold (1969). However, Freeman and Koelling (1971; 1972) contended that the 5*f* states were indeed delocalized and hybridizing with the 6*d* and 7*s* bands. These arguments were put to rest by Veal *et al.* (1973), where normal-incidence reflectivity measurements clearly showed that the 5*f* states of Th were itinerant and bonding, thus acting band-like. Later results by Weaver and Olson (1977) and Alvani and Naegele (1979) showed that the optical measurements by Veal *et al.* (1973) were partly inconclusive due to surface roughness of the sample. However, the fact remained that the 5*f* states of Th were proven to be delocalized and unlike the 4*f* states of the rare earth metals. Thus, the stage was set for the actinide metals to be different from the rare earth metals.

A pressing question at the time was how the fcc phase was possible in Th with delocalized 5*f* states. The answer was found in assuming entirely unoccupied 5*f* states, namely that the *s*, *p*, and *d* states achieved bonding of the metal. The first experimental indication that this was not the case was given by Baer and Lang (1980), where PES and BIS were employed to directly probe the density of states below and above the Fermi level, respectively. The BIS results for Th (reproduced in Fig. 19) clearly showed a shoulder at the Fermi level that indicated there must be a modest 5*f* occupation. The relativistic self-consistent calculations by Eckart (1985) also showed a modest *f* density of states at the Fermi level. Thus, at this point experiment and theory showed a small electron occupation in the 5*f* states.

Theoretical calculations by Skiver and Jan (1980) and Johansson *et al.* (1995) supported the fact that the 5*f* states of Th metal were delocalized and hybridizing with the *s*, *p*, and *d* valence bands. The results of Johansson *et al.* (1995) clearly showed that the metal was not a tetravalent *d* transition metal, since it would exhibit a bcc crystal structure given such an electronic configuration. What this meant was that even though Th was fcc, which was out of character with other light actinide metals that have low symmetry structures, there was active 5*f* bonding with some electron weight in the states due to hybridization. It should have come as little surprise that Th exhibits an fcc structure while having some electron weight in delocalized 5*f* states, since α-Ce exhibits an fcc structure having active bonding of about one 4*f* electron (Johansson 1974; Allen and Martin 1982).

With the roots of condensed-matter research set for Th and more ultimately the entire actinide series, we will now turn our attention to spectroscopy of the metal. The valence-band PE of Th by Fuggle *et al.* (1974), Veal and Lam (1974), Baer and Lang (1980), Baer (1980), and Naegele (1989) support a delocalized 5*f* band with a modest *f* electron count. The spectrum is shown in Fig. 20 (Baer and Lang, 1980), where the scale has been increased compared to the other metals. There is a double-peak structure that is primarily of *d* character with only a small *f* contribution. On the same scale, the Th valence-band PE would be much lower in intensity than U-Am because there is little *f* density of states at the Fermi level. The calculated density of states by Freeman and Koelling (1977) and Skriver and Jan (1980) are in good agreement, being almost identical to the experimental valence-band PE spectra.

The 4*f* PE spectra measured by Moser *et al.* (1984), shown in Fig. 21, contain asymmetric $4f_{5/2}$ and $4f_{7/2}$ peaks that are indicative of a delocalized actinide metal. There is also a satellite peak at the high-energy side of the $4f_{5/2}$ and $4f_{7/2}$ peaks. These satellite peaks are also present in the Th 4*f* PE spectra of McLean *et al.* (1982) and the 5*p* PE spectra of Sham and Wendin (1980), and are indicative of poor screening. Examining Fig. 21, we see that the satellite peaks are present in Th, but then entirely absent in U (Greuter *et al.*, 1980; Allen *et al.*, 1981) and almost entirely absent in Np (Naegele *et al.*, 1987). Th metal does not have localized 5*f* states, so why then are the satellite peaks present? Moser *et al.* (1984) argue that it is likely due to *sd* conduction electrons. Based on the Schönhammer and Gunnarson (1979) model, Fuggle *et al.* (1980) suggest that the shoulders result from the transfer of charge from the Fermi level to the unoccupied screening levels that are pulled down below the Fermi level when the hole is created in the localized core levels. The asymmetry of the $4f_{5/2}$ and $4f_{7/2}$ peaks in the U and Np spectra are due to the high density of 5*f* states at the Fermi level that cause a considerable number of low-energy electron-hole pairs to be generated during the photoemission process (Doniach and Sunjic, 1970). In Th, even though the 5*f* states are delocalized in the metal, there is not a large density of *f* states at the Fermi level to properly screen the core hole. This is clear from the fact that the valence-band PE spectrum for Th in Fig. 20 is mostly due to *d* states, with only a small *f* contribution, and the fact that the BIS spectrum by Baer and Lang (1980) in Fig. 19 shows most of the density of states above the Fermi level. A considerable amount of screening in the light actinides is due to the delocalized 5*f* electrons (Johansson *et al.*, 1980), but for Th there is simply not enough *f* electron density of states to effectively screen the core hole, producing poorly-screened satellite peaks.

The EELS and many-electron atomic spectral calculations for the $O_{4,5}$ and $N_{4,5}$ edges support the above picture, where Th is delocalized with some electron weight in the 5*f* states. The actual number of electrons in the 5*f* states, however, shows some variation. In the literature reviewed above, the 5*f* count is quoted between 0 and 0.5. In our EELS and many-electron atomic spectral calculations we observe either a 0.6 or 1.3 5*f* electrons, depending on the type of background removal utilized. In van der Laan *et al.* (2004) and Moore *et al.* (2004a; 2006c) a standard XAS background removal and peak fitting is used, yielding $n_f = 0.6$. This is of course close to the 0.5 count given by Baer and Lang (1980). Subsequent analysis by Moore *et al.* (2007a; 2007b) where the second derivative background removal is used yields $n_f = 1.3$. A 5*f* electron count of 1.3 for Th is high compared to literature, but this is the lowest number that does not yield a negative $j = 7/2$ occupation, which would be physically unrealistic. This higher 5*f* count may be due to how the second-derivative peak integration technique handles the rather peculiar background between the $N_4$ and $N_5$ peak in Th (see Fig. 13). Also, the uncertainty becomes larger when the branching ratio nears the statistical value 3/5, which is the case for Th metal. We feel that an *f* count near 0.5 is most accurate and in line with experiment and theory. Interestingly, *f* count variation between 0 and 1 makes no change in the angular momentum coupling scheme, since it take more than one electron for entanglement. Thus, up to $f = 1$ all three angular momentum coupling mechanisms are equivalent and even for $f = 1.3$ all three are almost identical.

Diamond-anvil-cell experiments show that fcc Th is stable up to 63±3 GPa (Ghandehari and Vohra, 1992), where it transforms in a gradual distortion to a body-centered tetragonal structure with the space group *I*4/*mmm* (Bellussi *et al.*, 1981; Akella *et al.*, 1988; Eriksson *et al.*, 1992a; Vohra and Akella, 1991). A Bain-type distortion allows the fcc crystal to transform to bcc (or body-centered tetragonal since it is only a slight distortion in one axis), where the transition is continuous and thermodynamically of second order (Ghandehari and Vohra, 1992). Thus, there is no dislocation motion needed for the transformation and the change in crystal structure is simply a distortion. Once the metal is pressurized to 63±3 GPa, the 5*f* band broadens and the *f* electron occupation increases, allowing the system to adopt a low-symmetry crystal structure indicative of actively bonding *f* states.

**B. Protactinium**

Protactinium has the express honor of having the least amount of data for the light actinides, which is mostly due to the fact that it has little or no technological applications. In fact, there are no spectra of the metal, besides the Mössbauer Spectroscopy of Friedt *et al.* (1978). Regardless of this lack of literature to review, we will discuss the electronic structure of Pa to the best of our abilities. Hopefully, this section will supply initial benchmarking for the behavior of the Pa 5*f* states, which can be utilized for future investigations of this heretofore neglected metal. Many of the crystallographic and physical properties of Pa are reviewed by Blank (2002).

Crystallographic determination shows that Pa adopts a bcc body-centered tetragonal structure (Zachariasen 1952a), as seen in Th under pressure (Ghandehari and Vohra, 1992). The metal also superconducts at 1.4 K (Fowler *et al.*, 1965; Smith *et al.*, 1979a). The low-symmetry nature of the body-centered tetragonal structure is the first suggestion that Pa metal has delocalized 5*f* states that are actively bonding. As shown in the previous section, there is ample evidence that Th has delocalized 5*f* states. Examining the spin-orbit analysis of the $N_{4,5}$ EELS edges in Fig. 17(a), we see that the metal shows little discernable difference between the three coupling mechanisms due to its low 5*f* occupancy. In the case of U, Fig. 17(a) clearly shows a pure *LS* coupling mechanism, which is due to the delocalized 5*f* states in the metal. Due to the delocalized nature of the 5*f* states in the light actinide metals, it seems safe to assume that the 5*f* states in Pa are delocalized and can be properly modeled with an *LS* coupling mechanism. Thus, adopting a 5*f* count of 2 for Pa (Söderlind and Eriksson, 1997), one could anticipate a spin-orbit expectation value of approximately –0.17, which would put the data point on the *LS* coupling curve.

When pressurized to 77±5 GPa in a diamond anvil cell, Pa metal transforms to the orthorhombic α-U structure with the space group *Cmcm* (Haire *et al.*, 2003). Associated with this phase transformation is a ~30% volume collapse, as predicted by Söderlind and Eriksson (1997), albeit at 25 GPa rather than the 77 GPa observed experimentally. Interestingly, many high-pressure experiments of light actinide metals result in the α-U structure being stable, as well as many rare-earth metals, such as Ce, Nd, and Pr (Ellinger and Zachariasen, 1974; McMahan and Nelmes, 1997; Chestnut and Vohra, 2000). Of course at 100s of GPa of pressures, close-packed metal structures, such as fcc, hcp, and bcc, are once again favored due to large electrostatic repulsions ruling out the more open and lower symmetry structures. This is indeed observed in Pa by Söderlind and Eriksson (1997) where the hcp structure becomes stable at extreme compressions.

## C. Uranium

Compared to all other actinides, uranium metal is rather well understood, mostly due to the fact that it is technologically relevant. There is a detailed review on U metal by Lander *et al.* (1994) and a shorter review by Fisher (1994). For this reason we will focus on the electronic structure of the 5*f* states, as well as some new physics of the metal that have evolved since 1994. Thus, we will first cover the angular momentum coupling mechanism, electron filling, and valence of the metal. We will then discuss the three charge density waves (CDWs) in U metal and how they relate to possible CDWs in α-Np and α-Pu. Finally, we will discuss intrinsic localized modes in α-U, which represent the first reported three-dimensional non-linear modes observed in a material. While these are not strictly an electronic effect, electron-phonon interactions deem them interesting and relevant for our purposes.

### 1. Why does U metal exhibit *LS* coupling?

The EELS and many-electron spectral calculation analysis in Secs. II and III clearly show that U metal falls on the *LS* coupling curve for the 5*f* states. It is the heaviest actinide metal to exhibit such a behavior. Indeed, valence-band (Baer and Lang, 1980; see also: Veal and Lam, 1974; Nornes and Meisenheimer, 1979; Baer, 1980; Schneider and Laubschat, 1980; Greuter *et al.*, 1980; McLean *et al.*, 1982; Allen and Holmes, 1988; Naegele, 1989; Molodtsov *et al.*, 2001; Opeil *et al.*, 2006) and 4*f* (Moser *et al.*, 1984; see also: Greuter *et al.*, 1980; Allen *et al.*, 1981; McLean *et al.*, 1982) PE spectra of α-U, which are shown in Figs. 20 and 21, respectively, support a metal with delocalized 5*f* states. The valence-band PE exhibits no discernable structure in the spectrum other than the asymmetric Doniach-Sunjic line shape indicative of a delocalized system, while the 4$f$ spectrum exhibits strong asymmetric 4$f_{5/2}$ and 4$f_{7/2}$ peaks with no apparent poorly-screened satellite peaks. In fact, valence-band and 4$f$ PE spectra remain structureless until less than a single monolayer of U is reached on free-standing thin films, at which point structure indicative of 5$f$ localization appears (Gouder and Colmenares, 1993; Gouder and Colmenares, 1995). The BIS spectrum of α-U in Fig. 19 (Baer and Lang, 1980) also illustrates delocalized 5$f$ states with a high density of $f$ states at the Fermi level. All this begs the question: if the 5$f$ states are known to exhibit strong spin-orbit interactions (Freeman and Lander, 1984), why then do we see pure $LS$ coupling from the EELS, XAS, and PE results?

The answer is that while U certainly has a strong spin-orbit interaction, the 5$f$ states in the metal are delocalized enough to create an $LS$-like situation due to mixing of the $j = 5/2$ and 7/2 levels. In other words, the delocalization of the 5$f$ states in α-U reduces the angular part of the spin-orbit interaction, even though there is indeed a strong radial part of the spin orbit interaction (van der Laan et al., 2004; Tobin et al., 2005). It is important to note that if the spin-orbit value falls on the $jj$-curve, then the atom has $jj$-coupling, because this is the only way to create a large spin-orbit interaction. However, if the spin-orbit value falls on the $LS$-curve, the atom does not necessarily need to have $LS$ coupling, although this will probably be often the case. There are other ways to couple spin and orbital moments that result in a reduced spin-orbit interaction of equal size as in $LS$ coupling. Therefore, a low spin-orbit value can not be uniquely assigned to $LS$ coupling. On the other hand, for given spin-orbit value and number of electrons, the partition into $j = 5/2$ and $j = 7/2$ states is unique. Broadening of these $j$ levels into bands due to hybridization results in increased mixing, increased $j = 7/2$ character, and hence reduced spin-orbit interaction. Once the 5$f$ states become slightly more localized than in α-U, they begin to exhibit the strong spin-orbit interaction. Evidence of this is given through itinerant $f$-electron magnetic materials and dimensional constraint of the metal, which causes localization of the 5$f$ states.

Numerous itinerant $f$-electron magnets exhibit a substantial orbital and spin magnetic moment (Brooks and Kelly, 1983; Fournier et al., 1986; Norman et al., 1988; Wulff et al., 1989; Severin et al., 1991). Neutron scattering experiments clearly show anomalous behavior of the $f$ magnetic form factor that is due to the strong orbital component. The field-induced magnetic form factor of pure α-U metal shows a monotonic decrease as a function of the scattering wave vector, which is rather normal behavior (Maglic et al., 1978). However, spin polarized electronic structure calculations show that the spin and orbital moments of U metal are aligned parallel when exposed to an external magnetic field (Hjelm et al., 1993). This is in contradiction to Hund's third rule, which specifies that the spin-orbit interaction of the 5$f$ states will cause the spin and orbital moments to be antiparallel in U (see Fig. 18(a) and (b)). This means that the magnetic field applied by Hjelm et al. (1993) in their spin polarized electronic structure calculations on α-U was enough to destroy the Hund's rule ground state with antiparallel spin and orbital moments. The spin-orbit interaction becomes large enough to mix higher $L$ and $S$ states into the ground state, yielding intermediate coupling.

Dimensional constraint of U atom in U/Fe multilayers results in more localized 5$f$ states. Using the U $M_{4,5}$ branching ratio, Wilhelm et al., (2007) show that 9 ML of U yield an XAS spin-orbit expectation value of -0.142, while 40 ML of U yield a spin-orbit expectation value of -0.215, which is similar to bulk α-U. This result means that as the U thickness in the multilayer is reduced, the 5$f$ states of the metal behaves more atomic-like, exhibiting a spin-orbit expectation value in accordance with intermediate coupling.

Although α-U metal exhibits an *LS* coupling mechanism, actinide elements have a pronounced tendency toward *jj* or intermediate coupling when in free atomic form.  This is clearly shown by the experimental and theoretical absorption spectra of Carnall and Wybourne (1964) where the actinide ion $U^{3+}$, $Np^{3+}$, $Pu^{3+}$, $Am^{3+}$, and $Cm^{3+}$ are studied in their dilute limit in various solution media. The spin-orbit coupling parameters of the actinides are approximately twice as large as found in the rare earths (Wybourne 1965), while the electrostatic parameters are two-thirds as large (c.f. Table II). This results in the actinide ions exhibiting a large departure from *LS* coupling.  The calculations of the composition of the states in both the *jj* and *LS* limit by Carnall and Wybourne (1964) show the coupling to be truly intermediate, as expected in a fully localized limit. Thus, while U metal exhibits *LS* coupling, it does in fact have strong spin-orbit interaction, which is masked by the degree of delocalization (band width) of the 5*f* states. Strong changes in the branching ratio have also been observed for Mn submonolayer thin films. The $L_{2,3}$ branching ratio in x-ray absorption increases dramatically when the electrons become localized in the Mn ultrathin film (Durr *et al.*, 1997). However, we need to distinguish this result from that of the actinides, because in the Mn case these changes originate from the *jj* coupling in the final state, which due to the 2*p*,3*d* electrostatic interactions affects the $L_{2,3}$ absorption edge. When the 3*d* electrons become itinerant the 2*p*,3*d* interaction reduces strongly. Since in the case of the actinides the *jj* coupling is small, the effects are here not due to changes in 2*p*,3*d* interaction, but arise from the change in the angular part of the spin-orbit interaction upon 5*f* (de)localization.

While oxides are an entirely different topic than the metals discussed in this review, we digress for a paragraph because an interesting and poignant issue can be raised – the localization of the 5*f* states.  When a delocalized state localizes, the electrons change from *LS*-like to intermediate coupling. An example of this is found in the 5*d* transition metals, which have delocalized and bonding 5*d* states, as show in Fig. 1. Accordingly, an *LS* coupling mechanism is appropriate. However, once these metals form dioxides, the 5*d* states localize and adopt intermediate coupling. Examining the EELS data by Moore *et al.* (2006b), one can see that the difference between the branching ratio of α-U metal and $UO_2$ is 3.0%.  Comparing α-Pu and $PuO_2$, the difference in branching ratio between the metal and dioxide drops to 1.8%.  In fact, unpublished EELS results by Moore and van der Laan reveal that the difference in branching ratio between the ground-state metal phase and dioxide for Th, U, Np, Pu, Am, and Cm reveal the largest difference for Th and U, a smaller difference for Np, a still smaller difference for Pu, then no difference for Am and Cm. We interpret this as direct evidence for the degree of localization of the 5*f* states in each metal species.  The states are delocalized in Th and U, less so for Np, even less for Pu (which will be discussed in the subsequent Pu section), then are localized for Am and Cm.  As a result, the coupling mechanism for the 5*f* metals changes from *LS* to intermediate when moving across the actinide series in Fig. 17(a). The actinide oxides require much more EELS research given that the current interpretations of valence electron count for $UO_2$ and $PuO_2$ by Moore *et al.* (2006b) have caused controversy. However, the branching ratio difference between the actinide metal and dioxide phases is measured directly from EELS with no interpretation.  The difference has meaningful implications for the degree of 5*f* localization in U, as well as Th, Np, Pu, Am, and Cm. Lastly to note, the DFT results of Prodan *et al.* (2007) show a 5*f* – O 2*p* orbital degeneracy that leads to significant orbital mixing and covalency in the $PuO_2$, $AmO_2$, and $CmO_2$. Their results also show a strong Hund's rule exchange opposing spin-orbit coupling, which yields an unexpected ground state in $CmO_2$. This strong emergence of exchange interaction in the 5*f* states in $CmO_2$ also occurs in Cm metal, as will be discussed subsequently in the section on Cm.

The extreme stability of α-U due to delocalized and strongly bonding 5$f$ electrons is clearly illustrated by the diamond-anvil-cell studies by Akella *et al.* (1997), Weir *et al.* (1998), and Le Bihan *et al.* (2003). In these experiments, α-U is pressurized to 100 GPa, revealing no change in the orthorhombic *Cmcm* crystal structure and only a relaxation of the orthorhombic axial ratios. Given the fact that the α-U crystal structure repeatedly appears in rare earth and actinide metals under pressure, it is clear that it is the archetypal structure for active and strong $f$-electron bonding when the 5$f$ occupation is about 2-4 (Söderlind *et al.,* 1994b). If this is true, then U metal should act as a model system to study the unique physical behavior of strongly bonding $f$ electrons. Two examples of this are charge density waves and intrinsic localized modes, both of which occur in α-U.

## 2. Superconductivity in Th, Pa, and U

Examining Fig. 6(a), we see that the ground state of Th, Pa, and U are in the superconducting region (blue). Indeed, the superconducting transition temperatures, $T_c$, for Th = 1.4 K (Wolcott and Hein, 1958; Gordon *et al.*, 1966), Pa = 1.4 K (Fowler *et al.*, 1965; Smith *et al.*, 1979a), and U = 0.7 or 1.8 K depending on crystal structure (orthorhombic α-U $T_c$ = 0.7 K and fcc γ-U $T_c$ = 1.8 K. Chandrasekhar and Hulm, 1958). Under pressure, the superconducting transition temperature of Th drops from 1.4 K to ~0.7 K (Palmy *et al.*, 1971; Fertig *et al.*, 1972; Griveau and Rebizant, 2006). In an opposite manner, pressure increases the superconducting transition temperature of U from about 1.0 K at ambient conditions to 2.3 K at 10 kbar (Smith and Gardner, 1965). This fundamentally different behavior of $T_c$ with pressure between Th and U is not the only difference between the metals when considering superconductivity.

The superconducting transition curves for Th, Pa, and U in Fowler *et al.* (1965) show that while Pa and U are similar, Th is different. Th has an abrupt superconducting transition temperature that occurs over a narrow temperature range. However, Pa and U exhibit a broad superconducting transition temperature that occurs over a considerably wider temperature range. The transition temperature for U is somewhat more abrupt in Lashley *et al.* (2001a), but is still not as rapid as observed in Th (Fowler *et al.*, 1965). This could be due to the fact that the Th sample is very clean, whereas the Pa and U contain impurities, either elemental or defects. Alternatively, it could be due to the states that are responsible for superconductivity. As discussed above, Th has only ~0.5 $f$ electrons in the delocalized 5$f$ states, while Pa is ~5$f^2$ and U is ~5$f^3$. In other words, superconductivity in Pa and U may emanate in the 5$f$ states, which have enough electron occupation to dictate or influence $T_c$, whereas in Th it could emanate in the $s$, $p$, or $d$ states due to the low $f$ electron occupation. Indeed, there is no evidence of $f$ electrons in the superconductivity of Th (Gordon *et al.*, 1966), and there are arguments by Smith and Gardner (1965) that the superconductivity of U is intimately associated with the 5$f$ electrons.

Pure Np and Pu metal are non-superconducting down to 0.41 and 0.5 K, respectively (Meaden and Shigi, 1964). However, the Pu-bearing compound PuCoGa$_5$ exhibits a superconducting transition temperature at 18.5 K (Sarrao *et al.*, 2002), which is shockingly high. Also, Am metal superconducts below 0.8 K (Smith and Haire, 1978; Smith *et al.*, 1979) and exhibits a pressure-dependant $T_c$ that ranges from 0.7 to 2.2 K (Link *et al.*, 1994; Griveau *et al.*, 2005). Both PuCoGa$_5$ and Am will be discussed in subsections G and F, respectively.

## 3. Charge density waves in α-U; in α-Np and α-Pu also?

Charge density waves (CDWs) form in quasi-low dimensional materials and low-symmetry crystal structures where the Fermi surface is nested. They are typically observed in one- or two-dimensional materials; α-U produces the only known elemental system where a three-dimensional CDW forms (Lander, 1982). Elastic constant measurements to liquid helium temperature by Fisher and McSkimin (1961) show a large change in $c_{11}$ at 43 K, revealing a phase transformation at that temperature. The CDW is clearly observed in the phonon dispersion curves, where cooling to 30 K causes the $\Sigma_4$ branch in the [100] direction to condense, or reduce in frequency (Crummett *et al.*, 1979; Smith *et al.*, 1980; Smith and Lander, 1984). At this point, superlattice reflections appear due to charge ordering, which are incommensurate with the atomic lattice. Neutron diffraction experiments by Marmeggi *et al.* (1990) show there are three separate CDWs, one at 43 K ($\alpha_1$), one at 37 K ($\alpha_2$), and one at 22 K ($\alpha_3$), where $\alpha_1$ and $\alpha_2$ are incommensurate with the lattice and $\alpha_3$ is commensurate. First-principles calculations by Fast *et al.* (1998) show the strong nesting of the narrow 5*f* bands in the Fermi surface for the $\alpha_1$ phase

Given the fact that CDWs form in α-U, a delocalized actinide metal with a low-symmetry orthorhombic crystal structure, one may reasonably ask whether CDW(s) form in orthorhombic α-Np or monoclinic α-Pu. Unpublished research by Butterfield and Moore show no CDW forms in α-Pu to 8 K. Using a liquid-helium specimen holder, an α-Pu sample is cooled in a TEM, where no superlattice reflections are observed in electron diffraction. The use of a TEM on single-grain regions is of paramount importance to avoid problems due to polycrystalline samples. Measurements on α-U, such as specific heat, show clear peaks for single crystals, but no peaks for polycrystals (Mihaila *et al.*, 2006). In other words, the polycrystalline samples effectively wash out the signal over enough temperature range to loose the small peaks. Because Np and Pu have no known large single crystals, this creates the problem that peaks due to CDW(s) would be unresolved. However, using a TEM circumvents this issue by allowing electron diffraction to be recorded from single grains within a polycrystalline sample. Chen and Lander (1986) show this for α-U, where the superlattice reflections for all three CDWs are observed using electron diffraction in a TEM and imaged using dark-field techniques. To date, all our experiments on monoclinic α-Pu shown no superlattice reflections down to 8 K, indicating there is no CDW in the metal. Work is underway on α-Np, which is orthorhombic like α-U, albeit with a different space group.

## 4. Intrinsic localized modes

Defects cause a loss of periodicity in crystals; this is a tenant of crystallography. However, there are circumstances when periodicity can be broken in a perfect lattice that is free of defects. Such an idea was first proposed by Sievers and Takeno (1988), where the presence of strong quartic anharmonicity in a perfect crystal lattice leads to localized vibrational modes, or intrinsic localized modes (ILMs). A good analogy to explain ILMs is given by Minkel (2006): "*Suppose you throw a rock into a pond, but instead of circular waves spreading across the surface, only a single bit of the surface at the rock's entry point oscillates up and down continuously.*" In atoms, a similar circumstance occurs, where a single atom vibrates with a large amount of localized energy. While at first this may seem like only a matter of trivial interest, we will see that ILMs may have far-reaching implications for actinides and materials in general.

Using inelastic neutron and x-ray scattering, Manley *et al.* (2006) recorded the phonon dispersion curves on a large single crystal of α-U from 298 to 573 K. Both measurements show a softening in the longitudinal optic branch along [00ζ] above 450 K. At the same temperature, a new dynamical mode forms along the [01ζ] Brillouin zone boundary. The changes in the phonon spectrum coincide with no observable change in heat capacity (Oetting *et al.*, 1976) or structural change, i.e., there is no phase transformation. Manley *et al.* (2006) suggest this is the first evidence of an ILM forming in a three-dimensional crystal. Proof of the ILM is given by room temperature excitation of the mode via x-rays at the exact energy for which they are observed at elevated temperature.

One of the most fascinating aspects of the discovery of ILMs in α-U by Manley *et al.* (2006) is the fact that the temperature at which they occur, 450 to 675 K, coincides with anamolies in the mechanical properties of the metal. Between these temperatures, tensile tests show changes in the amount of deformation that can be imparted in α-U (Taplin and Martin, 1963). While this mechanical anomaly has been known for over 40 years, it has never been explained. Manley *et al.* (2006; 2008) suggest that the anomalous change in physical properties of U is influenced by the high-temperature formation of ILMs. In a normal metal, plastic deformation due to mechanical stress causes production and movement of defects in the crystal. However, when ILMs are present, they can act to impede dislocation motion, similar to vacancies, interstitials, grain boundaries, or secondary phases. Much like a speed bump on a road, an ILM or other defect may slow dislocation motion, in turn inhibiting plastic deformation. Indeed, similar to the well-known Hall-Petch relationship (Hall, 1951; Petch, 1953) that shows the strength of a metal is proportional to the density of grain boundaries (or inverse to the size of the grains), so might the mechanical response of a metal be proportional to the density of ILMs.

Physically, the observation of ILMs demonstrates the ability of a uniform and defect-free material to concentrate energy spontaneously. With ILMs centered on a single lattice site, they give rise to a configurational entropy analogous to that for vacancies (Sievers and Takeno, 1989). In turn, this affects the physics of the material in a new way and produces interesting and new mechanisms for mechanical responses. With all this, one has to ask why it is that α-U has the only known three-dimensional CDW and the only known three-dimensional ILM? We don't have the answer, but the question itself illustrates the extraordinary physics that the metal holds.

## D. Neptunium

The orthorhombic crystal structure of Np with space group *Pmcn* was first solved by Zachariasen (1952b). Shortly afterward, Eldred and Curtis (1957) published a letter entitled "Some properties of neptunium metal" in *Nature*. Besides clearly illustrating that what appears in the journal has vastly changed, the letter showed the density of the metal was between U and Pu with a Vickers hardness of 355. The basic physical properties and crystal structures of Np were soon elaborated on by Evans and Mardon (1959) and Lee *et al.* (1959), specifically examining the three allotropic phases of the metal, α, β, and γ. Subsequently, Mössbauer spectroscopy was performed using nuclear γ-ray resonance by Dunlap *et al.* (1970), revealing the highly anisotropic lattice vibrations in α-Np. A common theme in all these early investigations was that Np behaved in a manner even more erratic than U. As we shall see, this was the first indications that Np was not just like U, and that the 5$f$ states were beginning to change.

While it is often argued that Th to α-Pu have delocalized 5$f$ states and Am to Lr have localized 5$f$ states, experimental data reveal that the delocalized-localized change starts well before this. The spin-orbit sum rule analysis of the $N_{4,5}$ EELS data in Fig. 17 shows that Np is already becoming more localized. A pure *LS* coupling mechanisms is observed for α-U, whereas for Np the spin-orbit expectation value for Np is closer to intermediate coupling. In a fully localized actinide material, such as an oxide or fluoride, the 5*f* states are expected to exhibit intermediate coupling. Given this fact, the spin-orbit results for the $N_{4,5}$ EELS clearly show that the 5*f* states in Np are beginning to localize, albeit slightly. This is supported by calculations of Brooks *et al.* (1984), which show that while the 5*f* spin-orbit interaction can be neglected in the lightest actinides, they become important for Np. In other words, the very beginning of the transition from *LS* to intermediate coupling occurs in Np, one element prior to the crystallographic volume jump observed near Pu in Fig. 1.

The fact that Np metal is the first actinide element to have a small yet measurable degree of localization in the 5*f* states is further illustrated by PE spectroscopy. Examining the valence-band PE spectrum in Fig. 20 and the 4*f* PE spectrum in Fig. 21 for Np (Naegele *et al.*, 1987), it is apparent that a subtle fine structure is beginning to evolve that is not present on the U 5*f* PE spectrum. The 5*f* states in U are delocalized enough to produce saw-tooth shaped valence-band PE spectrum in Fig. 20. However, a peak develops at about 0.8 eV in the Np valence-band spectrum. This peak is larger and broader in α-Pu, then exceedingly pronounced in δ-Pu. The same is true for the 4*f* PE spectra in Fig. 21. Whereas U has a pair of clean asymmetric $4f_{5/2}$ and $4f_{7/2}$ peaks that are indicative of a delocalized actinide metal, Np exhibits a small amount of structure on the high-energy side of those peaks. This is the emergence of the poorly-screened satellite peak in the 4*f* PE spectra that becomes evident for α-Pu, then even more evident for δ-Pu (Arko *et al.*, 2000). These poorly-screened satellite peaks in the Np spectrum grow as the metal is allowed to oxidize and form $Np_2O_3$ (Naegele *et al.*, 1987).

Support for the fact that the 5*f* states in Np metal are beginning to localize, albeit very slightly, may also be found in the bulk modulus of the metals. The experimental bulk modulus of each of the light actinides are plotted in Fig. 23 where the data are: Th at 50-72 GPa (Bellussi *et al.*, 1981; Benedict, 1987; Benedict and Holzapfal 1993), Pa at 100-157 GPa (Birch, 1947; Benedict *et al.,* 1982; Benedict, 1987; Haire *et al.*, 2003), U at 100-152 GPa (Yoo *et al.*, 1998; Benedict and Dufour, 1985; Akella *et al.*, 1990), Np at 74 -118 GPa (Dabos *et al.*, 1987; Benedict, 1987; Benedict and Holzapfal 1993), Pu at 40-55 (Roof, 1981; Benedict, 1987; Benedict and Holzapfal 1993), Am at 30 GPa (Heathman *et al.*, 2000), Cm at 37 GPa (Heathman *et al.*, 2005), Bk at 35 (Haire *et al.*, 1984), and Cf at 50 (Peterson *et al.*, 1983). Th exhibits a low bulk modulus due to the small amount of electrons in the 5*f* states, as discussed in the previous section on the metal, as well as the relatively soft fcc structure. Pa and U exhibit the highest bulk moduli, which within the ±5 GPa error of diamond anvil cell, are the same. Looking at Np, we see there is a noticeable drop in the bulk modulus from Pa and U. Why? Given the EELS, XAS, and PE data, the drop appears to be due to a fractional emergence of localization in the 5*f* states, which reduces the bonding strength. Reality, however, seems always to throw in a wrench or two. The wrench is this: Np has about four 5*f* electrons, which means it has begun to fill anti-bonding states in the $j = 5/2$ level, and this also can reduce the bonding strength in the metal. Also, the bulk modulus is sensitive to the crystal structure, so comparing each structure is not apples-to-apples. While we can not say for certain which factor causes the decrease in bulk modulus (it probably is a combination of all), it does suggest a change in the actinide metal series has begun.

Another indication that the 5*f* states of Np are becoming slightly localized is given by temperature-dependent resistivity and electronic specific heat of the metal, which are shown in Fig. 24(a) and (b), respectively. Resistivity measurements for Th, Pa, U, Np, α-Pu, and Am as a function of temperature are shown in Fig. 24(a) (Müller *et al.*, 1978). The curves for Th, Pa, and U show lower resistivities with normal curvature that are similar to more ordinary metals. α-Pu shows anomalously large resistivity with a negative temperature coefficient of resistivity, which is unlike any of the other metals in the plot. Looking at Np, we see that while the curvature of the temperature-dependent resistivity is similar to Th, Pa, and U, the value is much higher than those metals, close to the values of α-Pu as a function of temperature.

Electronic specific heat measures the contribution to the specific heat of a metal from the motion of conduction electrons. It gives information about the density of states at the Fermi level and is a direct link to the strength of electron correlations and electron localization. Results for Th (Fournier and Troc, 1985), Pa (Fournier and Troc, 1985), U (Lashley *et al.*, 2001a), Np (Fournier and Troc, 1985), α-Pu (Elliot *et al.*, 1964; Fournier and Troc, 1985), δ-Pu (Wick, 1980; Lashley *et al.*, 2003; 2005; Javorsky *et al.*, 2006), and Am (Müller *et al.*, 1978) are shown in Fig. 24(b), where the ground-state phases are indicated by black circles. Note that while α-Pu has the highest electronic specific heat, Np is a close second. The resistivity and electronic specific heat of Np both suggest that a measurable degree of localization is present in the 5*f* states of the metal. Thus, while band-structure calculations are able to accurately account for the physical properties of Np because the 5*f* states are still fairly delocalized (Söderlind, 1998), the beginnings of *f*-electron localization are appearing throughout the spectra and bulk measurements of the metal.

When experimentally compressed to 52 GPa using a diamond anvil cell, Np shows no evidence of a phase transformation from the ground-state orthorhombic structure with space group *Pmcn* (Dabos *et al.*, 1987). This is theoretically supported by first-principles electronic structure calculations by Söderlind *et al.* (1994b; 1995b) that show α-Np is stable up such pressure. DFT shows that upon further compression the sequence α-Np → β-Np → bcc is observed, where the first transition is for a 19% compression of Np and the second transition is a 26% compression (Söderlind 1998; Pénicaud 2000 and 2002). The transformation of low-symmetry actinides to high-symmetry structures is a reoccurring theme, which is due to broadening of the valence bands with pressure. Once exceedingly high pressures are realized, even the narrow 5*f* bands can become broad enough to exhibit high-symmetry fcc, bcc, and hcp structures.

**E. Plutonium**

Plutonium is like an onion; while it is certainly not a vegetable, it has numerous layers of complexity. The metal exhibits six crystal structures between absolute zero and melting, has a negative coefficient of thermal expansion in the δ and δ′ phases, and is exceedingly sensitive to pressure, temperature, and chemistry. The considerable influence of temperature and pressure on the metal are conveyed in Fig. 25 (Liptai and Friddle, 1967; Hecker, 2000), which shows the phase equilibiria of Pu from 0 to 10 kbar and 0 to 600°C. Immediately noticeable is that the high-volume phases γ, δ, and δ′ are squeezed out with relatively little pressure, about 2-3 kbar. Figure 24 also reveals how unique the δ phase is, enhabiting a rather small equilibrium phase field in temperature-pressure space.

Many experimental and theoretical investigations have been performed on the metal, mostly focused on the ground-state monoclinic α phase and the high-temperature fcc δ phase. This large number of investigations is due to the unique positioning of the metal between localized and delocalized 5*f* states, as seen in Fig. 1, and the myriad of interesting physical properties that Pu metal, alloys, and materials exhibit. Accordingly, we will give considerable attention to the electronic, magnetic, and crystal structure of Pu in this review, more than any of the other actinide metals. First and foremost, we will examine the available spectroscopy data, deriving known characteristics of the metal. Then, we will move to an overview of density-functional theory and dynamical mean-field theory, considering how they relate to experiment. Lastly, we will cover interesting aspects of Pu, such as crystal lattice dynamics, changes in electronic structure due to self-induced radiation damage, and superconductivity.

**1. What we know**

Experimental and theoretical results on Pu metal clearly show: The 5*f* states contain approximately five electrons (Söderlind 1998, van der Laan *et al.*, 2004; Moore *et al.*, 2003; 2007a; 2007b; Shim *et al.*, 2007; Zhu *et al.*, 2007), intermediate coupling near the *jj* limit is the appropriate angular momentum coupling scheme for the 5*f* states (Moore *et al.*, 2007b), electron-correlation effects are present, to varying degrees, in both the α and δ phases (Shim *et al.*, 2007), and all six allotropic phases of Pu metal are nonmagnetic (Lashley *et al.*, 2005, Heffner *et al.*, 2006). In order to understand this, let us look at the spectroscopic data and bulk measurements in turn.

The EELS and many-electron atomic spectral calculations in Secs. II and III show that the valence is at or near $5f^{\,5}$ and the coupling mechanism is exactly intermediate, being very close to the *jj* limit. In fact, EELS results by Moore *et al.* (2006b) show that this is true for both α- and δ-Pu, since the results for each phase are quite similar. The sum rule analysis of the $N_{4,5}$ branching ratio therefore supports a strong spin-orbit interaction in the 5*f* states, pushing the intermediate coupling mechanisms almost to the *jj* limit. Even with some variation of the *f* count, say 0.4, the result is still robust, showing that the 5*f* states in Pu are far removed from the *LS* coupling limit. In fact, for the *LS* coupled ground state $\langle w^{110} \rangle / n_h$ never reaches values lower than -0.23 (van der Laan *et al.*, 2004), which is far away from the experimental value of -0.565.

Valence-band PE spectra for both α- and δ-Pu are shown in Fig. 20 (Gouder *et al.*, 2001; see also Cox and Ward, 1981; Courteix *et al.*, 1981; Baptist *et al.*, 1982; Naegele *et al.*, 1985; Cox, 1988; Havela *et al.*, 2002; Gouder *et al.*, 2005; Baclet *et al.*, 2007). The spectrum of α-Pu shows a relatively delocalized metal due to the Doniach-Sunjic saw-tooth shape. However, upon closer inspection one can see that fine structure is present in the spectra. This fine structure is slightly larger than observed in the Np spectra, meaning the 5*f* states of α-Pu are further localized, but again the degree of this localization is small. The fine structure becomes considerably larger in the δ-Pu spectra, giving evidence of a further movement towards localization of the 5*f* states. This behavior of the valence-band PE of α- and δ-Pu directly relate to the Pu phase diagram in Fig. 2 where the volume difference between the α and δ phase is upwards of 25%. The α- and δ-Pu spectra shown here are achieved using very clean thin films of deposited Pu and represent the most concise portrayal of the behavior of the valence-band PE (Gouder *et al.*, 2001; Havela *et al.*, 2002). The kinetics of the surface reconstruction strongly depend on temperature and only when cooled to 77 K does the surface of the Pu thin film remain in the monoclinic α structure. This is supported by first-principles calculations that show the surface of α-Pu will reconstruct to δ due to free bonds (Eriksson *et al.*, 1992b). For these reasons, the α-Pu spectra were collected at 77 K, ensuring the surface was the correct crystal structure. Besides temperature, the Pu valence-band PE is also highly dependent on thin-film thickness, as discussed in the PE secion (Gouder *et al.*, 2001; Havela *et al.* 2002). What is more, stepwise addition of Si to Pu cause the 5*f* states to localize and hybridize with the Si 3*p* states (Gouder *et al.* 2005). These results clearly demonstrate that temperature, dimensional constraints, and doping affect the degree of localization of the 5*f* states, further illustrating the sensitivity of the 5*f* states of Pu..

Examining the 4*f* PE spectra of α- and δ-Pu in Fig. 21 (Arko *et al.*, 2000; see also Larson, 1980; Cox and Ward, 1981; Courteix *et al.*, 1981; Baptist *et al.*, 1982;, Naegele *et al.*, 1984; Cox, 1988; Cox *et al.*, 1992; Gouder *et al.*, 2001; Havela *et al.*, 2002; Gouder *et al.*, 2005; Baclet *et al.*, 2007), we see that the α-Pu spectrum contains asymmetric $4f_{5/2}$ and $4f_{7/2}$ peaks that are indicative of a fairly delocalized actinide metal. However, once again the effects of localization are observed, since the poorly-screened satellite peak on the high-energy side of each main peak is subtly visible. In a similar manner to the valence-band PE, these poorly-screened peaks grow considerably larger in the δ-Pu spectra (Arko *et al.*, 2000; Gouder *et al.*, 2005; Baclet *et al.*, 2007), showing further localization and subsequent electron correlations effects. When 4*f* PE spectra are used to analyzed Pu thin film as a function of thickness (Gouder *et al.*, 2001), the spectra show that films of one or a few monolayers are δ-like, with a marked loss on intensity in the well-screened peak, while a single monolayer of Pu produces a spectrum that is almost identical to the Am spectrum in Fig. 21.

EELS, XAS, and PE clearly show that Pu metal has approximately a $5f^{\,5}$ configuration and exhibits intermediate coupling near the *jj* limit. Why then is there no experimentally observed magnetism in any of the six allotropic phases of the metal (Lashley *et al.* 2005, Heffner *et al.* 2006)? The lack of magnetism for Am is obvious, since it has a nearly filled *j*= 5/2 level (total angular momentum *J* = 0), but Pu, which is $\sim f^{\,5}$ and has at least one hole in the *j* = 5/2 level, is positively vexing! Some mechanism must be obfuscating the moment, such as Kondo shielding (Shim *et al.*, 2007) or electron pairing correlations (Chapline *et al.*, 2007), which are illustrated in Fig. 8. Indeed, recent magnetic susceptibility measurements by McCall *et al.* (2006) show that magnetic moments on the order of 0.05 $\mu_B$/atom form in Pu as damage accumulates due to self-irradiation. This suggests that small perturbations to the gentle balance of electronic and magnetic structure of Pu metal may destroy or degrade possible screening effects of a moment due to the hole in the *j* = 5/2 level. If indeed the Kondo shielding picture is correct, then Pu has most of the spectral weight in the Hubbard bands with a small Kondo peak. This configuration makes Pu appear localized-like at high frequencies when probed by EELS because the technique examines the integral of the valence density of states. This could also explain why α- and δ-Pu are 25% different in volume, but show a similar $N_{4,5}$ EELS branching ratio and spin-orbit analysis of the 5*f* states (Moore *et al.* 2006b). In order to further examine electron correlations and localization of the 5*f* states in Pu, we now turn our attention to bulk-sensitive measurements.

Resistivity curves as a function of temperature for Th, Pa, U, Np, α-Pu, and Am are shown in Fig. 24(a) (Müller *et al.*, 1978). Th, Pa, and U show lower resistivities that are closer to ordinary metals, whereas Am and Np are considerably higher. For a metal, α-Pu shows an astoundingly large resistivity and exhibits a negative temperature coefficient of resistivity where the resistance goes up with temperature decrease. What is more, the resistivity of α-Pu is strongly anisotropic, showing higher resistivity for currents parallel to the (020) planes as compared to currents perpendicular to the (020) planes (Brodsky and Ianniello, 1964; Elliot *et al.*, 1964). It has been proposed that the anomalous low-temperature resistivity behavior for neptunium and plutonium are due to spin fluctuations, similar to UAl$_2$, a material where spin fluctuations are clearly present (Nellis *et al.*, 1970; Arko *et al.*, 1972). Spin fluctuations can be thought of as spin alignments that have lifetimes too short to see via specific heat, susceptibility, or nuclear magnetic resonance, in other words at times less than $\sim 10^{-14}$ seconds. However, the scattering time for resistivity is approximately $10^{-15}$ seconds, which is fast enough to catch the spin-flip contribution to resistance. A spin fluctuation model is used by Jullien *et al.* (1974) to calculate the resistivity curves for Np and Pu using a two-band model, reproducing the shape of the α-Pu curve. However, they employ a Stoner exchange enhancement value that is four times the experimental value of 2.5 (Arko *et al.*, 1972). Whether spin fluctuations are present in Np and Pu is still under debate. Interestingly, the resistivity of dopant-stabilized δ-Pu is lower than α-Pu, which is not expected (Smoluchowskii, 1962; Brodsky, 1965; Harvey *et al.*, 1965; Boulet *et al.*, 2003). Alternatively, the negative temperature coefficient of resistivity of Pu-Al and Pu-Ga δ-Pu alloys can be modeled using an LDA *ab initio* approach assuming ordinary electron-phonon interaction and its interference with electron-impurity interaction (Tsiovkin *et al.*, 2007).

Electronic specific heat gives information about the density of states at the Fermi level and is a direct link to the strength of electron correlations. Since electron-correlation effects are abundant on and near the itinerant-localized transition in Fig. 6(a), precisely where Pu is found, these measurements are of great value for understanding 5$f$ states of Pu in relation to the other actinide-series metals. It is often believed that electron correlations are present in δ-Pu, but absent or weak in α-Pu. However, this is clearly not correct given the results in Fig. 24(b). Even though δ-Pu is considerably higher than the rest of the phases, α-Pu does exhibit a sizable electronic specific heat when compared to Th. Indeed, we can go back to the PE spectra in Figs. 20 and 21 to see further evidence of electron correlation and localization. Subtle fine structure is observed in the valence-band PE spectra of α-Np and α-Pu, and this structure grows considerably in the spectrum of δ-Pu. This emergence of structure clearly shows that electron correlations are beginning to appear in the ground-state α phase of Np and Pu due to localization of the 5$f$ states.

Taken together, EELS, XAS, PE, resistivity, and electronic specific heat confirm that Pu metal has approximately a 5$f^5$ configuration, exhibits intermediate coupling near the *jj* limit, has electron correlation effects in both the α and δ phases, and shows no sign of bulk, long-range magnetism in any of the six allotropic phases. These factors strongly influence the current state of theoretical calculations and where they will proceed in the future. In order to understand this, we now digress to examine the progress of electronic-structure theory as applied to Pu, particularly the δ phase.

## 2. Density-functional theory

Density-functional theory is a ground-state theory for calculating the electronic structure of materials as well as bonding properties, such as crystal structures, equation-of-state, bulk modulus and elastic constants. The mathematical foundation for DFT was created in the mid 1960s and is described in great detail in many books (For example: Parr and Weitao, 1989; Dreizler, 1990; Nalewajski, 1996; Fiolhais, 2003), but can be summarized by the fact that it is completely material transparent and only relies on the total number of electrons for describing a specific component. For practical purposes, the complex behaviors of the electron interactions are often approximated by a local density approximation (LDA), which defines the electron potential given the electron density. Its early, minimal framework can handle the light actinide metals with delocalized 5$f$ states and simple crystal structures, but has great difficulty with Pu.

By the early 1970s various versions of the LDA converged to a level that was accurate enough that most calculations could be undertaken considering the limitations of computational power. An important step to further streamline the calculations was the concept of linear methods by Anderson (1975), which was quickly utilized for the actinides by Skriver *et al.* (1978). By the end of the 1970s these methods, combined with accurate algorithms for calculating the total energy of the system, opened the doors to study lattice constants and their pressure dependence (equation-of-state), magnetic moments, and structural properties for all but the most complex geometries. These advances aided in handling Pu electronic structure with theoretical computation, and a benchmark step forward came when Skriver (1985) applied DFT theory to metals throughout the periodic table and correctly predicted 35 ground-state phases out of 42 studied. At the same time, formulations of the relativistic spin-orbit interaction were implemented for its significant effect on magnetism as well as for its influence on the bonding in the actinide metals. Although remarkably successful, the calculations were still not sufficiently accurate for the study of elastic constants, distortions, low-symmetry crystal structures, or the complex electronic structure of Pu.

Fast forward a decade later and the development of the full-potential method with no geometrical approximations of the electron density or potential was developed, allowing effective and broad study of the actinides. At this point, the overall accuracy of the computational techniques had reached a level that revealed some deficiencies in the then two-decade-old LDA. Generally, it was found that LDA overestimated the strength of the chemical bonds in most materials. The so-called 'LDA contraction' was particularly severe for the light actinide metals. By the mid 1990s, an electron exchange and correlation energy functional was developed that consistently reduced the over binding displayed by the LDA and improved the description of the actinides in particular. The generalized gradient approximation (GGA) evolved to include dependencies of various gradients of the electron density for a better description of the non-local behavior. The formulation was carefully chosen so as not to violate rules for the exchange and correlation holes.

Armed with fully relativistic, full-potential, GGA methods, the actinide metals could finally be addressed theoretically with an accuracy that made meaningful comparisons with measured data. There were still, however, difficulties with δ-Pu, where nonmagnetic GGA calculations yielded equilibrium volumes 20-30% below that observed experimentally (Söderlind, 1998; Savrasov and Kotliar, 2000; Kutepov and Kutepova, 2003). Spin-orbit coupling in the 6*p* states was addressed by Nordström *et al.* (2000), who showed the treatment of the 6*p* states affected the calculated volume resulted for δ-Pu. The findings showed that spin-orbit splitting of the low-energy-lying 6*p* states in the actinides, are unimportant for the bonding properties. If included, they cause problems due to the choice of basis functions.

Spin and orbital polarization were then included within the GGA to bring the calculated volume of δ-Pu in accordance with experiment (Söderlind *et al.*, 1994; Antropov *et al.*, 1995; Savrasov and Kotliar, 2000; Kutepov and Kutepova, 2003; Robert, 2004). One particular set of calculations by Söderlind and Sadigh (2004) have been able to achieve appropriately spaced energies and atomic volumes for all six allotropic phases of Pu, with the exception of the high-temperature bcc ε phase, which showed an energy that was slightly too large in the zero-temperature calculations**.** The results are shown in Fig. 26 and can be directly compared to the experimental phase diagram of Pu in Fig. 2. The agreement between the unit-cell volumes given by the calculated energy curves and the phase diagram of Pu are quite incredible. Examining the energy curves, an expansion is observed when moving from α-Pu to δ-Pu followed by a reduction in volume moving from δ-Pu to ε-Pu, exactly as observed in the phase diagram in Fig. 2. This approach also yields an equation-of-state, bulk modulus, and elastic constants that are in agreement for most all of the six allotropic phases.

While the calculated volumes and bulk properties of Pu by Söderlind and Sadigh (2004) are accurate, the underlying physics of the theory are incorrect due to the prediction of substantial magnetic moment. Pu metal exhibits *no experimentally observed magnetism* (Lashley *et al.*, 2005) and so the DFT calculations showing a magnetic moment of ~5 $\mu_B$/atom is not in agreement with the experiment. This was not isolated to the work of Söderlind and Sadigh (2004), since for a period spin polarization was used in DFT to handle the large volume expansion of δ-Pu, whether by GGA (Skriver, 1978; Solovyev *et al.*, 1991; Söderlind *et al.*, 1997) or by extensions, such as LDA+U (Bouchet *et al.*, 2000; Savrasov and Kotliar, 2000). At this point it was clear that Pu was non-magnetic and, accordingly, theorist began to perform calculations with the intent of finding a non-magnetic solution for Pu.

One approach has been to use LDA+U calculations of Pu with a 5*f* electron count that is above 5, sometimes being close to or exactly 6 (Shorikov *et al.*, 2005; Pourovskii *et al.*, 2005; Shick *et al.*, 2005; 2006; 2007). In such a situation, Pu is neither 5 nor 6, but non-integer, which means that δ-Pu is not in a single-Slater-determinant ground state. These calculations explain the three-peak structure in valence-band PE and the relatively high electronic specific heat. However, a 5*f* count at or near 6 is entirely out of step with EELS, XAS, and PE spectroscopy as well as other theory. The EELS and many-electron atomic spectral calculations presented in Secs. II and III, respectively, clearly show that Pu has a 5*f* occupation at or near 5 (Moore *et al.*, 2007a; 2007b). Further, the inverse, valence-band, and 4*f* PE data for the actinide series support $\sim 5f^5$ for Pu when theoretically analyzed. Lastly, DFT (Söderlind, 1998; Pourovskii *et al.*, 2007) and DMFT results (Shim *et al.*, 2007; Zhu *et al.*, 2007) clearly show a $\sim 5f^5$ configuration in Pu. Therefore, the combination of EELS, XAS, PE, DFT and DMFT suggest a 5*f* count near 5 with 5.4 being a reasonable upper limit; 5*f* counts between 5.5 and 6 are unrealistic and out of step with the bulk of theory and experiment. If the LDA+U approach can result in a 5*f* occupation between 5.0 and 5.4, then it will be in agreement with EELS, XAS, and PE spectroscopy. A $5f^6$ configuration is what is found in Am metal (Graf *et al.*, 1956; Söderlind and Landa 2005) and this should not be utilized to remove magnetism in Pu calculations.

Two other approaches to avoiding magnetism in DFT calculations of Pu have been developed, one which has an equal but opposite spin and orbital moment that results in cancellation of a total moment (Söderlind 2007), and one where the spin moment is zero but the orbital polarization and spin-orbit interaction are strong (Söderlind 2008). The first approach yields a total moment of zero, but the strong spin polarization needed to cancel the orbital moment is in disagreement with experiment. Examining Fig. 17 (a), it is clear that while spin polarization is important for Cm, it is weak for Pu and Am where the spin-orbit interaction dominates (Moore *et al.*, 2007a; 2007b). Thus, EELS and XAS do not support strong spin polarization in Pu. In reality, absolute tests of polarization, such as x-ray magnetic circular dichroism or polarized neutrons on a $^{242}$Pu sample ($^{239}$Pu absorbs too many neutrons), need to be performed to address spin and orbital polarization in Pu. However, while these absolute tests have yet to be performed, the available EELS and XAS experiments do not support a strong spin polarization.

The second approach to avoiding magnetism in DFT calculations of Pu assumes the spin moment is zero, but orbital polarization and spin-orbit interaction are strong (Söderlind 2008). In this study, a quantitative analysis of the spin polarization, orbital polarization, and spin-orbit interaction in δ-Pu is performed using a more physically plausible description of the electron correlations in the metal. This is achieved by expanding the DFT-GGA scheme to include explicit electron interactions corresponding to the electron-configuration rules of the atom, i.e., Hund's rules. The calculations by Söderlind (2008) show no magnetic moment, a $5f^5$ configuration, a volume for δ-Pu that was correct within error, and a one-electron spectrum that was consistent with the valence-band PE spectra of Arko *et al.* (2000a) and Gouder *et al.* (2001). The results also show that spin-orbit coupling and orbital polarization are far stronger than spin polarization and thus more important in δ-Pu. As mentioned above, this is in agreement with EELS and XAS experiments that show spin polarization is not important for Pu, but spin orbit-coupling is strong (Moore *et al.*, 2007a; 2007b). Orbital polarization of the 5*f* states in Pu is experimentally unknown, and must be addressed using x-ray magnetic circular dichroism or polarized neutrons. Recent polarized neutron experiments on a $PuCoGa_5$ single-crystal showed that orbital polarization of the 5*f* states in Pu is much stronger than spin polarization (Hiess *et al.*, 2008); whether this is also true for metallic Pu waits to be determined through experiment.

## 3. Dynamical mean-field theory

The seeds of DMFT can be traced to the first investigations of the Hubbard model in the infinite limit of spatial dimensions by Metzner and Vollhardt (1989). Over time this has evolved into a technique that has the ability to handle strong electron correlations while avoiding the problem of long-range ordered magnetism. In short, DMFT offers a minimal description of the electronic structure of correlated materials, treating both the Hubbard and quasiparticle bands on equal footing. The technique is based on a mapping of the full many-body problem of solid-state physics onto a quantum impurity model, which is essentially a small number of quantum degrees of freedom embedded in a bath that obeys a self-consistency condition (Georges and Kotliar, 1992; Georges *et al.*, 1996). In DMFT, the spin and orbital moments occur at short time scales with site hopping and *s*, *p*, and *d* hybridization. To understand this, consider the atomic calculation in Fig. 18(a) and (b). It is seen that the spin and orbital moments for Pu are almost equal in magnitude and oriented in *opposite* directions due to spin-orbit interaction, where there is a close to complete cancellation. A similar configuration occurs in DFT calculations of Pu, where the spin and orbital moments are close to a complete cancellation (Söderlind, 2007). In both the atomic calculations and DFT the spin and orbital moments are temporally locked, resulting in a static result. On the other hand, when performing DMFT calculations there is not only the LDA eigenstates that contain the static terms, but also the DMFT that contains the additional dynamical terms (Georges *et al.*, 1996). A recent and detailed review is given by Kotliar *et al.* (2006), which not only fully describes the method, but also has sections on application of DMFT to Ce and Pu.

Applied to Pu, DMFT successfully matches the experimental photoemission spectra of α- and δ-Pu (Savrasov *et al.*, 2001), predicts the phonon dispersion curves for single-crystal δ-Pu (Dai *et al.*, 2003), provides insight into the 5*f* valence of the metal (Shim *et al.* 2007; Zhu *et al.*, 2007), and matches the experimental electronic specific heat of α- and δ-Pu (Pourovskii *et al.*, 2007). The calculated phonon dispersion curves for δ-Pu by Dai *et al.* (2003) are shown in Fig. 27(a) as a blue dashed line, and these can be compared to the experimental data points of Wong *et al.*, (2003). Notably, the DMFT calculations predict the Kohn-like anomaly in the $T_1$[011] branch, and the pronounced softening of the [111] transverse modes. Another success of DMFT has been the ability to simulate the $N_{4,5}$ optical spectra of the actinides, extract the branching ratio of the white line peaks, and analyze them with the spin-orbit sum rule. At the 2006 Plutonium Futures Conference, Haule *et al.* (2006) presented Fig. 27(b), which is a plot of the spin-orbit interaction as a function of the number of *f* electrons. The *LS*, *jj*, and intermediate coupling curves are shown, as calculated using an atomic model. The points correspond to the spin-orbit analysis of $N_{4,5}$ optical spectra calculated using DMFT. This plot may be directly compared to Fig. 17(a), where $N_{4,5}$ EELS data are shown against our atomic calculations. The match between the DMFT and EELS results is exceedingly good. The DMFT results show slight differences in the spin-orbit analysis between α- and δ-Pu, similar to the results of Moore *et al.* (2006b) where α is closer to the *LS* limit and δ is closer to the *jj* limit. In addition, the DMFT results show that α-Pu has a higher 5*f* count than δ-Pu, but this can not be verified with EELS and spin-orbit analysis due to the fact that the *f* count is not an output. The power of the DMFT approach is that the 5*f* count is indeed an output of the calculations, a variable that must be achieved by educated conjecture or literature in the spin-orbit analysis of the EELS spectra.

The importance of fluctuating valence in δ-Pu is pointed out by Shim *et al.* (2007), where fluctuations between $f^4$, $f^5$, and $f^6$ result in an average 5*f* occupancy of 5.2. This, however, should not come as a surprise. In the case of the localized rare-earth metals, the atoms are usually in a unique ground state; however, this is not the case for the light actinides where delocalization and *d-f* mixing causes mixed-valence to occur. Indeed, as far back as the late 1970's this was understood, as evidenced by the following quote by Brodsky (1978): "*In the case of the actinides (not unlike nickel) the ground state is nearly always a mixture of configurations, and it is only on rare occasions that the nearly equal energies of the 5f, 6d and 7s (also 7p) become separated and permit a single configuration to be the ground state.*" From Am on, the actinides exhibit a unique ground state due to localization of the 5*f* states. This mixed valence is important in Pu, since it results in a non-single-Slater-determinant ground state.

DMFT calculations are complicated by nature and are still in their youth compared to DFT, resulting in several difficulties. First, DMFT cannot easily handle large unit cells, which means low-symmetry structures with many unique atomic sites must be either assumed as high-symmetry (Savrasov *et al.*, 2001) or replaced by so-called 'pseudo structures' (Bouchet *et al.*, 2004; Pourovskii *et al.*, 2007). This limitation, however, is quickly becoming overcome by increased computational power and better mathematical arguments. Second, LDA+DMFT has the issue of 'double counting'. When adding electron-electron interactions to the Hamiltonian, the LDA eigenstates contain the static terms while the DMFT contain the additional dynamical terms (Georges *et al.*, 1996). Ensuring that terms are not double counted is imperative for accurate results. In reality, we have no idea how this is done, and simply trust in the razor-sharp decision making ability of DMFT theorists.

In the end, DFT and DMFT both have their powers and weaknesses. Over time the community has fought, fought some more for good measure, and are now coming to convergence on understanding actinide metals, particularly Pu. While there is still a considerable way to go to fully understand the physics of the localized-delocalized transition near Pu in the actinide series, great strides have been, and continue to be made by DFT and DMFT. Both techniques, however, must be guided by four facts about Pu: the metal has approximately a $5f^5$ configuration, it exhibits intermediate coupling near the *jj* limit, it has no bulk magnetism in any of the six allotropic phases, and electron correlation effects are important in both the α and δ phases. It is up to future experiments and theory to resolve why Pu has a $5f^5$ configuration and exhibits intermediate coupling near the *jj* limit, but shows no long-range magnetic order (Lashley *et al.*, 2005; Heffner *et al.*, 2006). Approaches that can discern between Kondo shielding (Shim *et al.*, 2007), electron pairing correlations (Chapline *et al.*, 2007), and a Mott transition (Johansson, 1974) will be paramount.

## 4. Crystal lattice dynamics

Once again we digress from electronic structure to look at crystal lattice dynamics, in this case that of δ-Pu. The reason is two-fold: First, recording the single-crystal phonon dispersion curves for fcc δ-Pu was decades in the making, and second, the strong electron-phonon interactions that occur in the metal dictate that they are of importance to the electronic nature of the metal (Skriver and Mertig, 1985; Skriver *et al.*, 1988). For example, calculations by Tsiovkin and Tsiovkina (2007) suggest that interference between electron–impurity and electron–phonon interactions causes the negative coefficient of resistivity as a function of temperature, a rather unusual feature of Pu metal.

One of the most frustrating and plaguing problems of Pu metal is the inability to produce large single crystals for experiments. The primary reason for this is the unsurpassed six solid allotropic phases the metal exhibits in pure form (Fig. 2, also see Hecker, 2002; 2004). When alloyed with Al, Ce, or Ga, the fcc δ phase can be retained to room temperature, but the mixture still goes through the ε phase when cooling from a melt to solid (Moment, 2000). The end result is always a polycrystalline sample with grains no larger than a few hundred micrometers. There has, however, been one known exception.

From the early sixties to mid seventies, Roger Moment attempted to grow large single crystals of gallium-stabilized δ-Pu by solidification without success (Moment, 2000). He then tried strain annealing (Lashley *et al.*, 2001b), a technique where a metal or alloy is deformed by a few percent at room temperature. During the deformation, a new crystal will form and when annealed at high temperature, this new crystal will, in theory, grow at the expense of the other grains. The process is driven by the reduction of internal stresses and a lowering of the total energy by the reduction in grain boundary area, accomplished by grain-boundary migration. Using strain annealing, one large single crystal was produced, but was not substantial enough to extract. The next attempt utilized transformation-induced stress as the facilitator for grain growth. A rod was taken through three heat treatment cycles from the δ phase to the ε phase, then back to δ. The rod was then annealed at 500°C to promote grain growth. After repeated cycles, a single grain was produced that was 7 mm long and of high quality, as evidenced by back-reflected Laue x-ray diffraction. The experimentally measured elastic constants showed that the response of δ-Pu to stress was highly anisotropic (Moment and Ledbetter, 1976). More precisely, Pu was found to be five times more stiff in the [111] direction than in the [001] direction.

Fast forward 30 years and no other large high-quality single crystals of δ-Pu have been made. Rather than struggle to produce the large crystals, several people at Lawrence Livermore National Laboratory have chosen another route, namely using experimental probes with spatial resolution high enough to circumvent the need for large single crystals. EELS in a TEM has been chosen for electronic-structure measurements due to the ability to form a 5 Å probe with which to collect spectra. Another approach is using inelastic x-ray scattering to record the phonon dispersion curves, rather than the usual neutron diffraction approach, due to the ability to focus the x-ray beam onto single micrometer-sized grains. Accordingly, inelastic x-ray scattering has enabled the collection of the first ever full phonon dispersion curves for δ-Pu, as shown in Fig 27(a) (Wong *et al.*, 2003; 2005). Numerous unusual features are observed, such as a large elastic anisotropy, a small shear elastic modulus, a Kohn-like anomaly in the $T_1$[011] branch, and a pronounced softening of the [111] transverse modes. The DMFT calculated phonon dispersion curves (Dai *et al.*, 2003) are in excellent agreement with experiment (Wong *et al.*, 2003), showing the same highly unusual features of δ-Pu crystal dynamics. The highly anisotropic and unusual behavior of the metal phase are further proven using polycrystalline samples of Ga-stabilized (Migliori *et al.*, 2006) and Al-stabilized (McQueeney *et al.*, 2004) δ-Pu. Finally, the phonon dispersion curves for δ-Pu have been used by Lookman *et al.* (2008) to show that the fcc δ → monoclinic α transformation occurs in a sequence of three displacive transformations: fcc → trigonal → hexagonal → monoclinic.

## 5. Effects of aging

In the general overview of the actinides we discuss the fact that many 5*f* metals change over time due to self-induced irradiation damage. This happens in Pu metal, and understanding the changes in the electronic structure as a function of time is of great interest. There are two spectroscopic investigations of the 5*f* states in 'new' and 'old' Pu material, Chung *et al.* (2006a) and Moore *at al.* (2006b), neither of which seems a reliable method to detect or define changes in the 5*f* states of Pu as a function of age. Chung *et al.* (2006a) show differences in $5d \rightarrow 5f$ resonant valence-band photoemission spectra between new and old Pu. They argue this differenceis a signature of aging based on the results of Dowben (2000) for the manganite system $La_{0.65}Ca_{0.35}MnO_3$. Dowben (2000) shows that the manganite exhibit a change in resonant PE as the material goes from a metallic phase at low temperature to a non-metallic phase at high temperature. Once the non-metallic phase is realized, a large increase in resonance is observed due to extra atomic decay channels. However, looking at the originally published resonant photoemission of Tobin *et al.* (2003), where data for α-Pu is shown (this spectra is omitted in Chung *et al.*, 2006a), there is practically no difference between α- and δ-Pu. Assuming the argument of Dowben (2000), how could new and old δ-Pu show a large difference in resonant photoemission while α- and δ-Pu are practically the same? The difference between the collapsed monoclinic α phase and the expanded δ phases should certainly exhibit some, arguably more, of a difference than the new and old δ-Pu, which are differentiated only by self-induced irradiation damage to the lattice. It is well understood that the surface of α-Pu reconstructs to δ-Pu due to the free bonds (Eriksson *et al.*, 1992b; Havela *et al.*, 2002). Certainly the α-Pu spectrum by Tobin *et al.* (2003) has contribution from a surface reconstruction of δ-Pu, since Havela *et al.* (2002) show the only way to avoid this is to thermally inhibit the reconstruction at low temperatures (~77 K). However, the penetration depth of the technique should be enough to reveal differences between α-Pu with a δ-reconstructed surface and pure δ-Pu. Thus, there should also be differences between the resonant photoemission of α- and δ-Pu, but none are observed, and this brings question to the results and/or interpretation of Chung *et al.* (2006a).

Using EELS in a TEM, Moore *et al.* (2006b) show small changes in the branching ratio and spin-orbit interaction of the $N_{4,5}$ edge between α-, new δ-, and old δ-Pu. The data trend as would be expected for varying *f* electron localization, with α-Pu behaving the most delocalized, aged δ-Pu behaving the most localized, and new δ-Pu falling in the middle. However, while the systematic trend of α-, new δ-, and old δ-Pu is repeatable, the error bars between phases are large enough to slightly overlap. Thus, EELS has not proven to be a robust and convincing method to detect changes in the 5*f* states of Pu as a function of age (lattice damage). It may simply be that the differences between new and old Pu are too small to detect within experimental uncertainties of spectroscopy. Alternatively, either significantly older specimens with more damage may be required to observe an effect, or very new samples given that results show the lattice constant rapidly changes in the first three months (Chung *et al.*, 2006b; 2007; Ravat *et al.*, 2007; Baclet *et al.*, 2007).

More bulk-sensitive techniques, such as transport measurements, may be the best hope for examining changes in electronic and physical properties of Pu as it ages. Fluss *et al.* (2004) measure the changes in electrical resistivity during isochronal annealing of self-induced radiation damage in Ga-stabilized δ-Pu. This approach is able to show clear changes in the resistivity as damage accumulated at low temperatures, revealing five stages of defect kinetics. Interestingly, the results of Fluss *et al.* (2004) show that most of the damage caused by the U recoil in the α decay event is removed by annealing well below room temperature. This result shows that a majority of self-induced radiation damage anneals out at room temperature, in turn suggesting that spectroscopic evidence for changes is in the electronic structure of Pu due to lattice damage may be more clearly observed at low temperature.

## 6. Superconductivity at 18.5 K

One of the most fascinating discoveries in the last few years pertaining to Pu is the discovery of superconductivity to 18.5 K in PuCoGa$_5$ (Sarrao *et al.*, 2002). While this review is focused on actinide metals, discussion of the PuCoGa$_5$ is justified, given the impact, understanding, and questions it has generated in the field of actinides and the behavior of the 5*f* states. The structure of PuCoGa$_5$ is a member of a large class of materials known as the 115 group, where the number represents the stoichiometric charcteer $X_1Y_1Z_5$. PuCoGa$_5$ is similar to fcc Pu-Ga, but elongated in the *z*-axis into a tetragonal crystal, allowing the accommodation of a layer of Co atoms. The inferred electronic specific heat coefficient for PuCoGa$_5$ is 77 mJ K$^{-2}$ mol$^{-1}$, which is even larger than the 35-64 mJ K$^{-2}$ mol$^{-1}$ observed in δ-Pu (Wick, 1980; Lashley *et al.*, 2003; 2005; Javorsky *et al.*, 2006). Thus, there is a small yet measurable increase in the quasiparticle mass between the δ-Pu metal and the superconductor.

A number of electronic-structure calculations are aimed at understanding how the *d* and *f* states in PuCoGa$_5$ affect superconductivity (Opahle and Oppeneer, 2003; Maehira *et al.*, 2003; Jutier *et al.*, 2007). Opahle and Oppeneer (2003) identified the superconductivity to be a direct consequence of Cooper pairing in the 5*f* states. This seems rather odd given that UCoGa$_5$ and URhGa$_5$ are nonsuperconducting (Ikeda *et al.*, 2002) and NpCoGa$_5$ orders antiferromagnetically below 47 K (Colineau *et al.*, 2004). What is even more interesting is that the isostructural compound PuRhGa$_5$ is reported to be superconducting below 8.5 K (Wastin *et al.*, 2003). Rather than just blanketly saying it is the 5*f* states responsible for superconductivity, the 5*f*-element 115 compounds above suggest that there is something special about Pu, and, possibly its location near the itinerant-localized transition in Fig. 6(a). If this is true, then Ce 115 should behave in a similar manner. In fact, the similarities between Ce and Pu 115 compounds are close; as an example, neither PuIrGa$_5$ nor CeIrGa$_5$ superconduct (Bauer *et al.*, 2004; Opahle *et al.*, 2004).This commonality of Pe and Ce suggests that the magic aspect of 115 superconductivity may be having *f* electron states that are near a localized-delocalized transition, and thus easily tunable with temperature, pressure and chemistry. To truly understand superconductivity in 115 compounds a considerable amount of work must still be done. However, it is already certain that the superconductive state in PuCoGa$_5$ is delicate and easily effected by impurities. Evidence of this is given by the fact that self-induced irradiation damage causes the superconducting transition temperature to drop ~0.2 K per month in PuCoGa$_5$, slowing in rate as defects accumulate over time. (Sarrao *et al.*, 2002; Booth *et al.*, 2007; Jutier *et al.*, 2007a and 2007b). PuCoGa$_5$ and the 115 compounds will be further discussed in the section on pressure-dependent superconductivity of Am metal.

## F. Americium

Technologically, Am is of passing importance, even though actinide scientists love to point out that it is an integral component of smoke detectors. In fact, one Eagle Scout with a merit badge in atomic energy took this to heart and attempted to create a home-made breeder reactor from the Am in smoke detectors (Silverstein, 2004). The scout purchased hundreds of smoke detectors, extracted the $^{241}$Am with help from an unbeknownst electronics company, then encapsulated the Am inside a hollow block of Pb with a small hole in one side to allow the release of alpha radiation. The boy then made a functional neutron gun by placing the Am-containing Pb block in front of an aluminum sheet, which effectively absorbs alpha emission and emits neutrons. The boy bought thousands of lanterns, removed the pouches, and used a torch to reduce them to ash. Lithium batteries, aluminum foil, and a Bunsen burner were utilized to extract the thorium, which was the correct element and isotope to be transferred into uranium. The reason why the small nuclear breeder reactor did not work? The neutron gun was simply not strong enough.

Scientifically, Am represents a changing point in the actinide series, the place where the 5$f$ states go from delocalized and bonding to localized and similar to the rare earth metals (Figs. 1 and 3). Moving from Pu to Am there is a myriad of changes in the physical properties, such as resistivity (Fig. 23(a)), electronic specific heat (Fig. 23(b)), lattice constant or atomic volume (Fig. 1), cohesive energy (Ward $et~al.$, 1976), superconductivity (Smith and Haire, 1978), high-pressure behavior (Roof $et~al.$, 1980; Heathman $et~al.$, 2000; Lindbaum $et~al.$, 2001), melting temperature (Fig. 3), and crystal structure, which is double hexagonal closest packed rather than the low-symmetry structures seen in Pa, U, Np, and Pu (Fig. 3). Temperature-independent magnetic susceptibility (Brodsky 1971; Kanellakopulos $et~al.$, 1975) shows Am has a $5f^{6}$ configuration, which is nonmagnetic with $J = 0$ due to coupling of the 5$f$ electrons.

Valence-band ultraviolet PE supplies clear evidence that Am is the first rare-earth-like actinide due to strongly localized 5$f$ states, as shown in Fig. 20 (Naegele $et~al.$, 1984). U, Np, and Pu exhibit a saw-tooth shaped valence-band spectrum with varying degrees of fine structure. However, in Am the density of states, which are dominated by 5$f$ character, are removed from the Fermi level by almost 2 eV with appreciable structure in the spectrum. This structure could arguably be due to surface effects from the surface sensitivity of UPS; however, it is probable that the spectrum contains both surface and bulk contributions. The small peak at 1.8 eV has been interpreted by Johansson (1984) and Mårtenson $et~al.$ (1987) as being due to a divalent ($spd^2$) surface layer, similar to observations made in Sm (Lang and Baer, 1979; Allen $et~al.$, 1980). Studies of AmN, AmSb, and Am$_2$O$_3$ in relation to Am metal conclude that the peak at 1.8 eV is attributed to a well-screened channel of photoemission from the $5f^6$ ground state, while the large peak at 2-3 eV is due to the poorly-screened channel (Gouder $et~al.$, 2005). Regardless of peak interpretation, the fact remains that the valence-band spectra clearly show that Am contains mostly localized 5$f$ states. The proof is in the original paper by Naegele $et~al.$ (1984), where PE taken at 1253.6 eV is also presented. The PE spectra taken with x-rays is practically identical to spectra taken with He II, albeit with little fine structure.

The 4$f$ PE spectrum of Am in Fig. 21 also show the large degree of 5$f$ localization in the metal (Naegele $et~al.$, 1984. See also Cox $et~al.$, 1992). In this spectrum, the 4$f_{5/2}$ and 4$f_{7/2}$ peak structure has changed such that the poorly-screened peak is almost entirely dominant and the well-screened peak is a weak feature on the leading edge. Of course, the absence of the well-screened peak is due to the localization of the 5$f$ states, which are not available to effectively screen the core holes produced by the photoemission process. The 3$d$ PE spectra of La, Ce, Pr, and Nd show similar behavior (Creselius $et~al.$, 1978), further supporting the rare-earth-like behavior of Am. Interestingly, while the small well-screened peaks are present in the spectrum for Am metal, they are absent in AmH$_2$ (Cox $et~al.$, 1992). This could be taken as the disappearance of any degree of delocalization of the 5$f$ states when moving from the metal to the hydride phase. In other words, the small well-screened peak on the leading edge of the 4$f_{5/2}$ and 4$f_{7/2}$ peak are proof that there is weak remnant 5$f$ bonding in Am metal.

EELS, XAS, and spin-orbit sum rule analysis show that Am has a very strong spin orbit interaction (Moore $et~al.$, 2007a; 2007b). This is due to the fact that the $j = 5/2$ level in the 5$f$ states holds six electrons. In the actinide metals up to Am, there is a preferential filling of the 5/2 states, as shown in Secs. III and IV. Once Am is reached, the spin-obit interaction acts as a stabilizing force by filling of the $j = 5/2$ level. In reality, exchange interaction is still present and in competition with the spin-orbit interaction, which results in intermediate coupling of the $5f$ states. However, intermediate coupling of the $5f$ states in Am is very close to the $jj$ limit, meaning only a small electron occupation is observed in the $j = 7/2$ level. The EELS and spin-orbit results by Moore *et al.* (2007a; 2007b) also clearly indicate a $5f^6$ configuration, which is in turn supported by PE results (Naegele *et al.*, 1984; Cox *et al.*, 1992; Gouder *et al.*, 2005), transport measurements (Müller *et al.*, 1978), thermodynamic measurements (Hall *et al.*, 1980), DFT (Söderlind and Landa, 2005), and DMFT (Shim *et al.*, 2007).

Examining the temperature-dependent resistivity along the actinide series in Fig. 23(a), the values peak for α-Pu, but then drop again for Am (Müller *et al.*, 1978). The reduction is partly due to the localization of the $5f$ states, which are no longer contributing strongly to the temperature-dependent resistivity curve. An anomaly at 50 K is reported by Schenkel *et al.* (1976) and Müller *et al.* (1978) in the temperature-dependent resistivity. An anomaly is also reported at the same temperature in specific heat of Am metal (Müller *et al.* 1978). It seems unlikely that this is a charge density wave due to the dhcp crystal structure of Am metal; CDWs prefer low-symmetry structures where Fermi surface nesting is viable. A similar anomaly is observed in Pu at 60 K (Miner, 1970; Lee and Waldren, 1972), but such peaks are shown to arise in resistivity measurements performed in exchange gas because the gas can be adsorbed onto the sample during the measurement (Taylor *et al.*, 1965; Gordon *et al.*, 1976). Thus, the anomaly at 50 K in the temperature-dependent resistivity and specific heat of Am are likely an artifact. It should be noted, however, that for low-temperature measurements on dopant-stabilized δ-Pu the α′ phase begins to grow in at ~140 K and continues down to ~110 K (Hecker *et al.*, 2004; Blobaum *et al.*, 2006; Moore *et al.*, 2007c). This will lead to one or more peaks in the low-temperature resistivity curves of stabilized δ-Pu.

Resistance of Am increases with self-induced radiation damage, as expected. When Am metal is held at 4.2 K for 738 hours, the metal goes from 2.44 μΩ-cm to a saturated value of 15.85 μΩ-cm by approximately 150 hours (Müller *et al.*, 1978). In a manner similar to the studies of Fluss *et al.* (2004) on Pu, Müller *et al.* (1978) perform isochronal annealing on Am and observe several stages of defect evolution through the release of stored energy. The effects of self-induced radiation damage are reduced as compared to the lighter actinides, presumably due to the lack of strong $5f$ bonding in Am. Nonetheless, this again shows clear evidence of the changes self-induced irradiation damages can cause in actinide metals, particularly at low temperatures.

Americium is a model system for pressure studies because it is the first actinide metal that is well described by localized $5f$ electrons with a $5f^6$ configuration and an almost full $j = 5/2$ level. Bonding is primarily achieved by the itinerant $spd$ band, resulting in the dhcp ground-state structure that is indicative of $d$-state bonding (Duthie and Pettifor, 1977). In other words, pressurizing Am and the actinide metals with higher atomic number, enables one to observe the changes in physical behavior of a metal as $5f$ bonding is 'turned on'. Diamond anvil cell experiments on Am up to 100 GPa by Heathman *et al.* (2000) and Lindbaum *et al.* (2001) show the metal has three phase transitions between four crystal structures: Am I (dhcp), Am II (fcc), Am III (orthorhombic, space group *Fddd*), and Am IV (orthorhombic, space group *Pnma*). The relative volume of Am metal as a function of pressure is shown in Fig. 28 (Heathman *et al.*, 2000; Linbaum *et al.*, 2001). Am III is the same crystal structure as γ-Pu, and Am IV is very close to the orthorhombic α-Np structure *Pmcn*, except for a rotation about the [111] axis. Based simply on the symmetry of the crystal structures, it is clear that $5f$ bonding becomes important for Am III and Am IV, meaning pressure induces active bonding of the 5*f* states. What is more, it seems the change to delocalized 5*f* bonding occurs over the two transformations, where Am II to Am III is an fcc → orthorhombic transformation with a 2% volume decrease and Am III to IV is an orthorhombic → orthorhombic transformation with a 7% volume collapse. Am IV is the least compressible, showing pressure-induced lattice changes that are similar to α-U. The pressure-induced phase transformations of Am are modeled using DFT (Pénicaud, 2002; Söderlind and Landa, 2005) and DMFT (Savrasov *et al.*, 2006), each yielding an equation-of-state that matches experiment exceptionally well.

## 1. Pressure-dependent superconductivity

Am metal superconducts below 0.8 K at ambient pressure (Speculation: Hill, 1971; Johansson and Rosengren, 1975. Experimental proof: Smith and Haire, 1978; Smith *et al.*, 1979). This is somewhat surprising given that both metals before Am in the actinide series, i.e., Np and Pu, show no signs of superconductivity (Meaden and Shigi, 1964), and the two elements after Am, i.e., Cm and Bk, exhibit magnetic moments (For Cm: Marei and Cunningham, 1972. For Bk: Peterson *et al.*, 1970). Furthermore, diamond-anvil-cell studies show that the pressure dependence of the superconducting transition temperature is rather unusual, ranging from 0.8 to 2.2 K (Link *et al.*, 1994; Griveau *et al.*, 2005). The superconducting transition temperature $T_c$ of Am is plotted as a function of the relative volume in the inset in Fig. 28. Notice that $T_c$ exhibits a rich and interesting dependence on the volume of the metal (pressure), showing two maxima. The strong dependence of resistivity as a function of pressure is not observed in lighter actinides and the results of Griveau *et al.* (2005) suggest that the *f* electrons play an important role in the transport properties, scattering the *spd* conduction electrons when they are localized and contributing to the transport when they are itinerant.

Am I and II can be safely described as a localized system, where the 5*f* electrons do not strongly participate in bonding. However, Am III and Am IV certainly have appreciable 5*f* bonding, which is evidenced by the increase in the bulk modulus from ~30 GPa in Am I to ~100 GPa in Am IV (Linbaum *et al.*, 2001), as well as the fact that Am IV exhibits an orthorhombic crystal structure indicative of *f* electron bonding. Thus, the two maxima in the inset in Fig. 28, one at the Am I - Am II transition and one just after the transition to Am IV, must be due to different influences. The first maxima in $T_c$ as a function of pressure occurs near the Am I - Am II phase transition, where the 5*f* states are localized and an itinerant $spd^3$ performs the bulk of bonding in the metal. With increased pressure, the energy of another configuration approaches the Fermi level, beginning to admix into the ground state. This accounts for the maximum of $T_c$ near the Am I - Am II phase transition, which in a mixed-valence fluctuation mechanism occurring when the configurations $5f^6$ and $5f^5sd^1$ (or $5f^6sd^1$ and $5f^7$ at the Fermi level) become degenerate (Griveau *et al.*, 2005). The second maximum of $T_c$ at $\Delta V/V_0 = 0.4$ in the inset to Fig. 28, which is just after the volume collapse, are described by two effects by Griveau *et al.* (2005): At lower pressures, the *f* electrons localize and do not participate in superconductivity, only acting to scatter the *spd* conduction electrons. At large pressures, the superconducting transition temperature decreases because the *f* kinetic energy becomes large compared to the pairing interactions. The maximum of $T_c$ at $\Delta V/V_0 = 0.4$ occurs between these two effects.

For decades, superconductivity was believed to be phonon-mediated and that magnetism was entirely immiscible with Cooper pairing of electrons. However, this idea was first questioned when both superconductivity and spin fluctuations were observed in CeCu$_2$Si$_2$ (Steglich *et al.*, 1979) and UPt$_3$ (Ott *et al.*, 1983; Stewart *et al.*, 1984). Conventional theory purports that magnetism destroys superconductivity by flipping the intrinsic spin of one of the paired electrons, thus breaking the Cooper pair. However, in UPt$_3$ magnetic forces appear to bind the conduction electrons into Cooper pairs, rather than electron-phonon interactions. Subsequently, more 'heavy fermion' compounds, such as CeCoIn$_5$ (Sidorov *et al.*, 2002), were discovered and examined, further showing the possibility of combined superconductivity and magnetism. Most recently, PuCoGa$_5$ was discovered with a superconducting transition temperature of 18.5 K (Sarrao *et al.*, 2002), as discussed in the Pu section of this review. The plutonium-based superconductor is *almost* magnetic with an unconventional pairing mechanism (Bauer *et al.*, 2004 Curro *et al.*, 2005), similar to that found in the heavy fermion superconductors. However, Jutier *et al.* (2007b) analyze the behavior of PuCoGa$_5$ at different ages in the framework of the Eliashberg theory, assuming electron-phonon coupling. They show that they can reproduce all the available measurements without invoking spin-fluctuations. In particular, they reproduce the *d*-wave symmetry of the gap, the local spin susceptibility in the superconducting phase, the spin-lattice relaxation rate, the electrical resistivity above $T_c$, the critical field, and the penetration depth. These conclusions are in contradiction to magnetic fluctuations playing a role in the superconductivity in these compounds. Thus, the question must be addressed: What role, if any, does magnetism play in the superconductivity of the 115 compounds.

The fact that the superconducting transition temperature in Am metal is dependent on the degree of *f*-electron bonding is intriguing and further suggests that the proximity of the 5*f* states to the localized-delocalized transition in the actinide series are integral to the PuCoGa$_5$ superconductor. This is simply one more example of the fascinating physics found at or near the itinerant-localized transition in Fig. 6(a). The 'tunability' of the valence electrons in these elements allows rapid and easy changes in the electronic structure of the material with temperature, pressure, and chemistry. Further, all these elements are good candidates for materials with quantum critical points due to the competition between localization and delocalization of conduction electrons, a balance that may hold the key to the co-existence of magnetism and superconductivity.

## G. Curium

Curium is the first actinide metal that exhibits magnetism due to the pronounced shift in the angular-momentum coupling mechanism of 5*f* states toward a more *LS*-like behavior. This creates a strong spin polarization though Hund's rule, producing an effective magnetic moment of ~8 $\mu_B$/atom in Cm metal (Marei and Cunningham, 1972; Huray *et al.*, 1980; Nave *et al.*, 1983; Huray and Nave, 1987). Pu, Am, and Cm all exhibit intermediate coupling; however, while Pu and Am exhibit strong spin-orbit interactions, Cm is strongly shifted towards the *LS* limit. This is clearly shown in the spin-orbit sum rule analysis of the EELS spectra (Moore *et al.*, 2007a; 2007b), which is shown in Fig. 17(a).

The physical origin of the abrupt and striking change in spin-orbit expectation value in Fig. 17(a) is caused by a transition from optimal spin-orbit stabilization to optimal exchange interaction stabilization (Moore *et al.*, 2007a; 2007b). In *jj* coupling, the electrons first fill the $f_{5/2}$ level, which can hold no more than six, then carry on to fill the $f_{7/2}$ level. The maximal energy gain in *jj* coupling is therefore obtained for Am $f^6$, since the $f_{5/2}$ level is full. However, for Cm $f^7$ (Smith *et al.*, 1969; Milman *et al.*, 2003; Shim *et al.*, 2007) at least one electron must be relegated to the $f_{7/2}$ level. Having six electrons in the $f_{5/2}$ level and one electron in the $f_{7/2}$ level costs energy. Therefore, the maximal energy stabilization for the $f^7$ configuration is achieved through exchange interaction, where the spins are parallel in the half filled 5*f* shell, and this can only be achieved in *LS* coupling. Thus, the large changes observed in the electronic and magnetic properties of the actinides at Cm are due to this transition from optimal spin-orbit stabilization for $f^6$ to optimal exchange interaction stabilization for $f^7$. In all cases, the spin-orbit and exchange interaction compete with each other, resulting in intermediate coupling; however, increasing the *f* count from 6 to 7 shows a clear and pronounced shift in the power balance in favor of the exchange interaction, resulting in the large shift of the expectation value for the intermediate coupling curve in Fig. 17(a). The effect is in fact so strong that, compared to Am, not one but two electrons are transferred to the $f_{7/2}$ level in Cm, as shown in Fig. 17(b) and Table IV. Consistent with that, Table V shows a dramatic increase of the high-spin character in the intermediate coupled ground state when going from $f^6$ to $f^7$, namely 44.9% septet for $f^6$ to 79.8% octet spin for $f^7$. As it turns out, this shift of the the angular-momentum coupling mechanism of 5*f* states has considerable implications for the crystal structure of Cm metal.

    A diamond-anvil-cell study of Cm by Heathman *et al.* (2005) reveals that the metal undergoes four phase transformations between five different crystal structures Cm I – Cm V. These structures are shown in Fig. 29(a), where the atomic structures are: Cm I (dhcp), Cm II (fcc), Cm III (monoclinic, space group *C*2/*c*), Cm IV (orthorhombic, space group *Fddd*), and Cm V (orthorhombic, space group *Pnma*). The structure of Cm IV is the same as Am III and the structure of Cm V is the same as Am IV (Heathman *et al.* 2000, Lindbaum *et al.* 2001). However, the monoclinic *C*2/*c* structure of Cm III is not observed in any other actinide, with the only other monoclinic structures being the *P*2$_1$/*m* structure of α-Pu and the *C*2/*m* structure of β-Pu. The pressure of each phase transformation is shown in Fig. 29(b), along with the associated volume collapse between each structure. Immediately noticeable is that the change from a large-volume metal to a small-volume metal occurs over a ~90 GPa span of pressure and over several phases. This is precisely what is observed for Am, which is also shown in Fig. 29(b). In contrast, α-U retains its orthorhombic structure from 0 to 100 GPa.

    Heathman *et al.* (2005) also perform *ab initio* calculations of the total energy difference between Cm II, Cm III, Cm IV, and Cm V structures as a function of volume, which is shown in the inset of Fig. 29(b). These calculations illustrate that in order to achieve the proper series of phases as a function of pressure, magnetic interactions must be present. What is more, the calculations show that the Cm III phase *can only be stabilized by spin polarization*. In other words, the metal's own intrinsic magnetism stabilizes the atomic geometry (Söderlind and Moore, 2008). It turns out that the angular momentum coupling of the 5*f* states is the root cause of this magnetically-stabilized Cm phase, and that the strong shift of the intermediate coupling curve from near the *jj* limit for Pu and Am to near the *LS* limit for Cm is the key. In order to better understand this, we must return to atomic calculations.

    The spin and orbital magnetic moments, $m_s$ and $m_l$, respectively, from atomic calculations are plotted against $n_f$ in Fig. 18(a) and (b), respectively (van der Laan and Thole, 1996b; Moore *et al.*, 2007a). In each graph, the three different angular momentum coupling mechanisms are shown: *LS*, *jj* and intermediate. Examining the plots, we see that for some elements the choice of coupling mechanism has a large influence on the spin and orbital magnetic moments. This is most remarkable for Cm ($n_f$ = 7), where Fig. 18(a) shows that the spin magnetic moment is modest for the *jj* coupling limit, but is large for both *LS* and intermediate coupling. The fact that the spin magnetic moment for the intermediate coupling is almost as large as that for the *LS* limit is because the intermediate coupling curve moves strongly back towards the *LS* limit at Cm in Fig. 17(a). Thus, it is the pronounced shift of the intermediate coupling curve towards the *LS* coupling limit at Cm – in order to accommodate the exchange interaction – that creates a large

and abrupt change in the electron occupancy of the $f_{5/2}$ and $f_{7/2}$ levels shown in Fig. 18(c). This figure is a reproduction of Fig. 17(b), but has the EELS data for Th, Pa, U, and Np removed, as well as the calculated coupling curves for the *LS* and *jj* limits. We leave only the intermediate coupling curve, since Pu, Am, and Cm adhere to the curve most closely. In Fig. 18(c), the calculated $n_{5/2}$ and $n_{7/2}$ occupation numbers are shown by the black and red lines, respectively, and the spin-orbit analysis of the experimental EELS spectra as blue points. Examining Fig. 18(a), (b), and (c) together clearly shows that if the intermediate coupling curve remained near the *jj* limit for Cm, the spin (and total) magnetic moment would be much smaller than the observed ~8 $\mu_B$/atom magnetic moment (Marei and Cunningham, 1972; Huray *et al.*, 1980; Nave *et al.*, 1983; Huray and Nave, 1987) and have little or no effect on the crystal structure of the metal.

    These results mean that spin polarization is an integral part for calculations of Cm, but should be weaker for Pu and Am. Indeed, recent DFT (Söderlind, 2007a) and DMFT (Shim *et al.*, 2007) have acknowledged this point, conducting calculations that show this large change in angular-momentum state occupancy between Am and Cm. These results also have considerable ramifications for Pu. Many calculations have used spin polarization to mimic electron correlations in Pu (For example: Skriver, 1978; Solovyev *et al.*, 1991; Söderlind *et al.*, 1997; Bouchet *et al.*, 2000; Savrasov and Kotliar, 2000; Söderlind and Sadigh, 2004), resulting in long-range magnetic order that is not observed experimentally (Lashley *et al.*, 2005). However, the EELS and spin-orbit analysis results (Moore *et al.*, 2007a; 2007b) clearly show that spin polarization is not strong in Pu.

    High pressure studies on rare earth metals provide insight into *f*-electron bonding that points in a direction for the study of magnetism as a function of 5*f* bonding. In a diamond-anvil cell-study, Maddox *et al.* (2006) pressuriz Gd to 113 GPa and analyze the metal through a 59 GPa volume collapse using resonant inelastic x-ray scattering and x-ray emission spectroscopy. Gadolinium has an $f^{\,7}$ configuration in the rare earth series and thus has similarities to Cm, which is $f^{\,7}$ in the actinides. Their x-ray emission spectroscopy data clearly show that as the metal is pressurized and the volume collapses, the 4*f* moment persists to the high-density phase above 59 GPa where *f* electron delocalization is believed to occur. A low-energy satellite due to intra-atomic exchange interactions between the 4*f* and core states is present in the $4d \rightarrow 2p$ $L\gamma_1$ x-ray emission spectra. The strength of the satellite in relation to the main peak reflects the size of the 4*f* moment. Maddox *et al.* (2006) found no significant change in the satellite peak up to 106 GPa, suggesting that the moment persisted well across the 59 GPa volume collapse. Based on high-energy neutron scattering from α-Ce, Murani *et al.* (2005) and Maddox *et al.* (2006) argued that x-ray emission spectroscopy revealed the bare 4*f* moment and that any loss of moment in the magnetic susceptibility must arise from screening by the valence electrons. Performing a similar diamond-anvil-cell experiment on Cm, where the metal is pressurized across the localized-delocalized transition, could yield great insight into the magnetic behavior of the 5*f* states. Such an experiment may give insight into the potential screening of the moment that should be in Pu given the $5f^{\,5}$ like configuration of the metal.

## H. Berkelium

Berkelium is named after the illustrious town of Berkeley, California due to the pioneering role of the Berkeley campus of the University of California in the early days of actinide research, where the first transuranium elements were first isolated under the leadership of Glenn Seaborg (Thompson *et al.*, 1950; Hoffman *et al.*, 2000). There are no spectra of Bk metal, to the best of our knowledge, but valence-band and 4*f* PE of Bk oxide can be found in Veal *et al.* (1977).

The many-electron atomic spectral calculations in Sec. III show that Bk should have similar, albeit slightly weaker magnetic behavior as Cm. The actinides from Pu to Cm closely follow the intermediate coupling curve for the 5*f* states, making the assumption that Bk will follow this curve also quite reasonable. This assumption is further supported by the atomic volume in Fig. 1, which shows Bk has localized 5*f* states, as observed in Am and Cm. The fact that Bk has localized 5*f* states is supported by the ground-state dhcp crystal structure it exhibits (Peterson and Fahey, 1971), which is indicative of *d* bonding states (Duthie and Pettifor, 1977). Having convinced ourselves that Bk will exhibit intermediate coupling for the 5*f* states, we can turn our attention to Fig. 17(a). In intermediate coupling, Bk will fall almost directly in the middle of the *jj* and *LS* coupling limits, meaning the spin-orbit and exchange interactions are equally important. Examining the calculated spin and orbital magnetic moments for Bk $f^8$ in Fig. 18(a) and (b), respectively, it is clear that the moments are not much different than observed in Cm. The spin magnetic moment is ~5 $\mu_B$/atom, while the orbital magnetic moment is ~2 $\mu_B$/atom. Due to the spin-orbit interaction of the 5*f* states, the spin and orbital moments will align parallel, resulting in a total moment of ~7 $\mu_B$/atom. Thus, as compared to Cm, atomic theory shows that Bk has a slightly smaller total magnetic moment with more contribution from the orbital component. Indeed, magnetic susceptibility measurements on Bk metal show an effective magnetic moment of ~8 $\mu_B$/atom (Peterson *et al.*, 1970). The electron occupation numbers for the *j* = 5/2 and 7/2 levels in Fig. 18(c) and Table IV show that more weight is gained in the *j* = 5/2 level than the 7/2 when moving from Cm $f^7$ to Bk $f^8$. This is precisely why we see the intermediate coupling curve move incrementally back towards the *jj* limit in Fig. 17(a) when moving from Cm to Bk.

When compressed using a diamond anvil cell, Bk exhibits three different crystallographic phases up to 57 GPa (Haire *et al.*, 1984; Benedict *et al.*, 1984). It exhibits a bulk modulus of 35±5 GPa (Haire *et al.*, 1984), as shown in Fig. 22. The ambient dhcp phase (Bk I) is transformed to the fcc phase (Bk II) at 8 GPa, which then transforms to a structure that was first believed to be α-U crystal structure (Bk III) with space group *Cmcm* at 22 GPa. However, subsequent analysis of the data shows the structure to be inconclusive. In addition, electronic-structure calculations suggest the Cm IV is a better candidate for Bk III than the α-U *Cmcm* structure (Söderlind 2005). Regardless of the controversy of the exact phase, this means that Bk behaves much like Am when compressed, going through several phase transformations with small volume collapses, rather than a single large one. In other words, the transition from localized to delocalized 5*f* states occurs over a large pressure range and numerous crystal structures. This is counter to theoretical predictions by Johansson *et al.* (1981) that suggested Bk would undergo a large volume collapse when the 5*f* electron transition from strongly localized to strongly delocalized and bonding. Bk-Cf (Itie *et al.*, 1985) and Cm-Bk (Heathman and Haire, 1998; Heathman *et al.*, 2007) alloys have also been studied using diamond anvil cell experiments, each exhibiting pressure-induced phase transformations that are similar to pure Bk and Cf.

## V. COMMENTS AND FUTURE OUTLOOK

The localized-delocalized transition at Pu in Fig. 1 appears abrupt; however, looking over the spectroscopy and transport data in the previous sections, we see a more gradual transition that spans Np to Am . Np and α-Pu both show evidence in their valence-band and 4*f* PE as well as EELS spectra of the onset of localization effects in the 5*f* states. The branching ratio and spin-orbit sum-rule analysis of the $N_{4,5}$ EELS spectra reveal Np is between *LS* and intermediate coupling and α-Pu is on or near intermediate coupling, close to the *jj* limit. In addition, The Am 4*f* PE data show a small, but present, well-screened peak on the leading edge of the $4f_{5/2}$ and $4f_{7/2}$ peaks, which can be taken as proof that there is weak remnant 5*f* bonding in Am metal. Further, when Am, Cm, and Bk are compressed in a diamond anvil cell, the transition from localized to delocalized 5*f* states occurs over a wide pressure range and numerous crystal structures. There is no large and abrupt volume collapse, but rather a gradual change with several small volume collapses. Taken all together, it seems that the transition from delocalized and bonding 5*f* electrons to localized and atomic-like occurs over many elements. In other words, while the large and noticeable crystallography and volume effects are observed at Pu, changes in 5*f* localization begin appearing in Np and delocalization persists, albeit weakly, in Am. A better understanding of the localized-delocalized transition and how it occurs over the allotropic phases of Np, Pu, and Am is paramount to understanding the change in the 5*f* states.

Throughout this review we see that the *LS* (or Russell-Saunders) coupling scheme is appropriate when the spin-orbit interaction is weak compared to the electrostatic interactions. This is observed in U, where the delocalized 5*f* states exhibit an *LS* angular momentum coupling scheme. Further, while the 5*f* states of Cm in fact exhibit intermediate coupling, they are strongly shifted towards the *LS* limit due to exchange interaction. Thus, in the case of U and Cm we see that the electrostatic interactions play an appreciable role in relation to the spin-orbit interaction. On the other hand, a strong spin-orbit interaction is observed in Pu and Am, being evidenced by the EELS and atomic spectra and subsequent spin-orbit analysis. At the end, the intermediate coupling mechanism is appropriate for the 5*f* states in most actinide metals. This is because in all cases the spin-orbit and exchange interaction compete with each other, resulting in intermediate coupling. However, there are exceptions, such as α-U where the spin-orbit sum-rule analysis of the EELS spectra clearly shows a pure *LS* coupling mechanism.

Spin-orbit sum-rule analysis on 5*f* materials using EELS or XAS is favorable for the $M_{4,5}$ and $N_{4,5}$ edges of the actinides, given the small exchange interactions between the 3*d* and 4*d* core levels with the 5*f* valence states. Calculations show that $B_0$ varies between only 0.59 and 0.60 for the light actinides and, therefore, remains very close to the statistical ratio of 0.60 (van der Laan *et al.*, 2004). This means that the EELS and XAS branching ratios depend almost solely on the 5*f* spin-orbit expectation value per hole, thus affording an unambiguous probe for the 5*f* spin-orbit interaction in actinide materials. Spin-orbit analysis of the 5*f* states using the $O_{4,5}$ edge of the actinides is much more difficult, if not impossible, given the large exchange interactions between the 5*d* core levels and the 5*f* valence states in comparison to the 5*d* spin-orbit interaction. While this makes for complicated edges, there is still the possibility of analyzing edges as a function of bonding environment.

The lack of magnetism in Pu metal is perplexing. Regardless of whether the moment due to the $5f^5$ configuration is obfuscated by Kondo shielding, pairing correlations, spin fluctuations, or some other mechanisms, experiments need to be performed to clarify this issue. Testing the Kondo shielding hypothesis could be done via spin-polarized resonant photoemission. Two opportunities are available for $d \to f$ resonant PE: The $O_{4,5}$ edge, which involves the 5$d$ state, and the $N_{4,5}$ edge, which involves the 4$d$ state (The $M_{4,5}$ edge that involves the 3$d$ state is at ~3.5 keV, which is too high for reasonable energy resolution). Using the $O_{4,5}$ edge for resonant photoemission will result in spectra that are plagued by surface effects due to the fact that $h\nu \approx$ 100 eV. However, the resonant photoemission at the $N_{4,5}$ edge will be highly bulk sensitive because $h\nu \approx$ 795 and 840 eV for the $N_4$ or $N_5$ edge, respectively. This has been clearly shown in 3$d$ and 4$d$ resonant photoemission of CeRu$_2$Si$_2$ and CeRu$_2$ by Sekiyama *et al.* (2000). In this case, Ce is the element under investigation and it has an $N_{4,5}$ (4$d \to$ 4$f$) and $M_{4,5}$ (3$d \to$ 4$f$) edge that are near the same energies as the $O_{4,5}$ and $N_{4,5}$ edges of Pu. When resonant valence-band PE spectra are collected from the $N_{4,5}$ of Ce at 120 eV, a large $f^0$ peak is observed that is due to surface electronic structure (free bonds that cause localized 4$f$ states). However, when resonant valence-band PE spectra are collected from the $M_{4,5}$ of Ce at 880 eV, a spectra is achieved that is indicative of the bulk electronic structure. Following this logic, the best experiment for Pu would be spin-resolved resonant photoemission using the $N_{4,5}$ near 800 eV. This would offer energy high enough to probe the bulk electronic structure and would give the spin resolution to detect transient spin polarization of the 5$f$ states. Furthermore, using circularly polarized x-rays it is possible to measure the magnetic circular dichroism in non-resonant photoemission at the 4$d$ core level. Any possible magnetic polarization of the 5$f$ electrons will show up in the 4$d$ multiplet structure due to the 4$d$,5$f$ electrostatic interactions (Thole and van der Laan, 1991; van der Laan *et al.*, 2000).

It is possible that the moments in Pu are *dynamic* in nature (Arko *et al.*, 1972; Shim *et al.*, 2007). Indeed, this can be experimentally tested using inelastic neutron scattering as done by Murani *et al.* (2005) in Ce. However, the moment cannot be on the small energy scale as this is excluded by the Lashley *et al.* (2005). This begs the question whether the moment could occur on a much larger scale in energy. Here, in fact, one touches base with the high-$T_c$ materials, where the interesting point is the dynamics of the Cu spins. An experiment of paramount importance that will further advance our understanding of Pu electronic structure and the lack of magnetism is angle-resolved photoemission. Having a two-dimensional band mapping of δ-Pu is surely one of the kingpin experiments waiting to be performed.

The spin pairing correlations hypothesis can be tested using Andreev reflection experiments (Deutscher, 2005). This experiment tests how electrons pass through an interface between two materials. If two normal metals are adhered, like the close to free-electron metal Al, electric current would pass though the interface with little disruption. However, if a normal metal and a material with electrons that are paired are adhered, something different would happen. As a current is passed though a metal-superconductor sandwich, an electron incident from the metal that has lower energy than the superconductor energy gap would be converted into a hole at the metal-superconductor interface and move backward with respect to the electron current. Because a hole is created at the metal-superconductor interface and moves backward in the metal, a current of 2e$^-$ now moves forward though the superconductor. The result is a measured doubling of current of the superconductor side due to the electron-hole conversion at the interface. This technique could be used for Pu, where an Al-Pu sandwich is made and current passed from the Al side to the Pu side. If there are no spin pairing correlations in Pu, then the current would remain unchanged; however, if there are spin pairing correlations in Pu, then the current would double, or at least grow due to some fraction of electron-hole conversion at the Al-Pu interface.

Beyond Pu, the magnetic configurations of the heavier actinide metals must be addressed. In particular both Cm and Bk have strong effective magnetic moments, as evidenced by Figs. 17 and 18. A particularly useful experiment would be to examine the magnetic susceptibility of these metals as a function of pressure in a diamond anvil cell. Heathman *et al.* (2005) determined the crystallographic structure of Cm as a function of pressure and Moore *et al.* (2007a) calculated the phase stability of Cm as a function of pressure using magnetic and non-magnetic DFT. However, knowing how the magnetism changes as a function of pressure in the *experiment* would be highly desirable. Indeed, from Cm III to V the 5*f* states transition from localized to delocalized, meaning one would be examining the magnetic structure of the metal across the localized-delocalized transition. This, of course, would give more insight into Pu.

A myriad of low-temperature experiment should be done, including de Haas van Alphen on single-crystal samples. As stated previously, some metals are quite resistant to the production of large single crystals, such as Pu; however, if produced, de Haas-van Alphen experiments can be used to map the band structure contour at the Fermi level of the metal. A benefit of de Haas van Alphen is that it can be performed with single crystals on the order of 100 μg, whereas e.g., angle-resolved photoemission demands larger single crystal samples. Further, low-temperature diffraction experiments on Np and Pu for the purpose of searching for charge density waves is of great interest. Finding another three-dimensional charge density wave in an elemental metal would indeed be great find for actinide physics.

The possibility of a quantum critical point associated with Np or Pu should also be addressed. We know the effective mass of the electrons in Np, $\alpha$-Pu, and especially $\delta$-Pu is high given the large electronic specific heat. Doniach (1977) suggests the low-temperature phase diagram of heavy fermion materials emanates due to the competition between the Kondo coupling that screens the local *f*-electron moments and the magnetic exchange interaction between neighboring *f*-electron moments that is mediated by the surrounding polarized electron cloud. If the exchange interaction dominates, there is long range magnetic order, but the moments are screened and the material is paramagnetic to low temperatures. Accordingly, there is likely an order-disorder transition at 0°K separating two ground states. In the actinide series, this is most likely to occur in Pu when the lattice is expanded, or in Am as the lattice is compressed. For this reason, Pu-Am alloys represent a good place to look for a quantum critical point. Further, Pu-Am forms an fcc solid solution over the entire composition range, excluding complications due to phase change (Ellinger *et al.*, 1966). Magnetic field is another option; however, given that the Kondo energy is computed to be ~ 800°K (Shim *et al.*, 2007), and 1°K ≈ 1 Tesla, that makes *H* a tall order!

As illustrated in the first sections of this review, temperature, pressure, and chemistry rapidly affect the state and behavior of actinide metals. For this reason, geometric configurations that alter the bulk properties, such as thin films and multilayers, can provide interesting avenues of research for the actinides. This is because reduced size and/or dimensionality causes surface or interfacial energies to be closer in magnitude to bulk energies, in turn creating a new thermodynamic equilibrium. The work by Gouder *et al.* (2001) on Pu thin films show how the thickness of the film affect the localization of the 5*f* states, and the subsequent work of Gouder *et al.* (2005) illustrates how stepwise addition of Si to Pu cause the 5*f* states to localize and hybridize with the Si 3*p* states. In each of these cases pressure and temperature, respectively, are utilized to bring about a desired effect. Presently, the crystallographic and magnetic structure of U in thin film and multilayers is being examined with the interesting result of a stabilized hcp phase of pure U metal (Wilhelm *et al.*, 2007; Springell *et al.*, 2007a; 2007b). Using density-functional theory, Rudin (2007) shows that in nano-scale Pb-Pu superlattices there are two competing phases separated by a Mott transition between itinerant and localized 5*f* electrons.

Thus, the use of dimensional constraints remains a fruitful and rather untouched area of actinide physics.

Besides performing experiments on actinide materials themselves, surrogate materials should be considered. Not only does this avoid the problems of handling the materials, but it also opens our eyes to important results outside of the immediate community. For example, the semi-metal bismuth has proven to be a plausible surrogate for Pu in many regards (Du *et al.*, 2005; Murakami 2006). The specific heat of bismuth changes appreciably as the strength of the spin-orbit interaction changes, meaning that thermodynamic properties can be affected by spin-orbit interactions of the bonding electrons (Díaz-Sánchez *et al.*, 2007). Given that the 5*f* spin-orbit interaction is strong in many actinide metals, particularly Pu and Am, it is important to understand how this affects the physics of the materials. Examining Fig. 24(b) we see that the electronic specific heat increases from U to Np to Pu, in step with the increase in the spin-orbit interaction observed in Fig. 17(a). The electronic specific heat does drop for Am even though the 5*f* states have a strong spin-orbit interaction, but at this point the 5*f* states have become mostly localized.

Working on surrogate materials can save time, effort, and money as well as give an appreciation of the physics from a viewpoint outside of the actinides themselves; however, experiments on actinides must continue to truly understand the materials and the unique behavior of the 5*f* states. There is expanding interest in next-generation nuclear reactors (Clery, 2005) and an ever-aging nuclear stockpile to monitor. Further, many exceptional material behaviors are observed in the actinide series due to the 5*f* states and their extreme sensitivity to temperature, pressure, and chemistry. This opens the door to new and exciting frontiers in physics, ones that must be investigated if we are to understand the entire Periodic Table and all the secrets that it holds.

**Acknowledgements**

Covering all topics within a field, both accurately and inclusively, is exceedingly difficult for one or two people. For that reason, we greatly thank the following people for critical review of this manuscript: Ladia Havela, Gerry Lander, Jason Lashley, Michael Manley, Chris Marianetti, Scott McCall, Adam Schwartz, and Per Söderlind. We also thank Stephen Heathman for help ensuring the high pressure research is presented correctly and Richard Haire for synthesis of Am and Cm samples. We also wish to thank the following colleagues for prior and/or current collaborations: Mark Wall, Adam Schwartz, Per Söderlind, Theo Thole, Richard Haire, Brandon Chung, Simon Morton, David Shuh, Roland Schulze, Sorin Lazar, Frans Tichelaar, and Henny Zandbergen. This work performed under the auspices of the U.S. Department of Energy by Lawrence Livermore National Laboratory under Contract DE-AC52-07NA27344.

**TABLES:**

TABLE I. The expected number of 5f electrons, $n_f$, the experimental branching ratio, $B$, of the $N_{4,5}$ EELS spectra, and the expectation value of the 5f spin-orbit interaction per hole, $<w^{110}>/(14-n_f) - \Delta$, for the α phase of Th, U, Np, Pu, Am, and Cm metal. Each branching ratio value is an average of between 10 to 20 EELS spectra, with the standard deviation given in parenthesis. The sum rule requires a small correction factor, which is $\Delta$ = -0.017, -0.010, -0.005, 0.000, 0.005, and 0.015 for $n_f$ = 1, 3, 4, 5, 6, and 7, respectively. The experimental electron occupation numbers $n_{5/2}$ and $n_{7/2}$ of the $f_{5/2}$ and $f_{7/2}$ levels are obtained by solving Eqs. (49) and (50).

| Metal | $n_f$ | $B$ | $<w^{110}>/(14-n_f)-\Delta$ | $n_{5/2}$ | $n_{7/2}$ |
|---|---|---|---|---|---|
| Th | 1.3 | 0.646 (003) | -0.115 (008) | 1.28 | 0.02 |
| U | 3 | 0.686 (002) | -0.215 (005) | 2.35 | 0.65 |
| Np | 4 | 0.740 (005) | -0.350 (013) | 3.24 | 0.76 |
| Pu | 5 | 0.826 (010) | -0.565 (025) | 4.32 | 0.68 |
| Am | 6 | 0.930 (005) | -0.825 (013) | 5.38 | 0.62 |
| Cm | 7 | 0.794 (003) | -0.485 (008) | 4.41 | 2.59 |

TABLE II. Comparison of the radial parameters for the Coulomb interaction, $F^k(\ell,\ell)$, and spin-orbit interaction, $\zeta_\ell$, for actinides with rare earths (Thole *et al.*, 1985b) and 3*d* transition metals (van der Laan and Kirkman, 1992). All values in eV. The Slater integrals have been reduced to 80% of the atomic Hartree-Fock values.

|  | $F^2$ | $F^4$ | $F^6$ | $\zeta_\ell$ |
|---|---|---|---|---|
| $^{25}$Mn$^{2+}$ 3$d^5$ | 8.25 | 5.13 | - | 0.040 |
| $^{64}$Gd$^{3+}$ 4$f^7$ | 11.60 | 7.28 | 5.24 | 0.197 |
| $^{96}$Cm$^{3+}$ 5$f^7$ | 8.37 | 5.46 | 4.01 | 0.386 |

TABLE III. Parameter values for the Coulomb interaction, $F^k$, and spin-orbit interaction, $\zeta$, for the ground state of the trivalent actinides (Ogasawara *et al.*, 1991). The Slater integrals have been reduced to 80% of the atomic Hartree-Fock values. The 5*f* spin-orbit splitting is 7/2 $\zeta$. All values are in eV.

|  | $n$ | $S$ | $L$ | $J$ | $F^2$ | $F^4$ | $F^6$ | $\zeta_{5f}$ |
|---|---|---|---|---|---|---|---|---|
| Th$^{3+}$ | 1 | 0.5 | 3 | 2.5 |  |  |  | 0.169 |
| Pa$^{3+}$ | 2 | 1 | 5 | 4 | 6.71 | 4.34 | 3.17 | 0.202 |
| U$^{3+}$ | 3 | 1.5 | 6 | 4.5 | 7.09 | 4.60 | 3.36 | 0.235 |
| Np$^{3+}$ | 4 | 2 | 6 | 4 | 7.43 | 4.83 | 3.53 | 0.270 |
| Pu$^{3+}$ | 5 | 2.5 | 5 | 2.5 | 7.76 | 5.05 | 3.70 | 0.307 |
| Am$^{3+}$ | 6 | 3 | 3 | 0 | 8.07 | 5.26 | 3.86 | 0.345 |
| Cm$^{3+}$ | 7 | 3.5 | 0 | 3.5 | 8.37 | 5.46 | 4.01 | 0.386 |
| Bk$^{3+}$ | 8 | 3 | 3 | 6 | 8.65 | 5.65 | 4.15 | 0.428 |
| Cf$^{3+}$ | 9 | 2.5 | 5 | 7.5 | 8.93 | 5.84 | 4.29 | 0.473 |
| Es$^{3+}$ | 10 | 2 | 6 | 8 | 9.19 | 6.02 | 4.42 | 0.520 |
| Fm$^{3+}$ | 11 | 1.5 | 6 | 7.5 | 9.45 | 6.19 | 4.55 | 0.569 |
| Md$^{3+}$ | 12 | 1 | 5 | 6 | 9.71 | 6.36 | 4.68 | 0.620 |
| No$^{3+}$ | 13 | 0.5 | 3 | 3.5 |  |  |  | 0.674 |

TABLE IV. The electron occupation numbers $n_{5/2}$ and $n_{7/2}$ of the $j = 5/2$ and $j = 7/2$ levels for each of the three different coupling schemes, $jj$, $LS$ (Hund's rule), and intermediate coupling for the atomic ground state of the actinides.

|   | LS | | IC | | jj | |
|---|---|---|---|---|---|---|
| n | $n_{5/2}$ | $n_{7/2}$ | $n_{5/2}$ | $n_{7/2}$ | $n_{5/2}$ | $n_{7/2}$ |
| 1 | 1 | 0 | 1 | 0 | 1 | 0 |
| 2 | 1.71 | 0.29 | 1.96 | 0.04 | 2 | 0 |
| 3 | 2.29 | 0.71 | 2.79 | 0.21 | 3 | 0 |
| 4 | 2.71 | 1.29 | 3.45 | 0.55 | 4 | 0 |
| 5 | 3 | 2 | 4.23 | 0.77 | 5 | 0 |
| 6 | 3.14 | 2.86 | 5.28 | 0.72 | 6 | 0 |
| 7 | 3 | 4 | 4.10 | 2.90 | 6 | 1 |
| 8 | 3.86 | 4.14 | 5.00 | 3.00 | 6 | 2 |
| 9 | 4.57 | 4.43 | 5.57 | 3.43 | 6 | 3 |
| 10 | 5.14 | 4.86 | 5.82 | 4.18 | 6 | 4 |
| 11 | 5.57 | 5.43 | 5.89 | 5.11 | 6 | 5 |
| 12 | 5.86 | 6.14 | 5.96 | 6.04 | 6 | 6 |
| 13 | 6 | 7 | 6 | 7 | 6 | 7 |

TABLE V. Spin state character of the actinide ground state in intermediate coupling (using the calculated results in Gerken and Schmidt-May, 1983).

| 2S+1 = | 8 | 7 | 6 | 5 | 4 | 3 | 2 | 1 |
|---|---|---|---|---|---|---|---|---|
| $f^1$ | - | - | - | - | - | - | 100 | - |
| $f^2$ | - | - | - | - | - | 77.5 | - | 22.5 |
| $f^3$ | - | - | - | - | 84.1 | - | 15.9 | - |
| $f^4$ | - | - | - | 80.9 | - | 17.8 | - | 1.0 |
| $f^5$ | - | - | 67.2 | - | 26.7 | - | 3.5 | - |
| $f^6$ | - | 44.9 | - | 38.1 | - | 14.6 | - | 2.2 |
| $f^7$ | 79.8 | - | 18.1 | - | 2 | - | 0.1 | - |
| $f^8$ | - | 78.3 | - | 20.3 | - | 1.3 | - | 0.1 |
| $f^9$ | - | - | 75.4 | - | 23.2 | - | 1.2 | - |
| $f^{10}$ | - | - | - | 74.8 | - | 23.6 | - | 1.6 |
| $f^{11}$ | - | - | - | - | 89.3 | - | 10.7 | - |
| $f^{12}$ | - | - | - | - | - | 96.8 | - | 3.2 |
| $f^{13}$ | - | - | - | - | - | - | 100 | - |

TABLE VI. Relation between *LS*-coupled tensor operators $w^{xyz}$ and the standard ground-state operators (van der Laan, 1998). For the *f* shell: $\ell = 3$

|  | $w^{xyz}$ | $\ell$-shell |
|---|---|---|
| Number operator | $w^{000}$ | $n$ |
| Isotropic spin-orbit coupling | $w^{110}$ | $(\ell s)^{-1} \sum_i \mathbf{l}_i \cdot \mathbf{s}_i$ |
| Orbital moment | $w_0^{101}$ | $-\ell^{-1} \sum_i \ell_{z,i} = -\ell^{-1} L_z$ |
| Spin moment | $w_0^{011}$ | $-s^{-1} \sum_i \mathbf{s}_{z,i} = -s^{-1} S_z$ |
| Charge quadrupole moment | $w_0^{202}$ | $3[\ell(2\ell-1)]^{-1} \sum_i (\ell_z^2 - \tfrac{1}{3}\mathbf{l}^2)_i$ |
| Anisotropic spin-orbit coupling | $w_0^{112}$ | $3\ell^{-1} \sum_i (\ell_z s_z - \tfrac{1}{3}\mathbf{l}\cdot\mathbf{s})_i$ |

TABLE VII. Expectation value of $\langle w^{110}\rangle = \tfrac{2}{3}\langle \mathbf{l}\cdot\mathbf{s}\rangle$ for the *LS*-, intermediate-, and *jj*-coupled ground state. The parameters $F^k$ and $\zeta$ used to calculate the intermediate-coupling values are given in Table III. The spectroscopic notation $^{2S+1}L_J$ is for the *LS*-coupled Hund's rule ground state. In intermediate- and *jj*-coupled ground state only $J$ is a good quantum number. All values are dimensionless.

|  | LS | IC | jj |
|---|---|---|---|
| $f^1$ $^2F_{5/2}$ | -4/3 | -1.333 | -4/3 |
| $f^2$ $^3H_4$ | -2 | -2.588 | -8/3 |
| $f^3$ $^4I_{9/2}$ | -7/3 | -3.562 | -4 |
| $f^4$ $^5I_4$ | -7/3 | -4.170 | -16/3 |
| $f^5$ $^6H_{5/2}$ | -2 | -5.104 | -20/3 |
| $f^6$ $^7F_0$ | -4/3 | -6.604 | -8 |
| $f^7$ $^8S_{7/2}$ | 0 | -2.812 | -7 |
| $f^8$ $^7F_6$ | -1 | -3.865 | -6 |
| $f^9$ $^6H_{15/2}$ | -5/3 | -4.106 | -5 |
| $f^{10}$ $^5I_8$ | -2 | -3.612 | -4 |
| $f^{11}$ $^4I_{5/2}$ | -2 | -2.754 | -3 |
| $f^{12}$ $^3H_6$ | -5/3 | -1.906 | -2 |
| $f^{13}$ $^2F_{7/2}$ | -1 | -1 | -1 |

TABLE VIII. The calculated values for $G^1(c,\ell)/\zeta_c$ and the correction term $\Delta$ for a range of different absorption edges in 3d, 4d, and 5d transition metals, lanthanide, and actinides (van der Laan *et al.*, 2004). The linear relation between both quantities is evident from the numbers.

|  | $c \to \ell$ | $G^1(c,\ell)/\zeta_c$ | $\Delta$ |
|---|---|---|---|
| Ti $3d^0$ | $L_{2,3}$ ($2p \to 3d$) | 0.981 | -0.89 |
| La $4f^0$ | $M_{4,5}$ ($3d \to 4f$) | 0.557 | -0.485 |
| Th $5f^0$ | $N_{4,5}$ ($4d \to 5f$) | 0.041 | -0.020 |
| Th $5f^0$ | $M_{4,5}$ ($3d \to 5f$) | 0.021 | -0.018 |
| Zr $4d^0$ | $L_{2,3}$ ($2p \to 4d$) | 0.018 | -0.015 |
| Hf $5d^0$ | $L_{2,3}$ ($2p \to 5d$) | 0.002 | -0.002 |

**FIGURE CAPTIONS**

FIG. 1. (Color online) Wigner-Seitz radius of each metal as a function of atomic number $Z$ for the 5$d$, 4$f$, and 5$f$ metal series (from Boring and Smith, 2002). The upper-left insets schematically illustrate localized and delocalized 5$f$ states between adjacent actinide atoms (from Albers, 2003).

FIG. 2. (Color online) Atomic volume of Pu as a function of temperature, including the liquid phase. The crystal structure of all six solid allotropic crystal structures of the metal is given in the lower-right side of the figure (after Hecker, 2002). Note the structure changes from low-symmetry monoclinic to high-symmetry fcc, which occurs with exceeding large volume changes over a short temperature span.

FIG. 3. (Color online) A 'pseudo-binary' phase diagram of the light to middle 5$f$ actinide metals as a function of temperature (after Smith and Kmetko, 1983). Near Pu the melting temperature reaches a minimum, the number of phases increases to a maximum, and the crystal structures become exceedingly complex for a metal, exhibiting tetragonal, orthorhombic, and even monoclinic geometries. This is not entirely a thermodynamically valid phase diagram as some phase boundaries are based on educated guesses, however, the diagram does offer great insight into the behavior and electronic structure of the metals across the series.

FIG. 4. (Color online) A 'pseudo-binary' phase diagram of the light to middle 5$f$ actinide metals as a function of pressure (from Lindbaum *et al.*, 2003 with the phase boundaries for Cm updated using Heathman *et al.*, 2005 and Cm-Bk alloys from Heathman *et al.*, 2007). The pressure behaviors of Np and Pu are not shown, but the ground-state crystals structure of each metal is indicated.

FIG. 5. (Color online) Plot of the calculated energies as a function of the calculated bandwidth for Al, Fe, Nb, and U (from Söderlind *et al.*, 1995). Note that for all metals, regardless of bonding states, the crystal structure adopts a low-symmetry geometry when the bandwidth becomes narrow. This clearly shows that the symmetry of a crystal structure depends on the bandwidth of the bonding electron states. Thus, the narrow 5$f$ bands that are actively bonding in the light actinides are directly responsible for the low-symmetry crystal structures observed.

FIG. 6. (a) (Color online) Rearranged Periodic Table where the five transition metal series, 4$f$, 5$f$, 3$d$, 4$d$, and 5$d$, are shown (after Smith and Kmetko, 1983; Boring and Smith, 2002). When cooled to the ground state, the metals in the blue area exhibit superconductivity while the metals in the red area exhibit magnetic moments. The white band running diagonally from upper left to lower right is where conduction electrons transition from itinerant and pairing to localized and magnetic. Slight changes in temperature, pressure, or chemistry will move metals located on the white band to either more conductive or more magnetic behavior. (b) Version of (a) where the number of solid allotropic crystal structures for each metal is indicated by gray scale. Lighter shades indicate more phases. Notice that a band of lighter shades mirrors the white band in (a), showing that metals on or near the transition between magnetic and superconductive behavior exhibit numerous crystal phases.

FIG. 7. (Color online) A 'Hill plot' for a large number of U compounds (after Hill 1980). The superconducting ($T_S$) or magnetic ordering ($T_N$ or $T_C$) temperature for each compound is plotted

as a function of U-U interatomic distance. The transition from superconductivity to magnetism occurs around 3.5 Å, with only a few exceptions.

FIG. 8. (Color online) Schematic drawing showing (a) Kondo shielding of the 5$f$ moment by $s$, $p$, and $d$ conduction electrons (McCall *et al.*, 2007) and (b) electron pairing correlations (Chapline *et al.*, 2007). Either mechanism may be responsible for masking the magnetic moment in Pu that should be present due to the 5$f^5$ configuration (Moore *et al.*, 2003; 2007a; 2007b; Shim *et al.*, 2007).

FIG. 9. (Color online) Schematic diagram showing the production of a U and He atom through α decay of Pu. This self-induced irradiation slowly damages the crystal structure over time, making the understanding of the physics of the metal even more challenging.

FIG. 10. (a) Bright-field TEM image of a $CmO_2$ particle contained in a dhcp α-Cm metal matrix. (b) An [0001] electron diffraction pattern of α-Cm metal and (c) an [001] pattern of $CmO_2$. Examining the scale bar in (a) attests to the fact that recording EELS spectra and diffraction patterns from a single phase is straightforward given the ability to form a ~5 Å electron probe in the TEM. Thus, spectral investigations can be performed on highly site-specific regions, such as interfaces, dislocations, and grain boundaries.

FIG. 11. The $N_{4,5}$ ($4d \rightarrow 4f$) and $M_{4,5}$ ($3d \rightarrow 4f$) transitions for γ-Ce metal as acquired by EELS, XAS, and many-electron atomic spectral calculations. Of particular importance is the fact that the XAS and EELS from a monochromated TEM are essentially identical in both resolution and spectral shape. This means that comparison between the techniques, as well as to multi-electronic atomic calculations, is entirely justified.

FIG. 12. The experimental $O_{4,5}$ ($5d \rightarrow 5f$) EELS edges for Th, U, Np, Pu, Am, and Cm metal. In each case the ground-state α phase was examined. Electron diffraction and imaging of the Am sample in the TEM showed that it contained heavy amounts of stacking faults, which can be argued produces a combination of α and β phases as it is simply a change in the 111 plane stacking. However, spectra taken from areas with varying amounts of stacking faults showed no detectable difference in branching ratio. Thus, α- and β-Am should have very similar $N_{4,5}$ spectra and, in turn, branching ratios.

FIG. 13. The $N_{4,5}$ ($4d \rightarrow 5f$) EELS spectra for Th, U, Np, Pu, Am, and Cm metal, each normalized to the $N_5$ peak height. Immediately noticeable is the gradually growing separation between the $N_4$ and $N_5$ peaks from Th to Cm, in pace with the increase in 4$d$ spin-orbit splitting with atomic number. Second and more importantly, the $N_4$ ($4d_{5/2}$) edge gradually decreases in intensity relative to the $N_5$ ($4d_{3/2}$) edge going from Th to Am, then increases again for Cm. This behavior gives the first insight into the filling of the $j = 5/2$ and $7/2$ angular momentum level occupancy, and will be analyzed in detail using many-electron atomic calculation in the subsequent section.

FIG. 14. (Color online) Calculated actinide $O_{4,5}$ absorption spectra with (black thick line) and without (red thin line) 5$d$ core spin-orbit interaction for the ground state configurations $f^0$ to $f^9$. Atomic values of the Hartree-Fock-Slater parameters were used as tabulated in Ogasawara (1991). The relative energy refers to the zero energy of the average of the total final state

configuration. The decay channels that give rise to the broadening were not taken into account, instead all spectral lines were broadened with the same Lorentzian line shape of $\Gamma = 0.5$ eV. The pre-peak region and giant resonance are expected to be below and above ~5 eV, respectively.

FIG. 15. Calculated actinide $O_{4,5}$ absorption spectra with $5d$ core spin-orbit interaction for the ground state configurations $f^0$ to $f^9$ using a Fano line shape broadening for the giant resonance. Calculational details are the same as in Fig. 14.

FIG. 16. XAS spectra calculated using many-electron atomic theory in intermediate coupling for the $4d$ absorption edges of $^{92}$U $5f^1$ to $f^5$ and $^{100}$Fm $5f^7$ to $f^{13}$. Convolution by 2 eV, which corresponds to the intrinsic lifetime broadening (from van der Laan and Thole, 1996b).

FIG. 17. (Color online) (a) Ground-state spin-orbit interaction per hole, $<w^{110}>/(14-n) - \Delta$ as a function of the number of $5f$ electrons ($n_f$). The three theoretical angular momentum coupling schemes are shown: $LS$, $jj$, and intermediate. The points indicate the results of the spin-orbit sum rule analysis using the experimentally measured branching ratios of each metal in Fig. 13. (b) Electron occupation numbers $n_{5/2}$ and $n_{7/2}$ calculated in the three coupling schemes as a function of $n_f$. The dots indicate the experimental results: the ground-state $n_{5/2}$ and $n_{7/2}$ occupation numbers of the $5f$ shell from the spin-orbit analysis of the EELS spectra in Fig. 13.

FIG. 18. (Color online) The ground-state atomic (a) spin magnetic moment $m_s = -2\langle S_z \rangle = -2\sum_i s_{z,i}$ and (b) orbital magnetic moment $m_l = -\langle L_z \rangle = -\sum_i l_{z,i}$ (in $\mu_B$) for the actinide elements against the number of $5f$ electrons ($n_f$). The total magnetic moment is equal to $m_s + m_l$ (not shown). In each frame the three theoretical angular momentum coupling schemes are shown: $LS$, $jj$, and intermediate coupling. The spin magnetic moment becomes exceedingly large for either $jj$ or intermediate coupling at $n = 7$, meaning that if Cm exhibit either of these coupling mechanisms it will produce a large spin polarization and subsequent magnetic moment. (c) The electron occupation numbers $n_{5/2}$ (solid black line) and $n_{7/2}$ (solid red line) in intermediate coupling as a function of $n_f$. The $n_{5/2}$ and $n_{7/2}$ occupation numbers from the spin-orbit sum rule analysis of the EELS spectra are indicated by blue dots.

FIG. 19. Bremsstrahlung isochromat spectra for α-Th and α-U (from Baer and Lang, 1980). This technique gives a measure of the unoccupied states above the Fermi level.

FIG. 20. Valence-band PE spectra for α-Th (Baer and Lang, 1980), α-U (Baer and Lang, 1980), α-Np (Naegele *et al.*, 1987), α- and δ-Pu (Gouder *et al.*, 2001), and α-Am (Naegele *et al.*, 1984). The spectrum for α-Th is scaled up compared to the other spectra so that it is easily visualized. In reality, it is much lower in intensity due to a small $f$ density of states at the Fermi level.

FIG. 21. The $4f$ PE spectra for Th (Moser *et al.*, 1984), U (Moser *et al.*, 1984), Np (Naegele *et al.*, 1987), α- and δ-Pu (Arko *et al.*, 2000b), and Am (Naegele *et al.*, 1984). The satellite peaks on the high-energy side of the $4f_{5/2}$ and $4f_{7/2}$ peaks are indicative of poor screening. The satellite peaks are present in Th, but entirely absent in U and weak in Np, where there appears to be a change in slope where the satellite peak should be. α-Pu begins to show strong satellite peaks, then the spectrum of δ-Pu is dominated by them. The Am spectrum consists almost entirely of the poorly-screened satellite peak, with a very small amount of weight where the well-screened

metallic peak should be. Th, U and Np are delocalized to varying extents, so why then is the shake-down peak present in Th metal? A considerable amount of screening in the light actinides is performed by the delocalized $5f$ electrons (Johansson *et al.* 1980) and because Th has little $5f$ weight the core-electron ionization is not effectively screened. The $4f$ PE spectra of the Th-Cf oxides can be found in Veal *et al.* (1977).

FIG. 22. Valence-band PE spectra acquired from a pure Mg substrate and for coverage with Pu of increasing thickness indicated in monolayers (ML) using He II radiation (hν = 40.8 eV). The vertical bars indicate the positions of narrow features, particularly peak D, indicative of more localized $5f$ states. Inset: Pu $4f$ PE spectra for increasing Pu layer thickness in ML. The positions of the 'well- and poorly-screened' features are indicated by the full and dashed vertical lines, respectively. The background corrected fit for the 1-ML spectrum is given for the well- and poorly-screened components by the black and white peaks, respectively (Gouder *et al.*, 2001; Havela *et al.* 2002).

FIG. 23. The experimental bulk modulus as a function of actinide element for Th at 50-72 GPa (Bellussi *et al.*, 1981; Benedict, 1987; Benedict and Holzapfal 1993), Pa at 100-157 GPa (Birch, 1947; Benedict *et al.*, 1982; Benedict, 1987; Haire *et al.*, 2003), U at 100-152 GPa (Yoo *et al.*, 1998; Benedict and Dufour, 1985; Akella *et al.*, 1990), Np at 74-118 GPa (Dabos *et al.*, 1987; Benedict, 1987; Benedict and Holzapfal 1993), Pu at 40-55 (Roof, 1981; Benedict, 1987; Benedict and Holzapfal 1993), Am at 30 GPa (Heathman *et al.*, 2000), Cm at 37 GPa (Heathman *et al.*, 2005), Bk at 35 (Haire *et al.*, 1984), and Cf at 50 GPa (Peterson *et al.*, 1983). Notice the striking similarity between bulk modulus and the melting temperature in the pseudo-binary phase diagram in Fig. 3.

FIG. 24. (Color online) (a) Resistivity as a function of temperature for the α phase of Th, Pa, U, Np, Pu, and Am as a function of temperature (Müller *et al.*, 1978). (b) Electronic specific heat of Th (Fournier and Troc, 1985), Pa (Fournier and Troc, 1985), U (Lashley *et al.*, 2001a), Np (Fournier and Troc, 1985), α-Pu (Elliot *et al.*, 1964, Fournier and Troc, 1985), δ-Pu (Lashley *et al.*, 2003; 2005; Javorsky *et al.*, 2006), and Am (Müller *et al.*, 1978). The ground-state α phase for each metal is indicated with black dots, while the δ phase of Pu is indicated with red dots. There are multiple reported values for the electronic specific heat of δ-Pu, which is why there are multiple results for the metal.

FIG. 25. (Color online) Three-dimensional plot showing the phase stability of Pu metal as a function of pressure and temperature (after Liptai and Friddle, 1967 and from Hecker, 2000). Notice that the high-volume γ, δ, and δ′ phases become rapidly unstable with small amounts of pressure. As seen in Fig. 2, temperature from ambient conditions to melting cause the metal to change through six allotropic crystal structures.

FIG. 26. (Color online) DFT calculated total energies of all six solid allotropic crystal structures of Pu metal (Söderlind and Sadigh, 2004). Notice the agreement of the energy curves with the phase diagram of Pu in Fig. 2. These calculations predict substantial spin and orbital moments. Since none of the six phases of Pu exhibit any form of magnetism (Lashley *et al.*, 2005), this was an issue. However, subsequent calculations (Söderlind, 2007a) illustrate that orbital correlations (spin-orbit interaction and orbital polarization) strongly dominate over spin (exchange) correlation, which is in agreement with XAS and EELS experiments (van der Laan *et al.*, 2004;

Moore *et al.*, 2007a). As a result, spin polarization can, with good approximation, be ignored in a completely non-magnetic model.

FIG. 27. (Color online) (a) Phonon dispersions along high symmetry directions in δ-phase Pu-0.6 wt% Ga alloy. The longitudinal and transverse modes are denoted L and T respectively. The experimental data are shown as circles with error bars. Along the $[0\xi\xi]$ direction, there are two transverse branches $[011]\langle 01\bar{1}\rangle$ (T$_1$) and $[011]\langle 100\rangle$ (T$_2$). Note the softening of the TA$[\xi\xi\xi]$ branch towards the *L* point in crystal momentum space. The lattice parameter of the δ-phase is $a = 0.4621$ nm. The red solid curves are the fourth-nearest neighbor Born-von Karman model fit. The blue dashed curves are calculated dispersions for pure δ-Pu based on DMFT results of Dai *et al.*, (2003). (b) The spin-orbit interaction of α-U, α- and δ-Pu, Am I, Am IV, and Cm metal, extracted by calculating the $N_{4,5}$ absorption spectra using DMFT. Note the exceptional agreement with the experimental EELS results in Fig. 13.

FIG. 28. The relative volume of Am metal as a function of pressure (Heathman *et al.*, 2000; Linbaum *et al.*, 2001). The experiments, performed in a diamond anvil cell, show that up to 100 GPa Am metal undergoes three phase transitions between four crystal structures: Am I (dhcp), Am II (fcc), Am III (orthorhombic, space group *Fddd*), and Am IV (orthorhombic, space group *Pnma*). Inset: The superconducting transition temperature $T_c$ of Am as a function of pressure, which is given in the relative volume of the metal. $T_c$ varies from 0.8 to 2.2 K with two distinct maxima (Link *et al.*, 1994; Griveau *et al.*, 2005).

FIG. 29. (a) Atomic models of the Cm I to Cm V phases, where the structures can be viewed as being composed of close-packed hexagonal planes designated with A, B, C, and D for there spatial orientation. The dhcp structure of Cm I is (A-B-A-C) and the fcc structure of Cm II is (A-B-C). The Cm III phase can be represented by an (A-B-A) sequence where the close-packed hexagonal planes have a slight rectangular distortion. The orthorhombic Cm IV structure with space group *Fddd* can be described by the sequence (A-B-C-D) where the planes are slightly distorted. Finally, Cm V with the space group *Pnma* can be represented by quasihexagonal planes with a stacking sequence (A-B-A). The planes in Cm III, IV, and V all have distorted planes, which reduces their symmetry from hexagonal to orthorhombic (Cm IV and V) or monoclinic (Cm III). (b) The relative volume ($V/V_0$) as a function of pressure for α-U (Le Bihan *et al.*, 2003), Am (Heathman *et al.*, 2000; Linbaum *et al.*, 2001) and Cm (Heathman *et al.*, 2005). For each metal the vertical lines designate the pressure range for each phase of Am and Cm. The percent value between each phase indicates the size of the collapses in atomic volume. (Inset) Calculated ab-initio total energy difference between Cm II, Cm III, Cm IV, and Cm V structures as a function of volume. The energy of the Cm II phase is taken as a reference level, shown as a horizontal line at zero. The vertical dashed lines indicate the crossover points for each phase. Figure reproduced from Heathman *et al.* (2005).

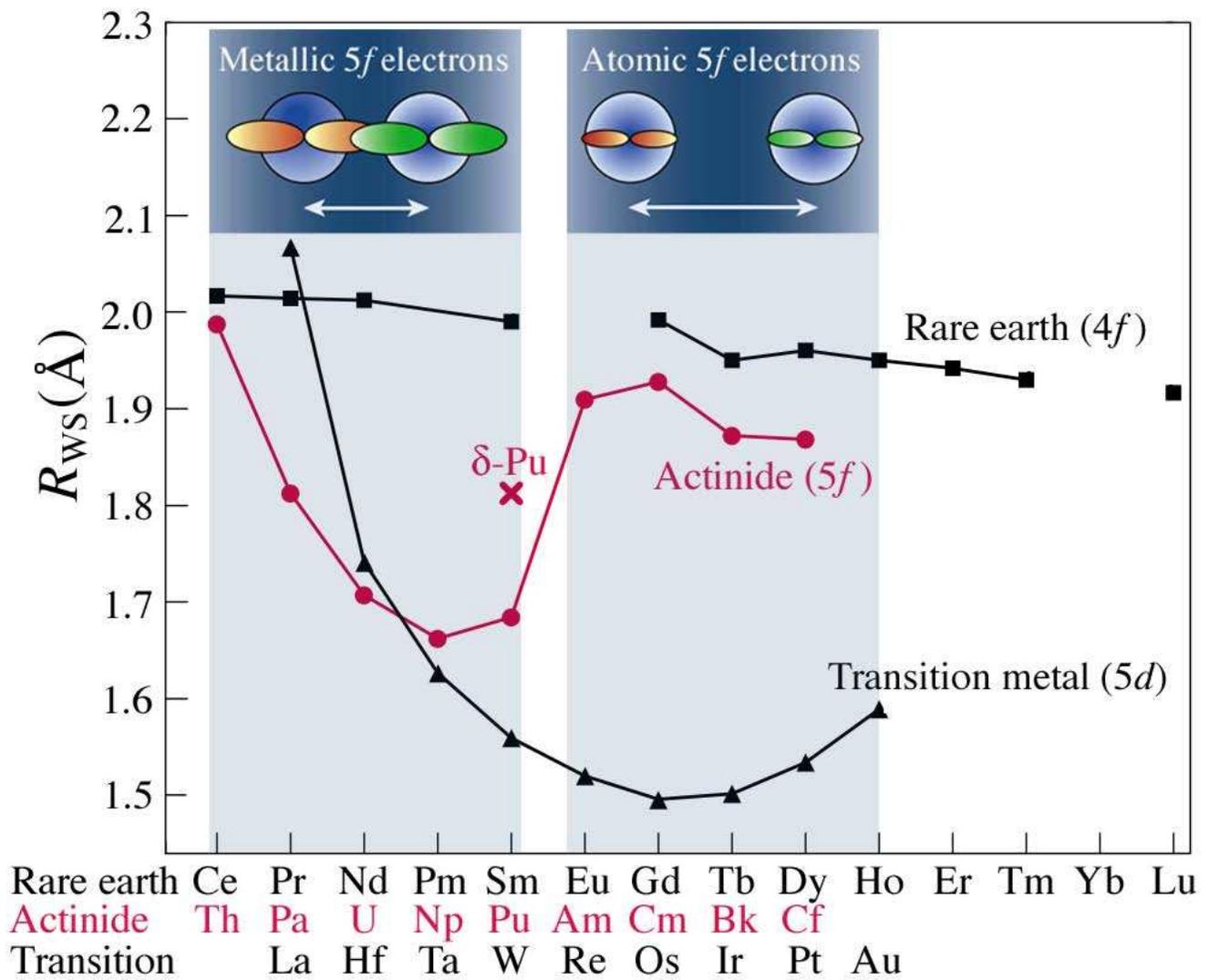

Figure 1   RG10015   06MAY2008

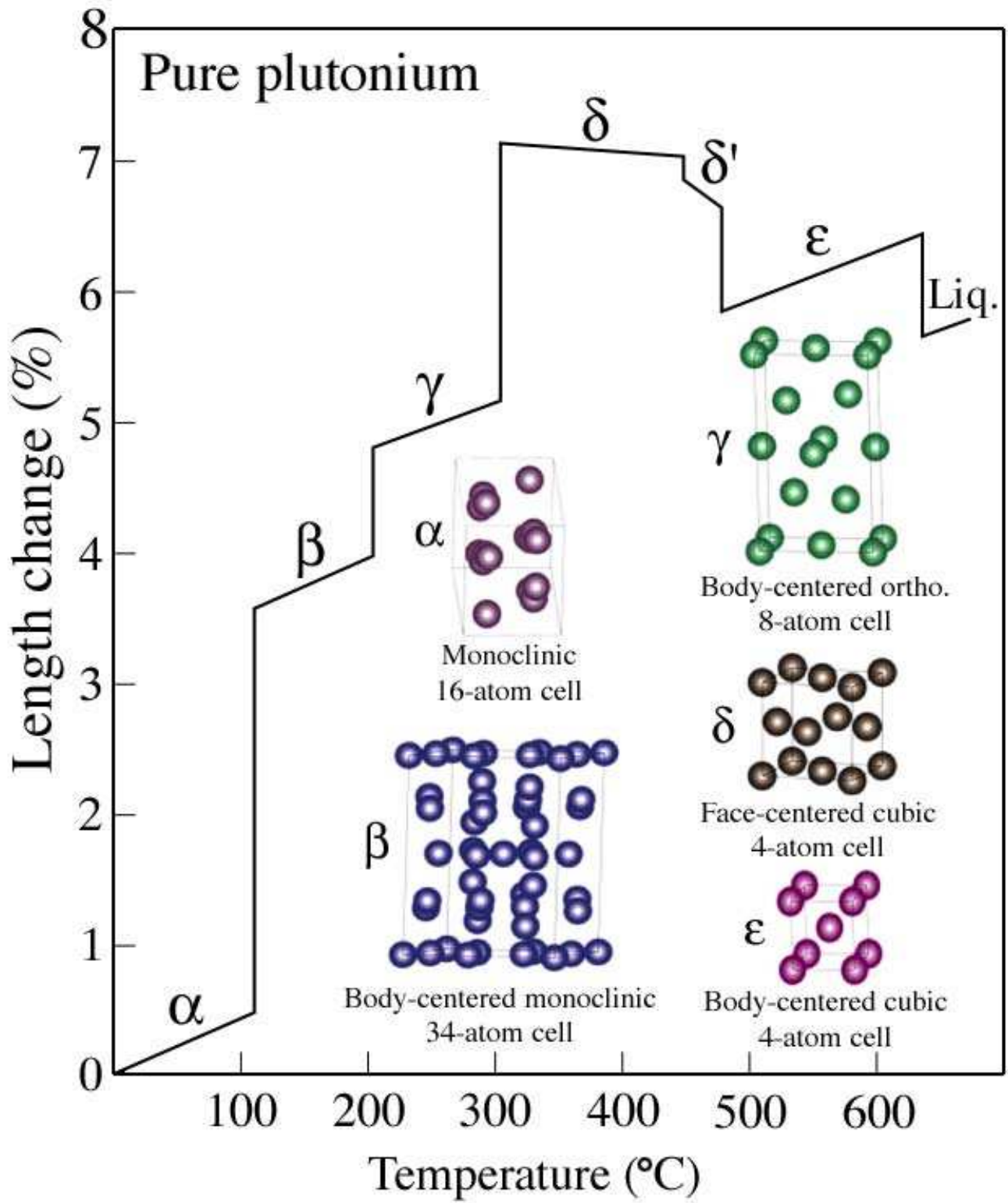

Figure 2     RG10015    06MAY2008

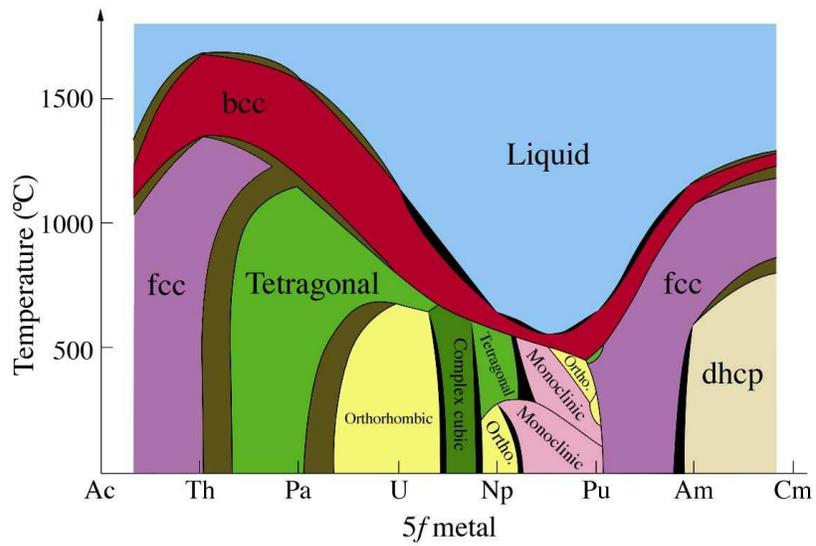

Figure 3    RG10015    06MAY2008

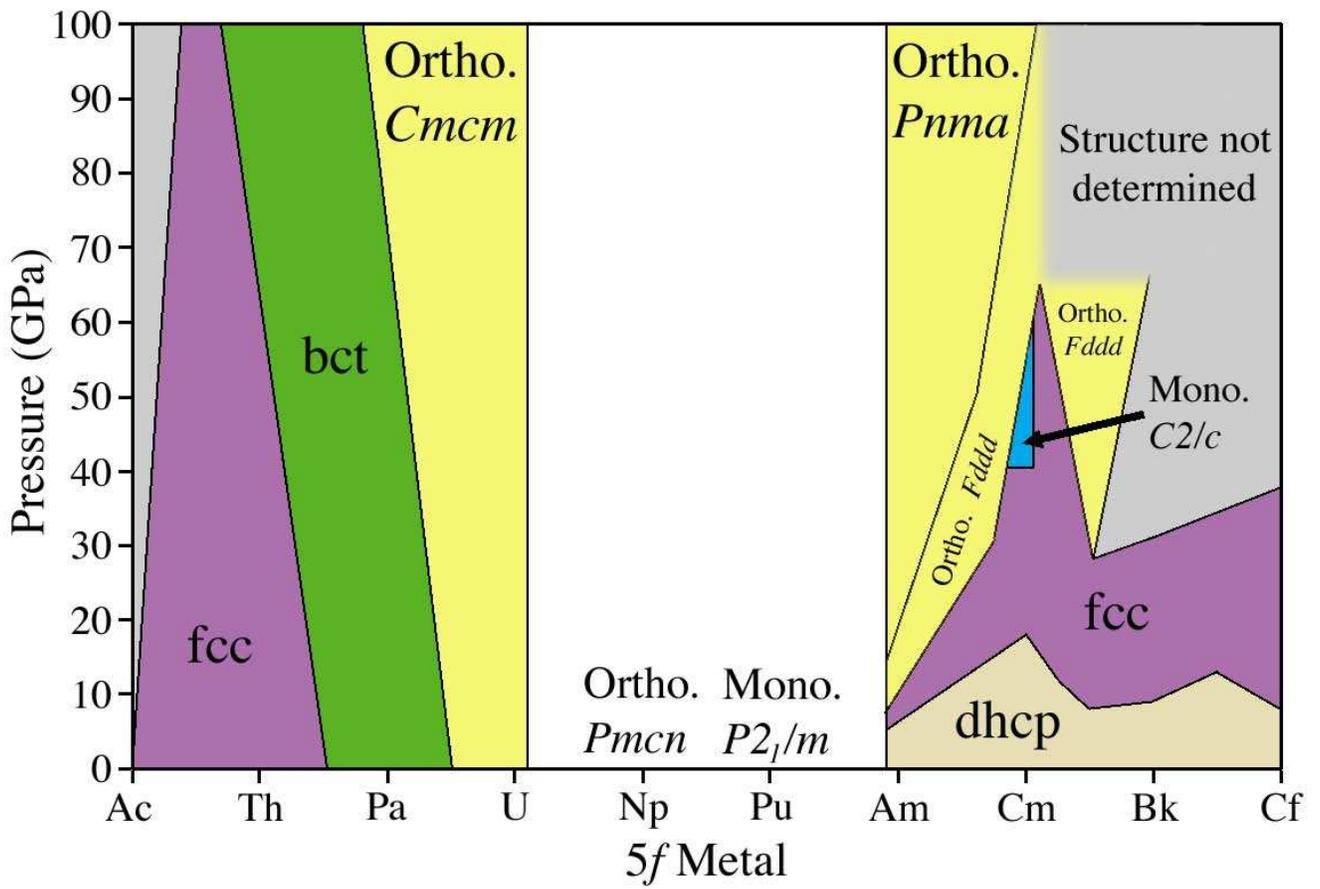

Figure 4   RG10015   06MAY2008

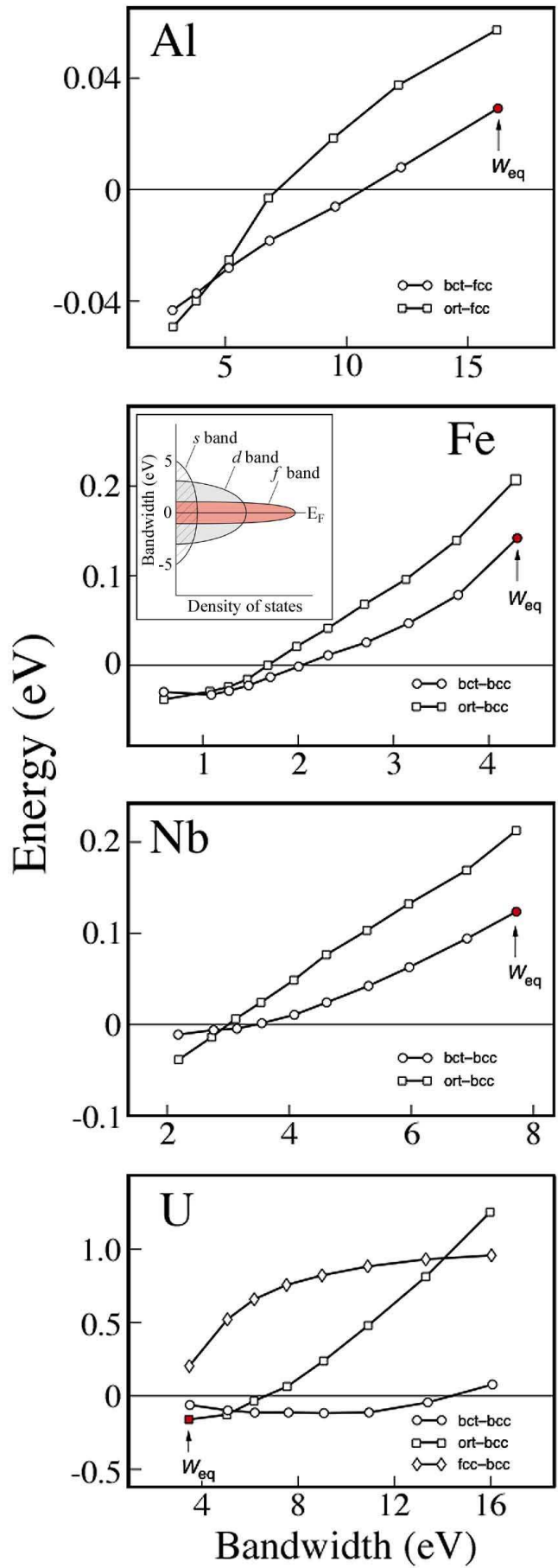

Figure 5 RG10015 06MAY2008

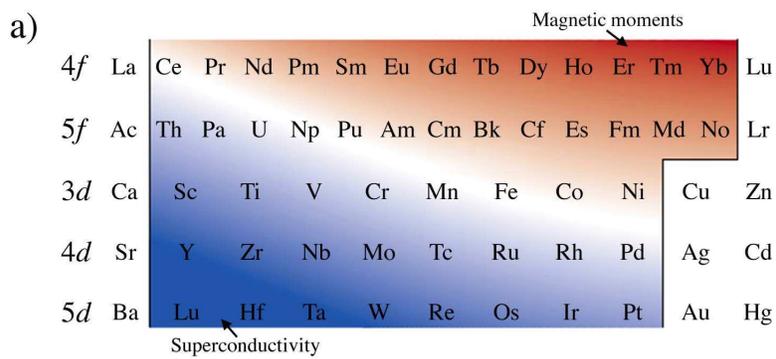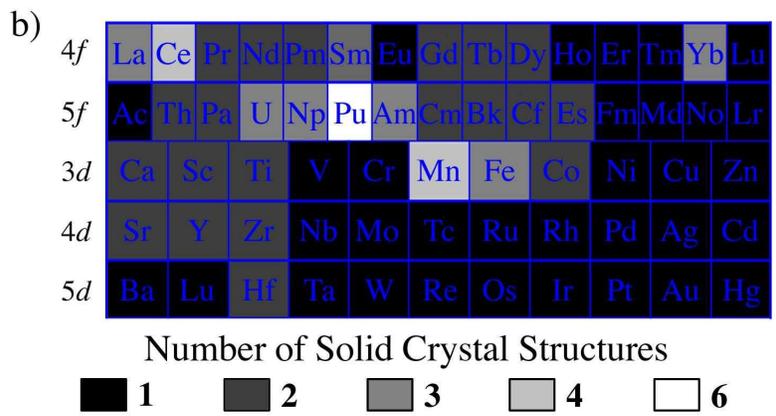

Figure 6 RG10015 06MAY2008

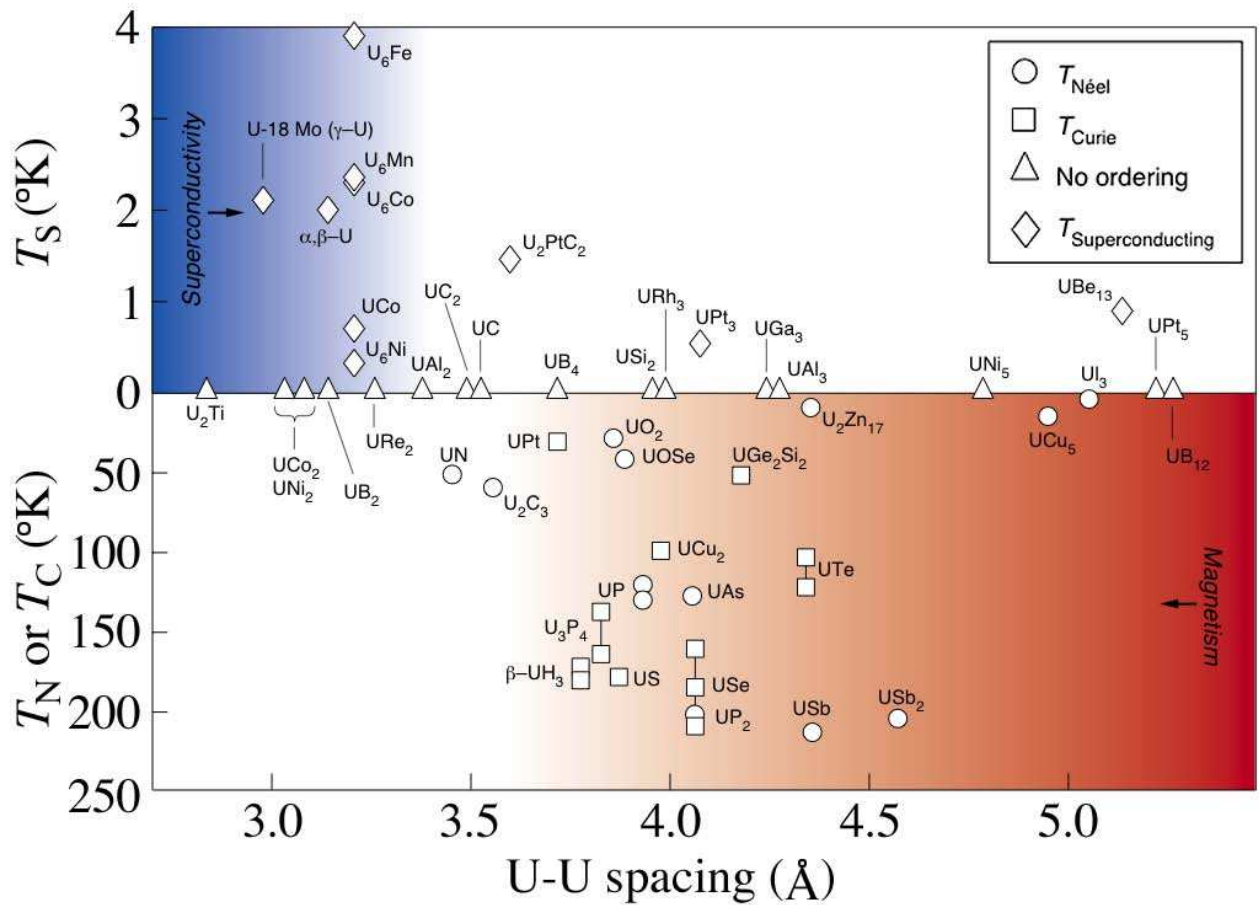

Figure 7    RG10015    06MAY2008

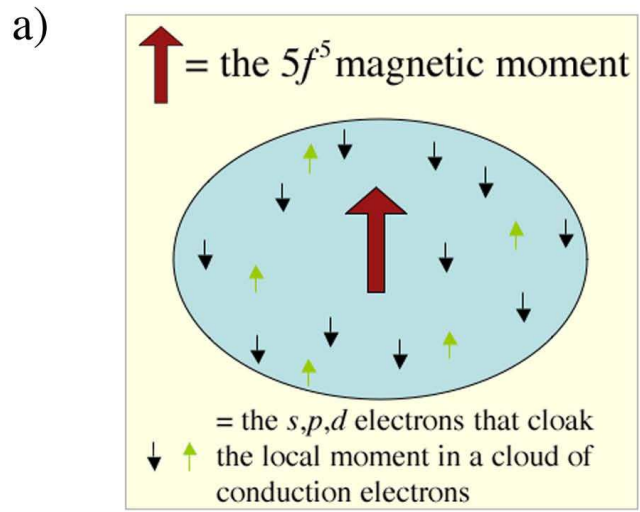

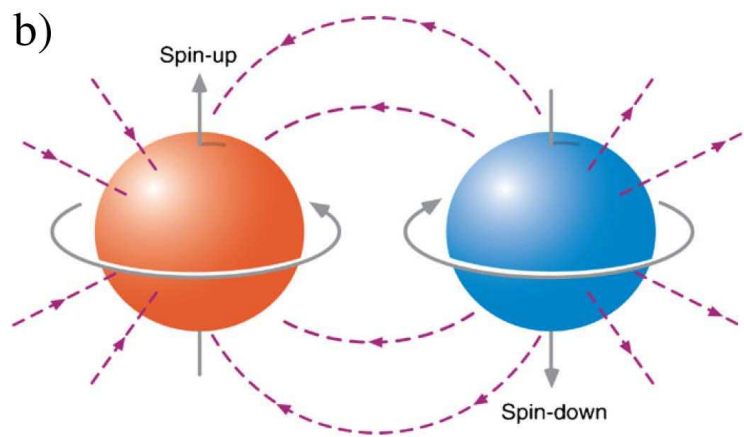

Figure 8 RG10015 06MAY2008

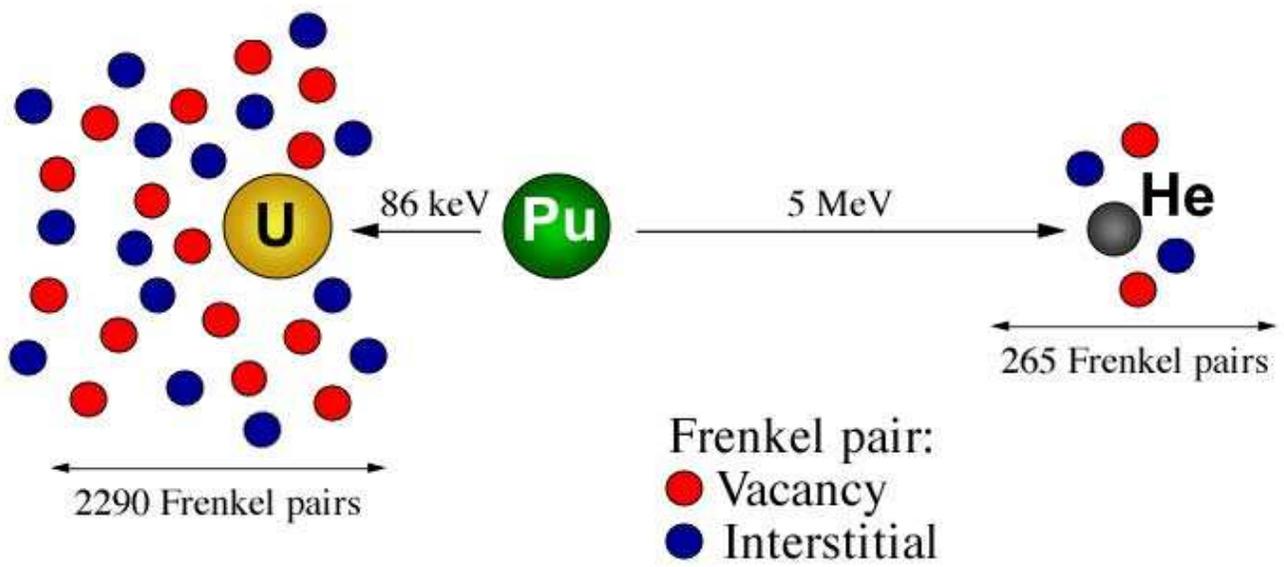

Figure 9    RG10015    06MAY2008

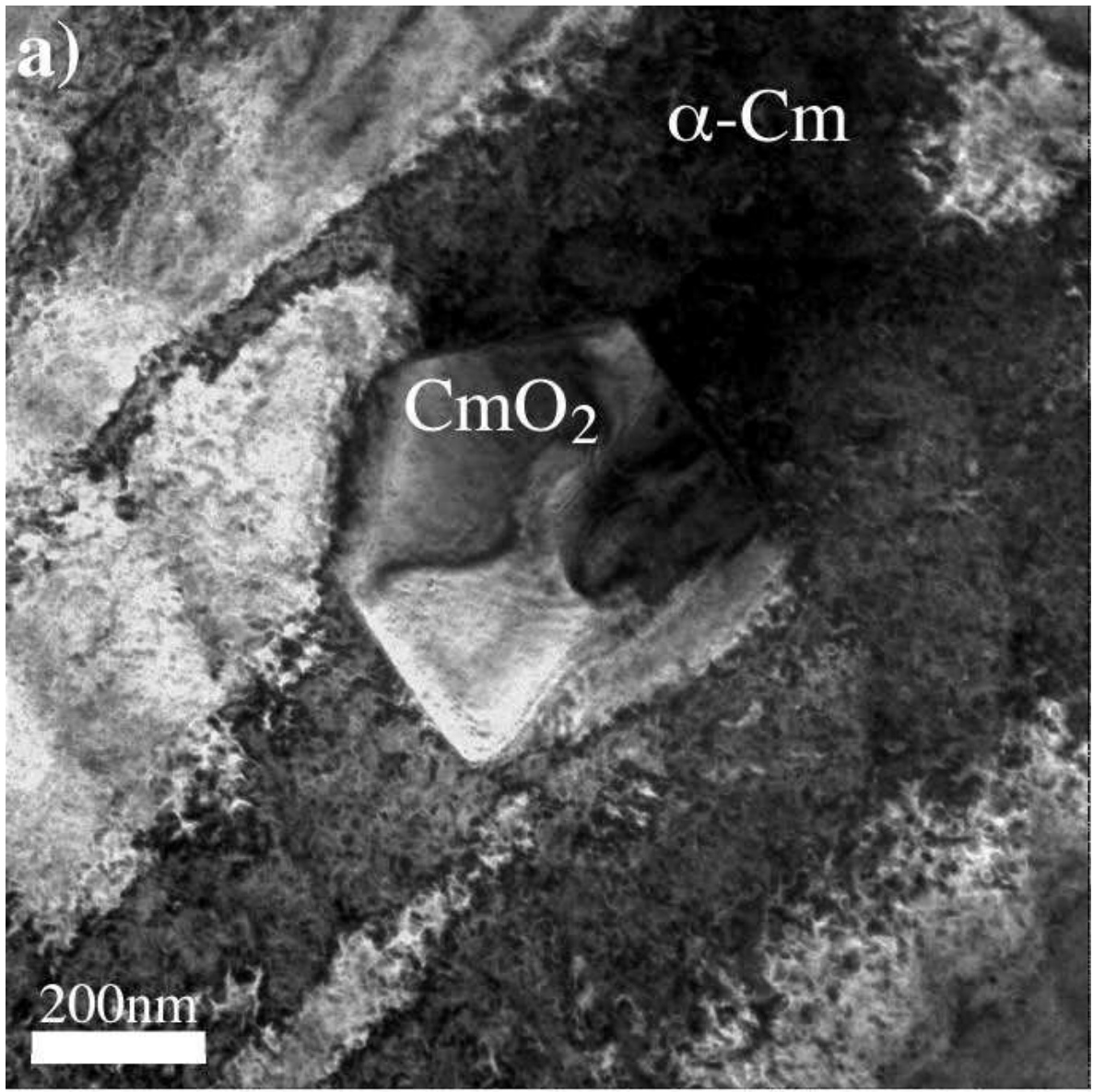
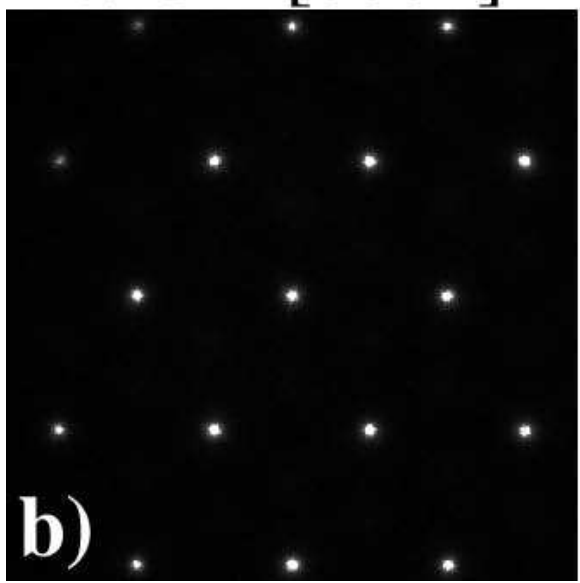
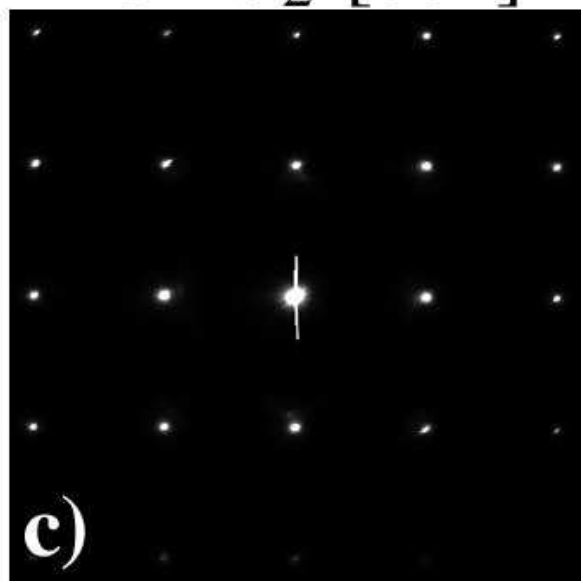

Figure 10    RG10015    06MAY2008

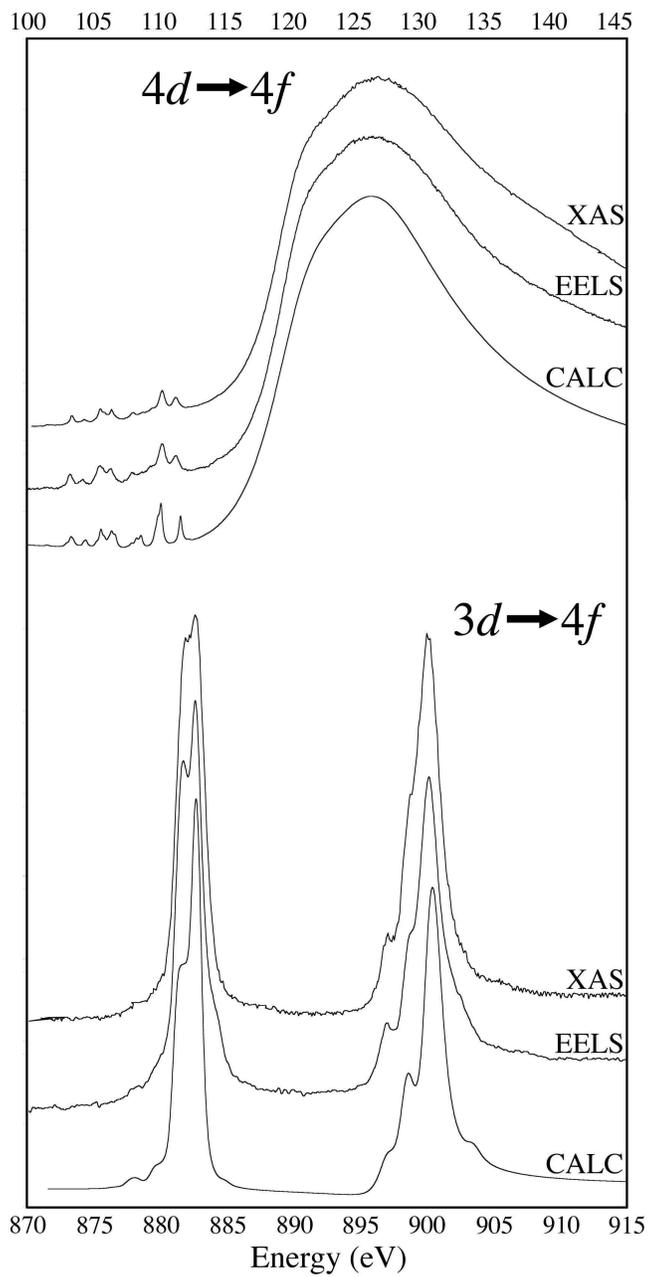

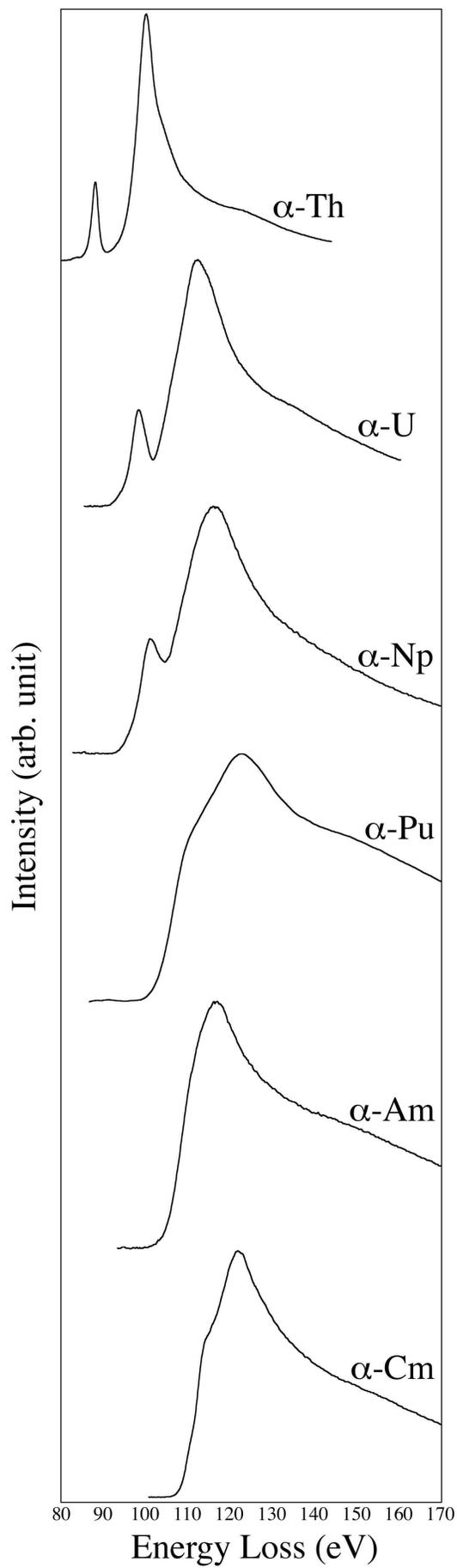

Figure 12      RG10015    06MAY2008

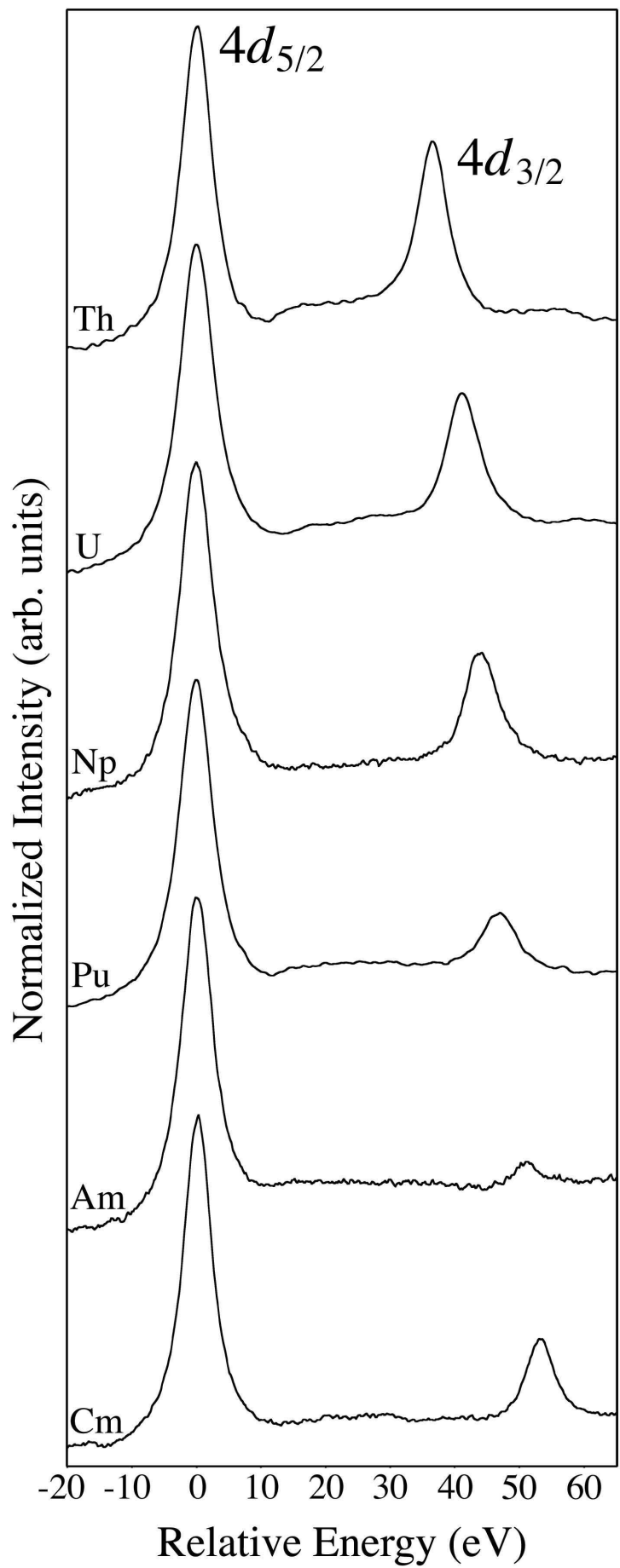

Figure 13       RG10015    06MAY2008

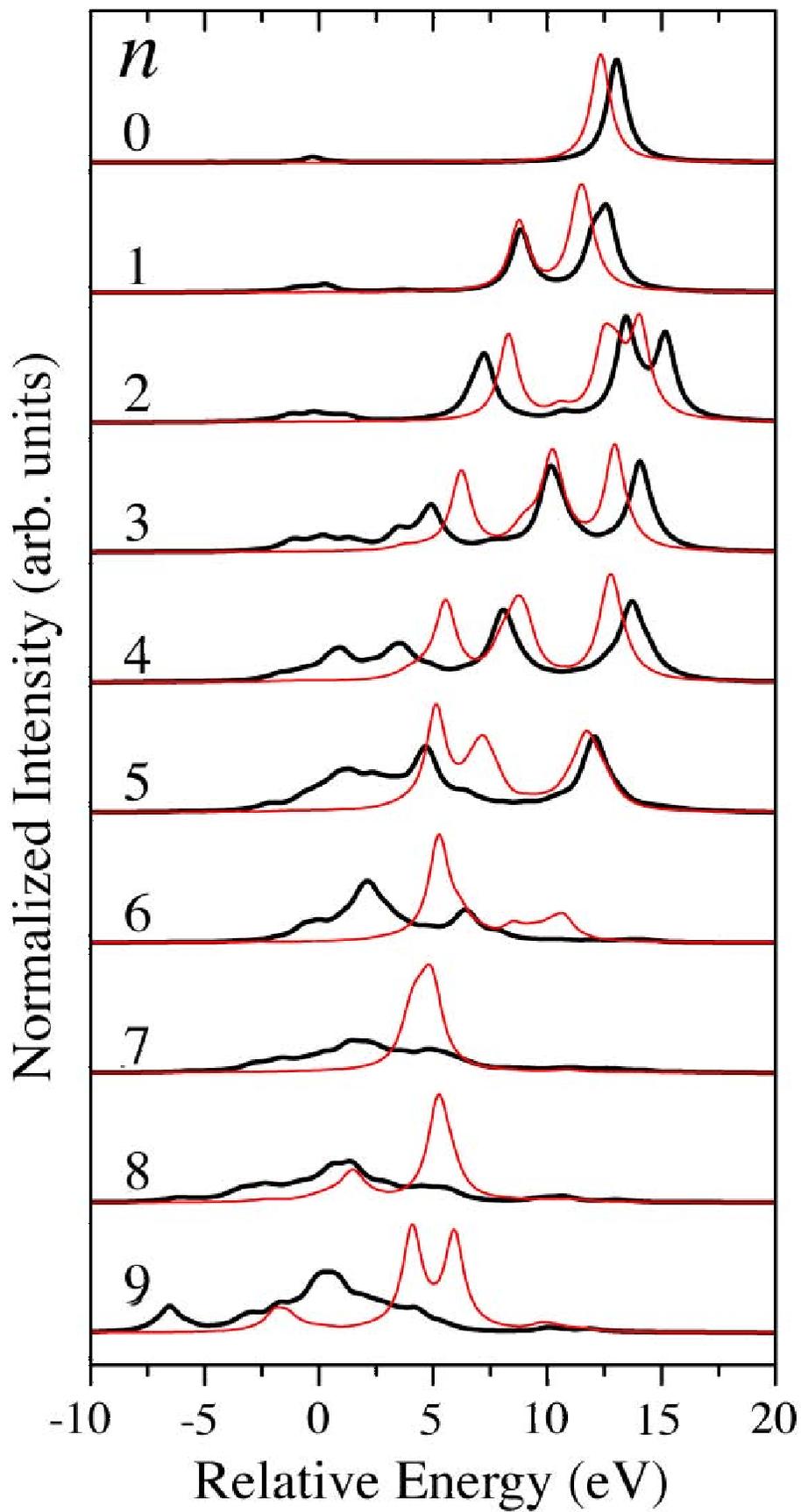

Figure 14   RG10015   06MAY2008

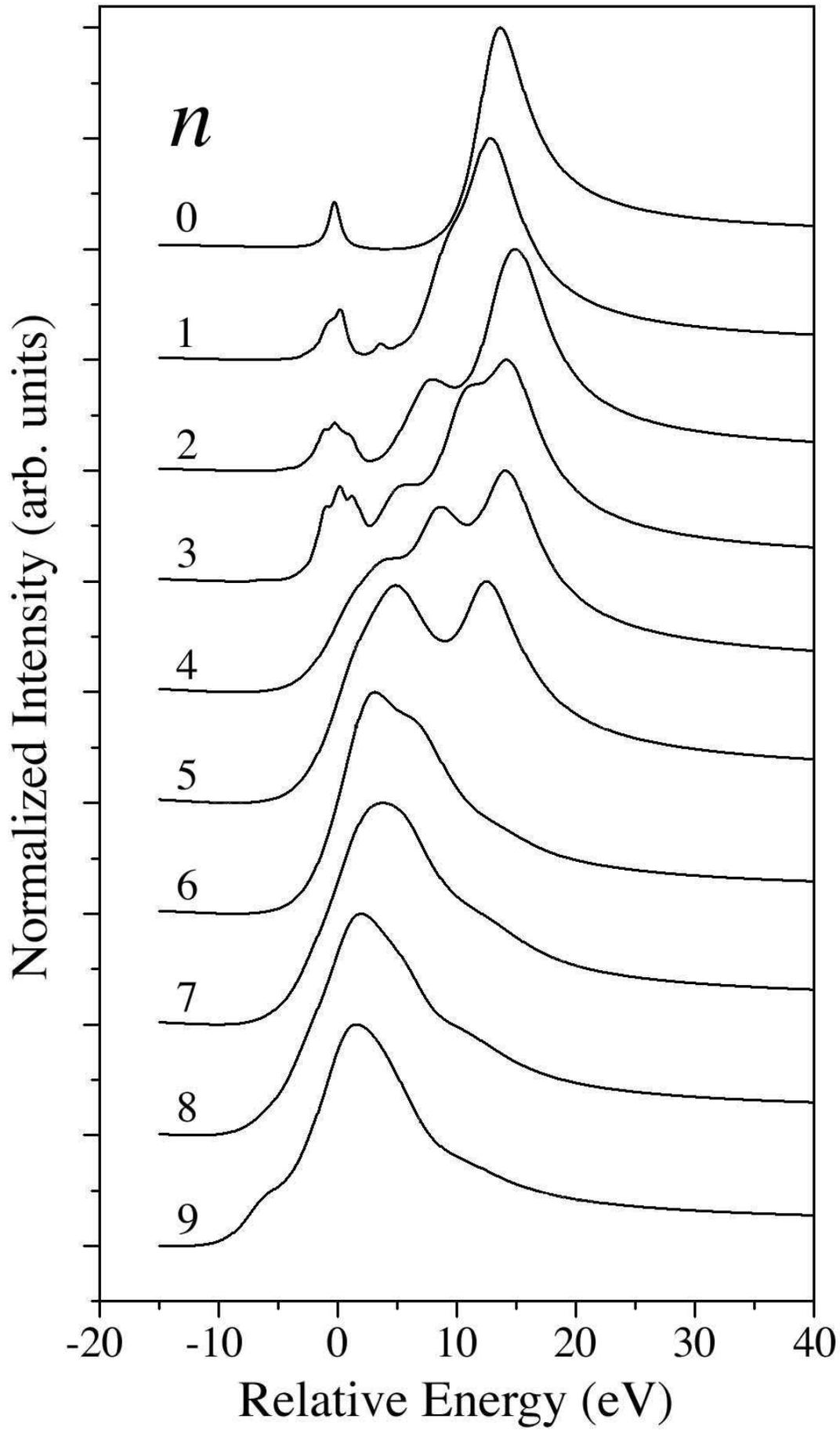

Figure 15   RG10015   06MAY2008

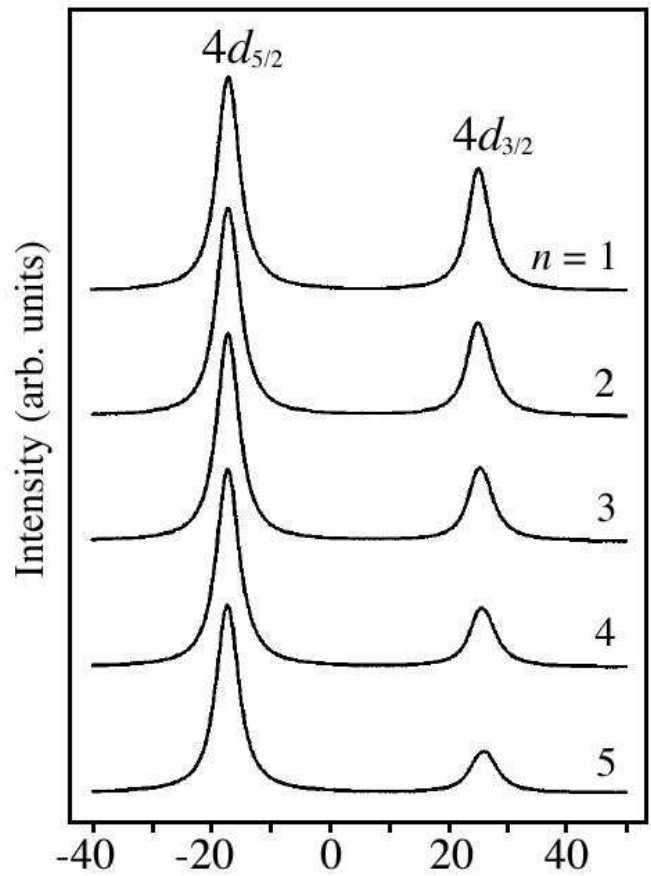
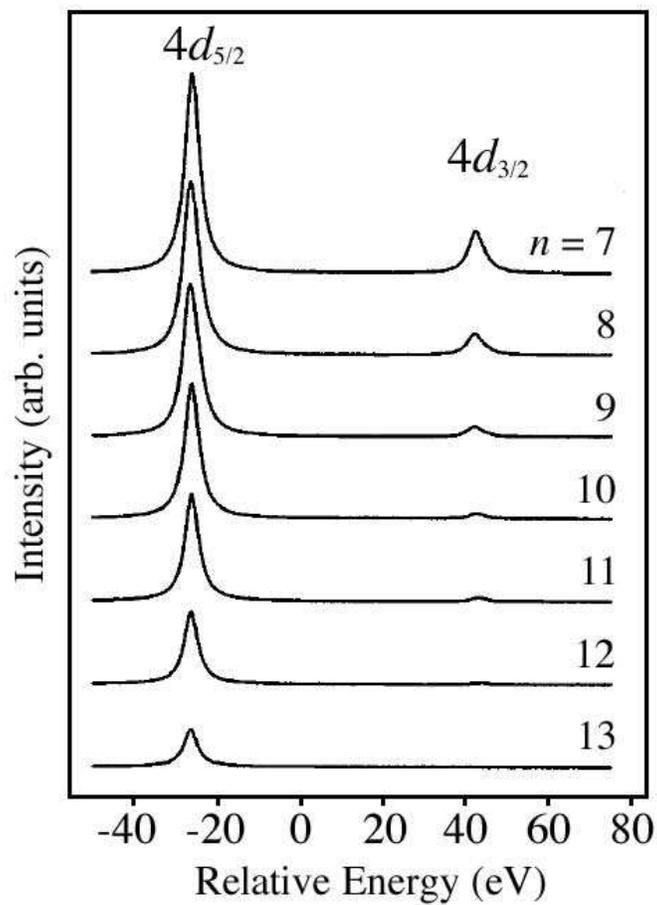

Figure 16    RG10015    06MAY2008

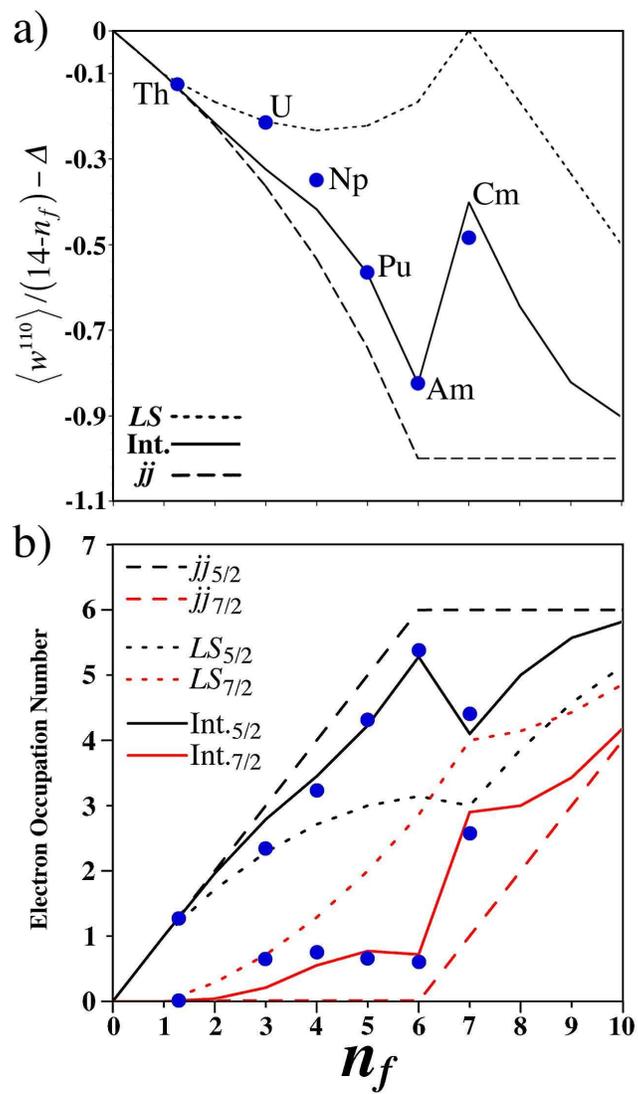

Figure 17     RG10015     06MAY2008

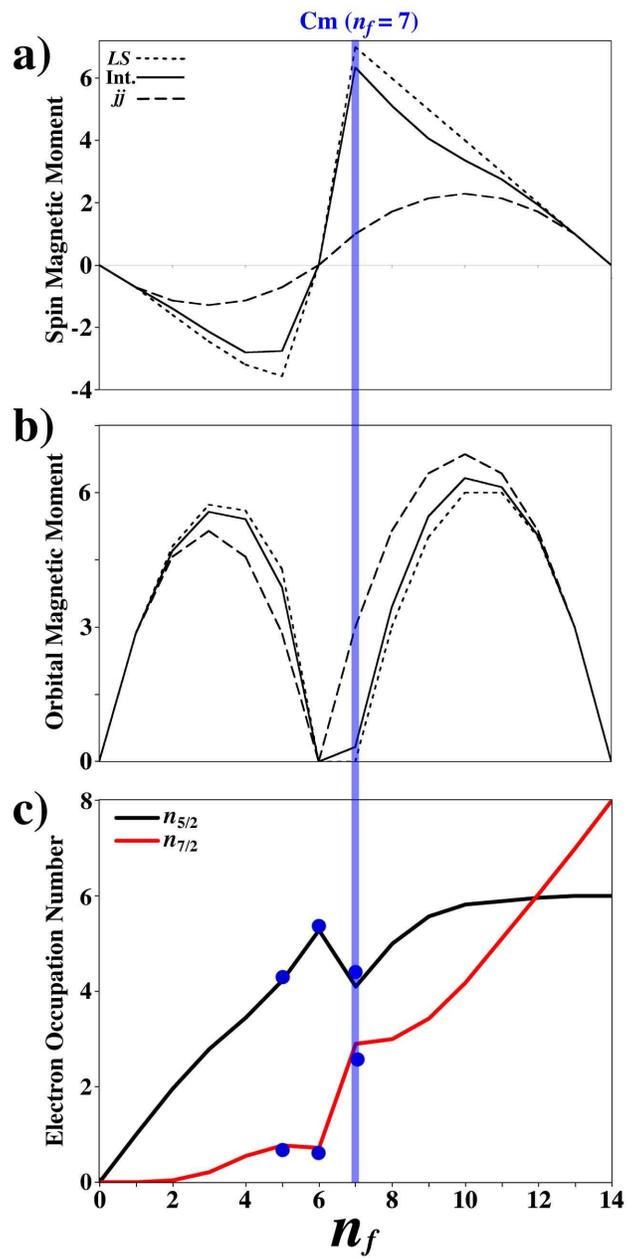

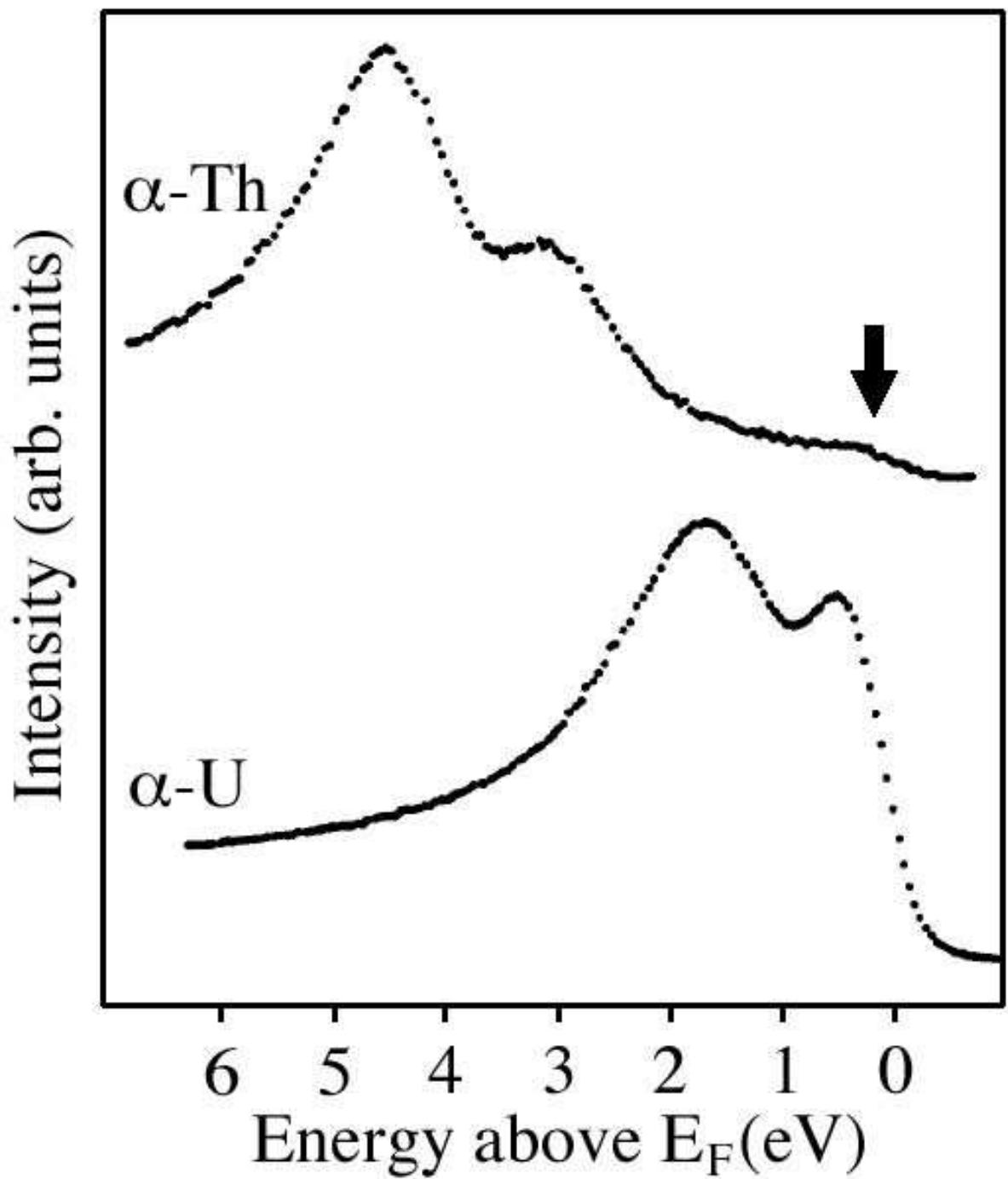

Figure 19     RG10015     06MAY2008

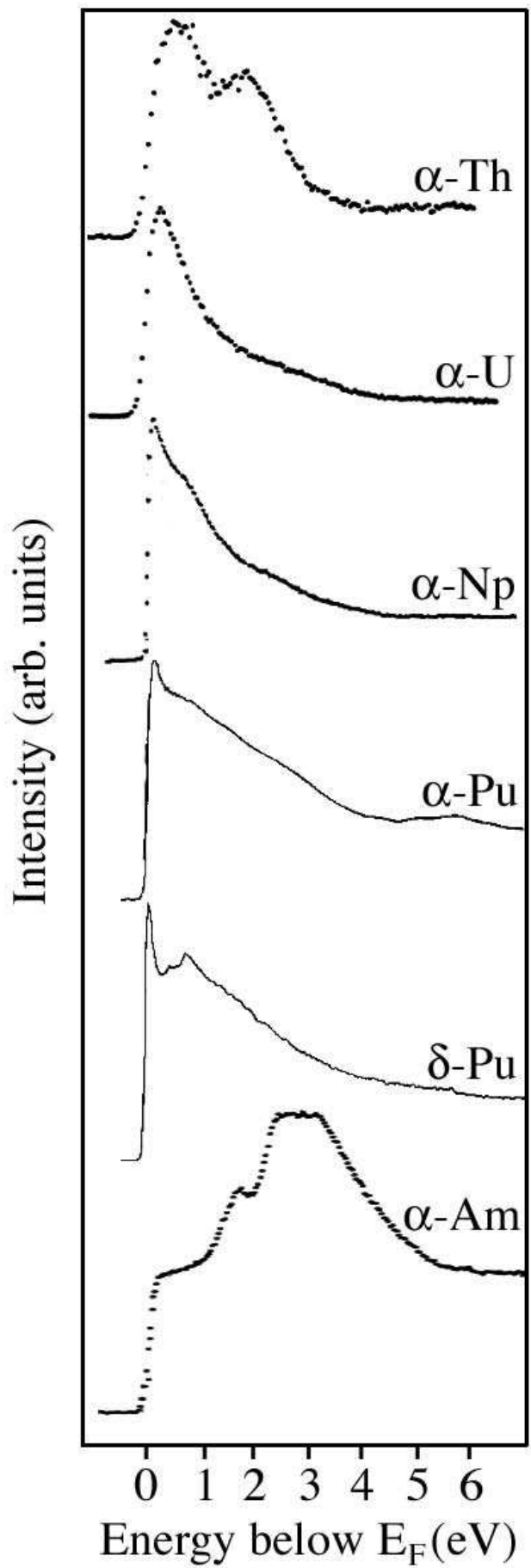

Figure 20    RG10015    06MAY2008

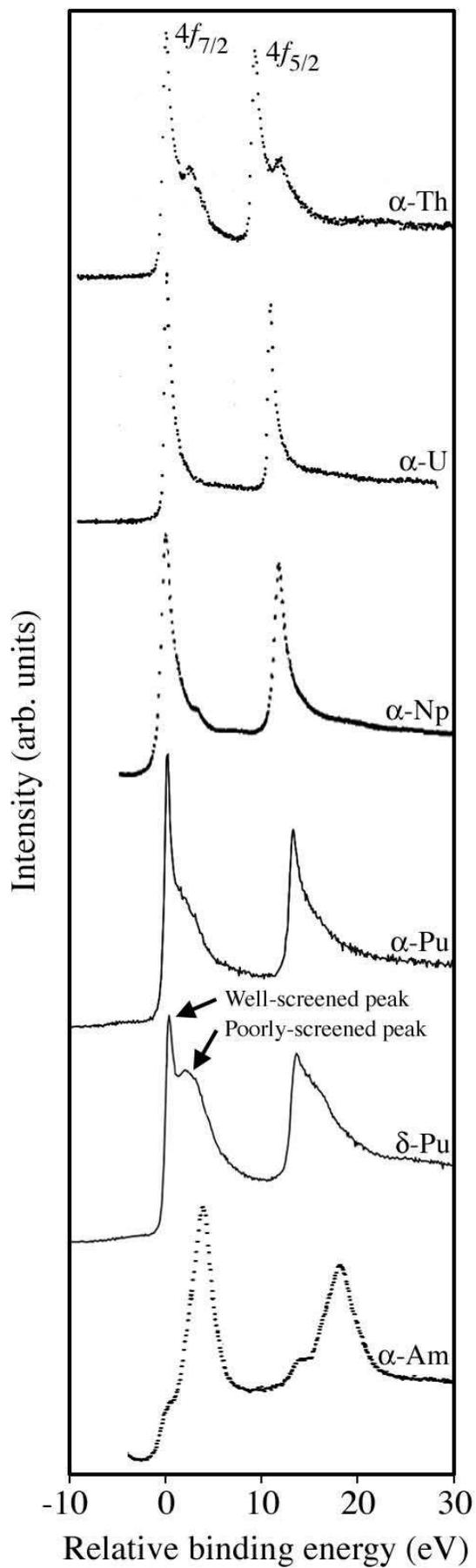

Figure 21     RG10015     06MAY2008

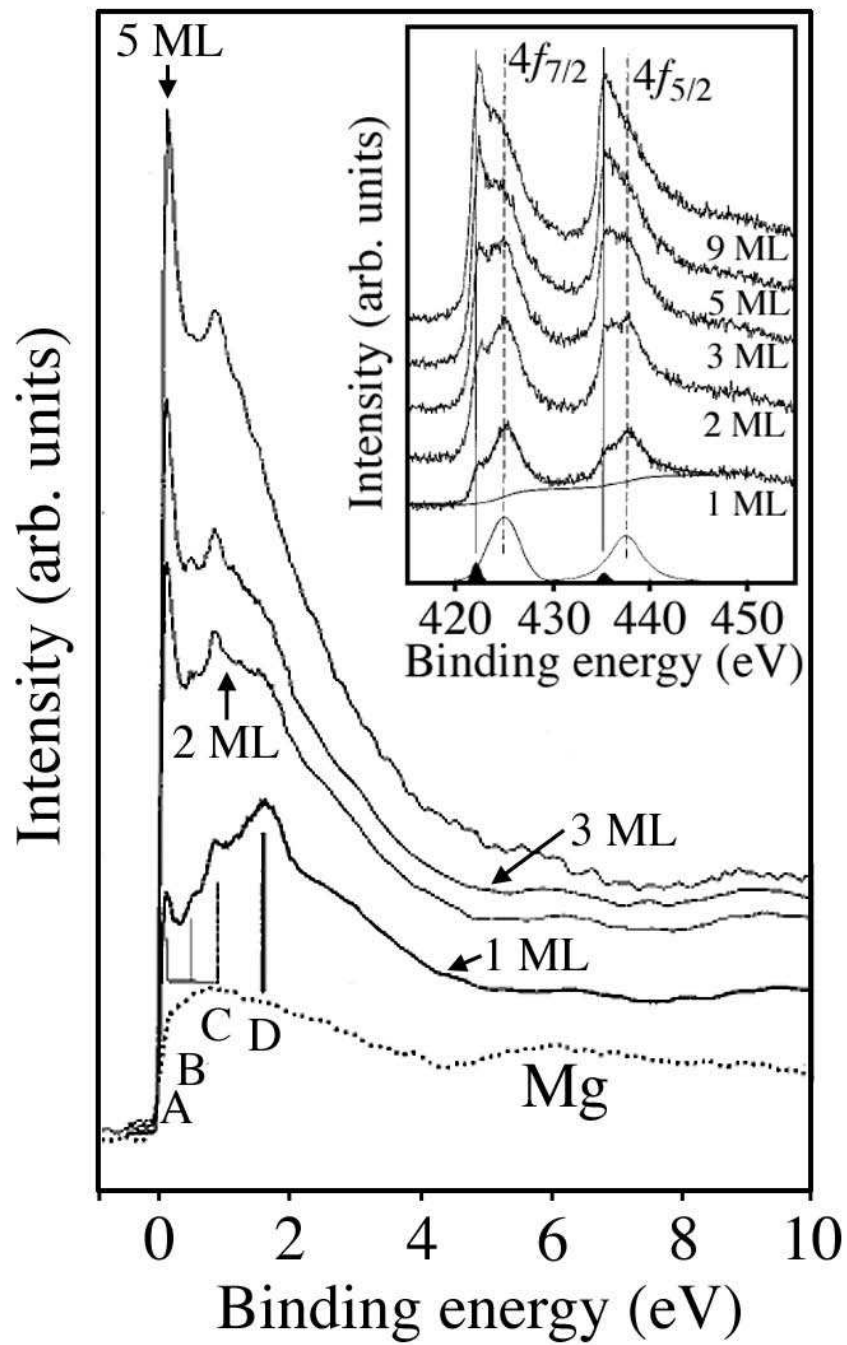

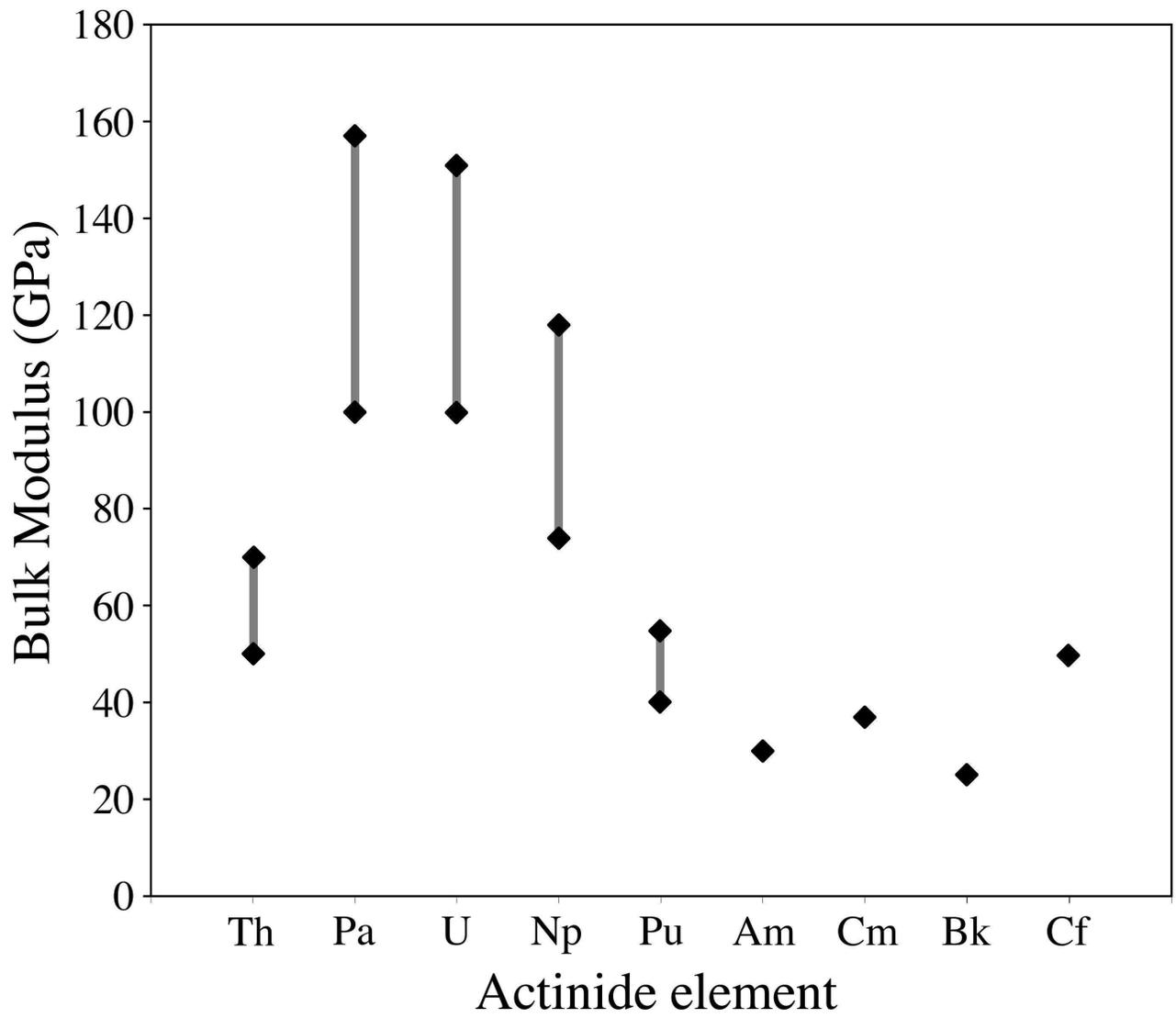

Figure 23      RG10015     06MAY2008

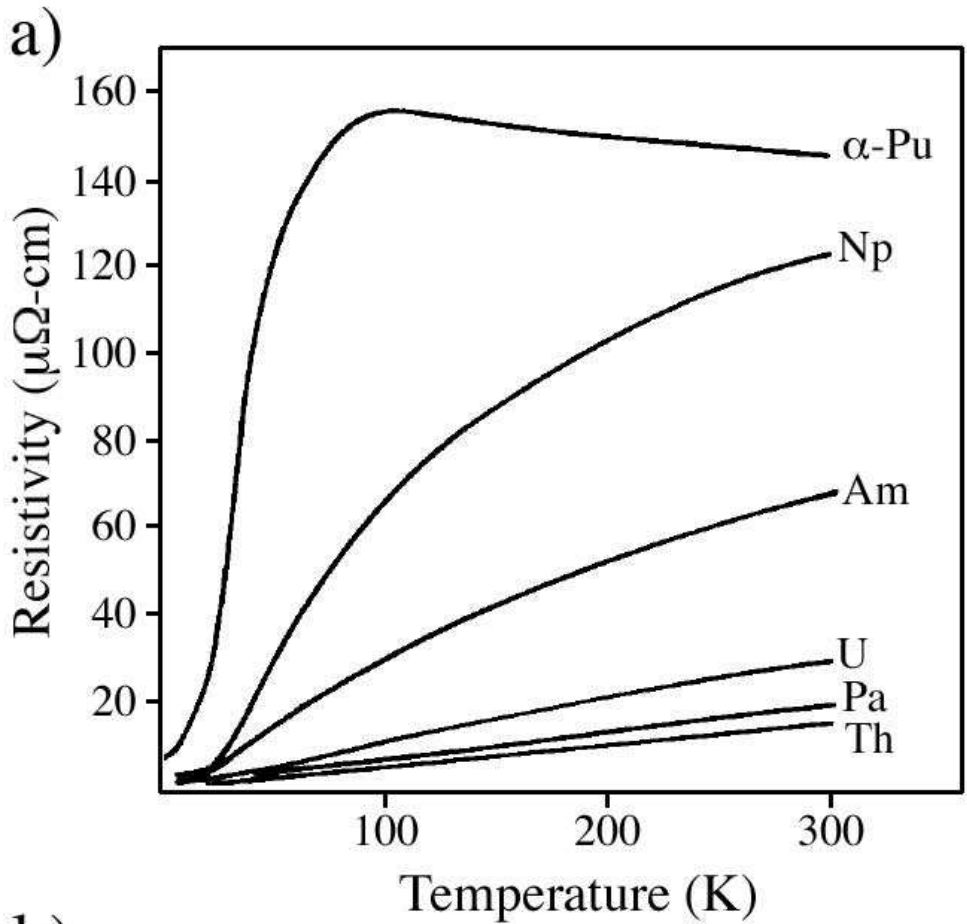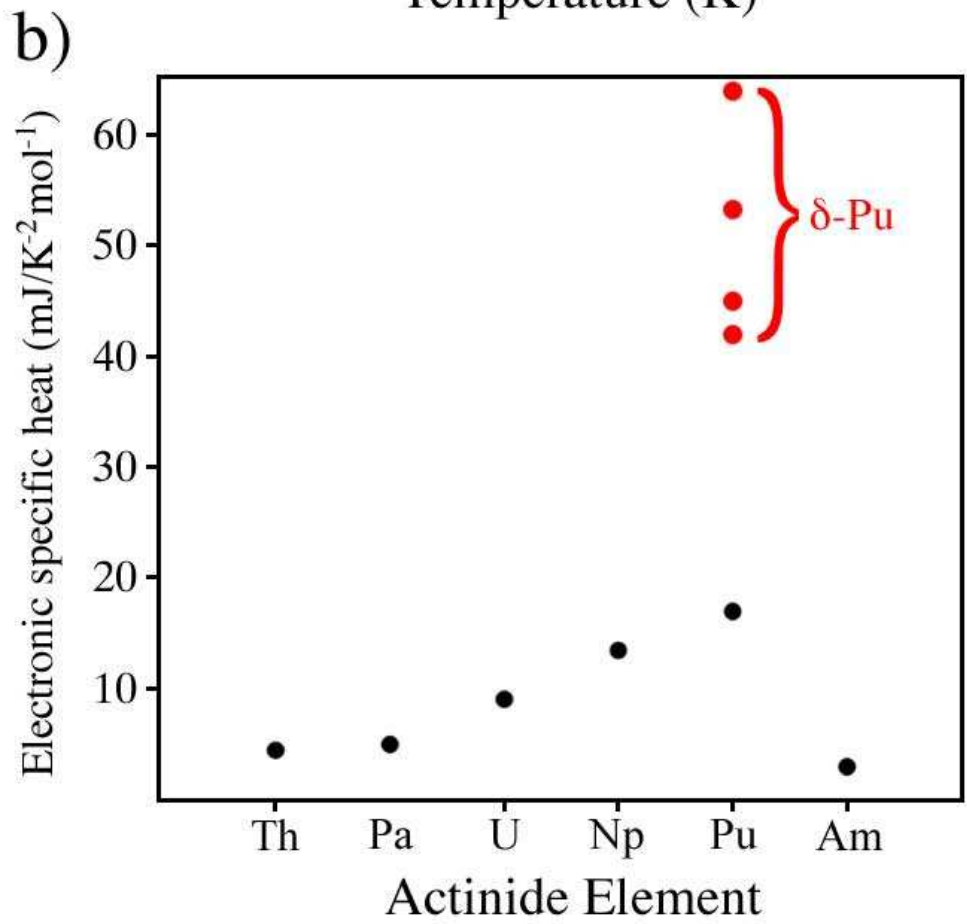

Figure 24 RG10015 06MAY2008

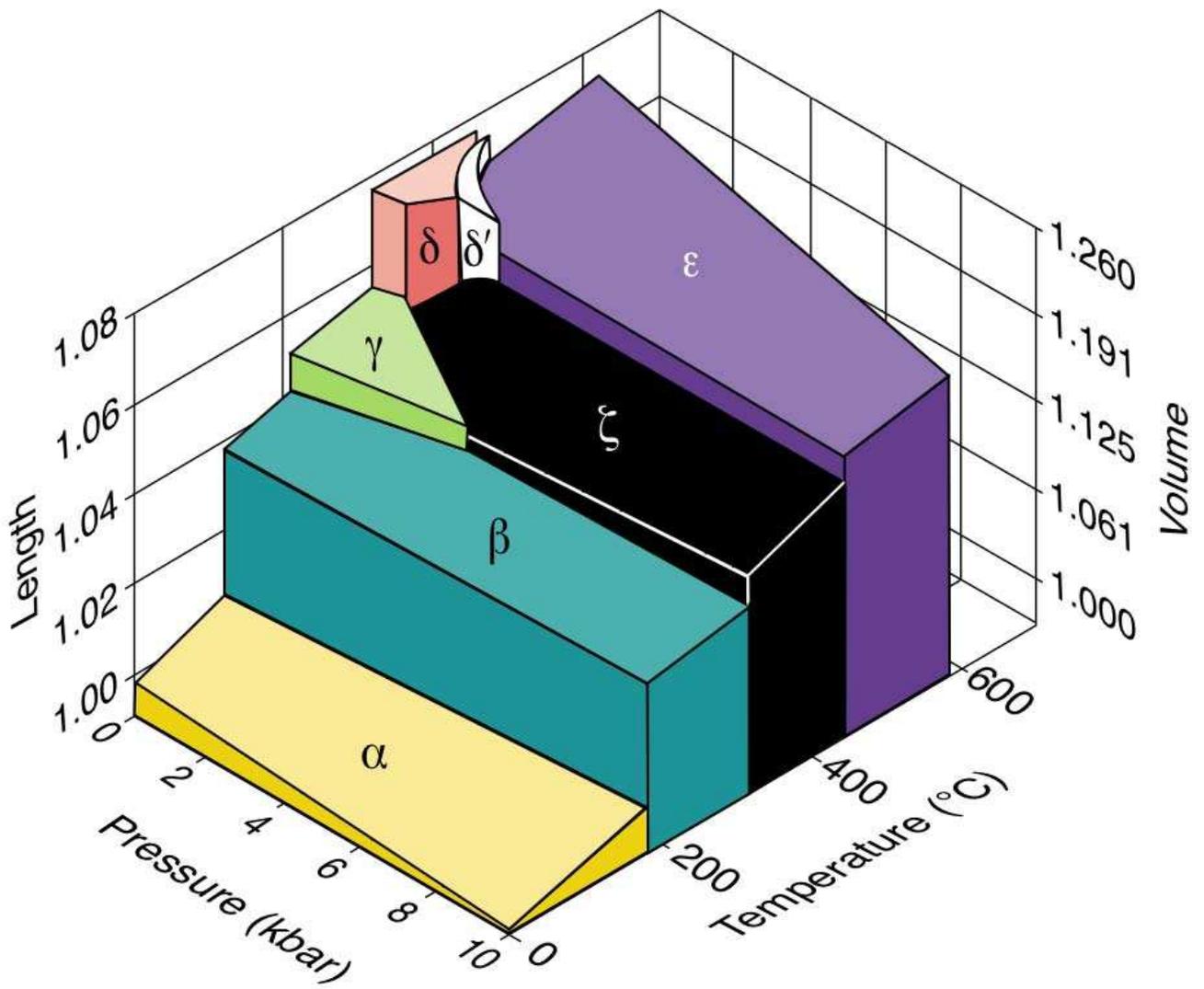

Figure 25      RG10015    06MAY2008

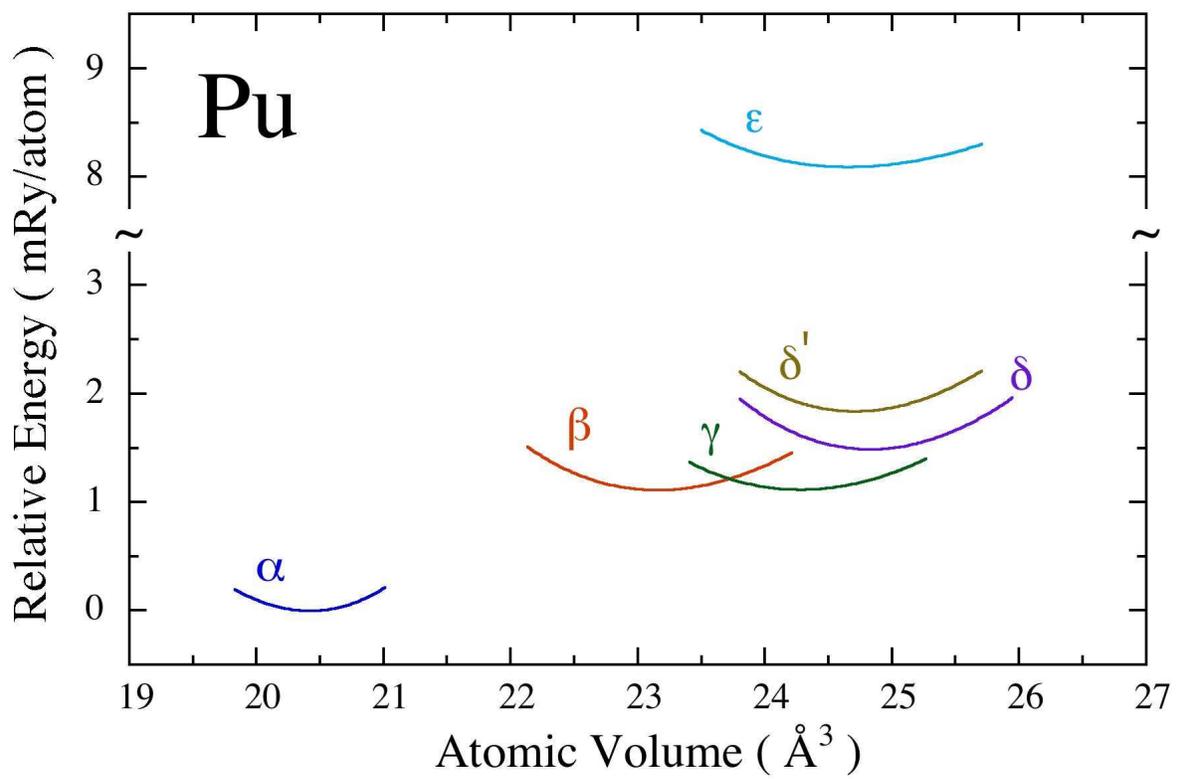

Figure 26   RG10015   06MAY2008

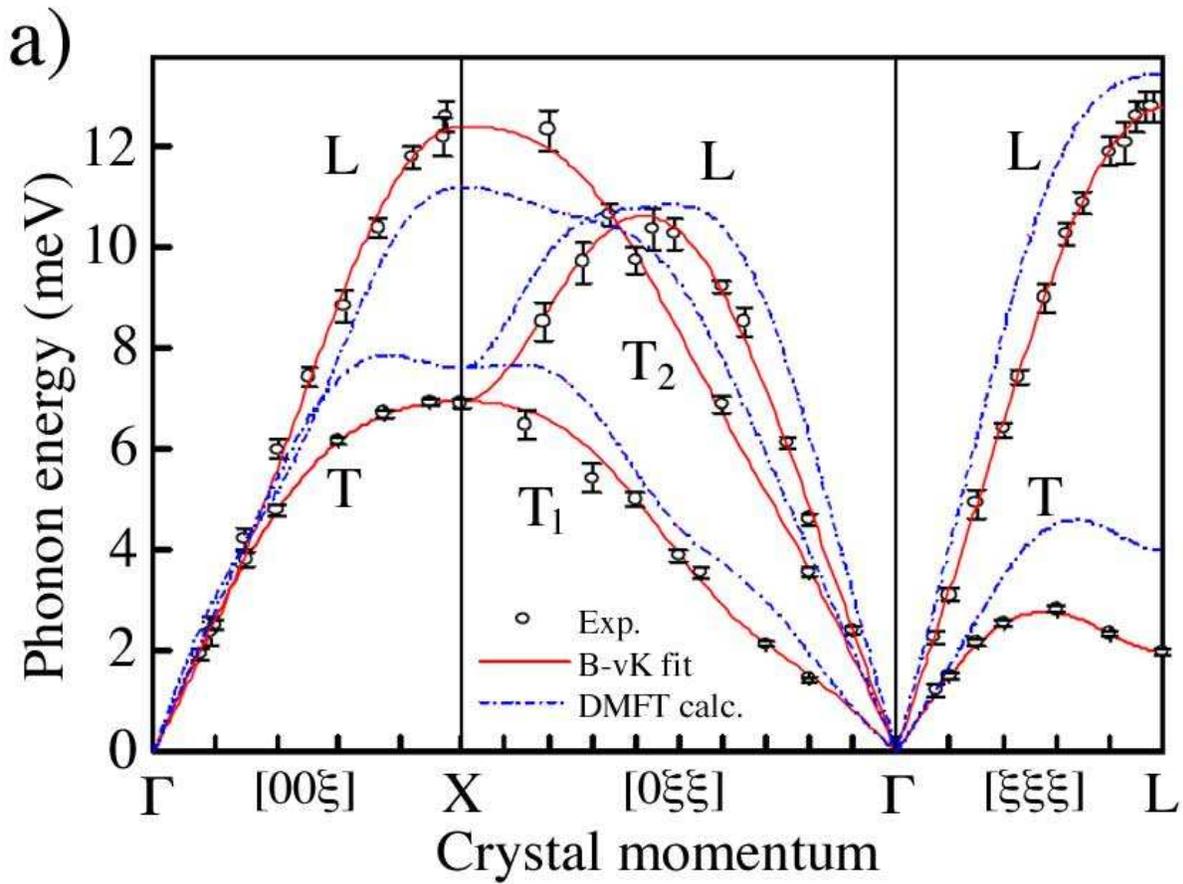

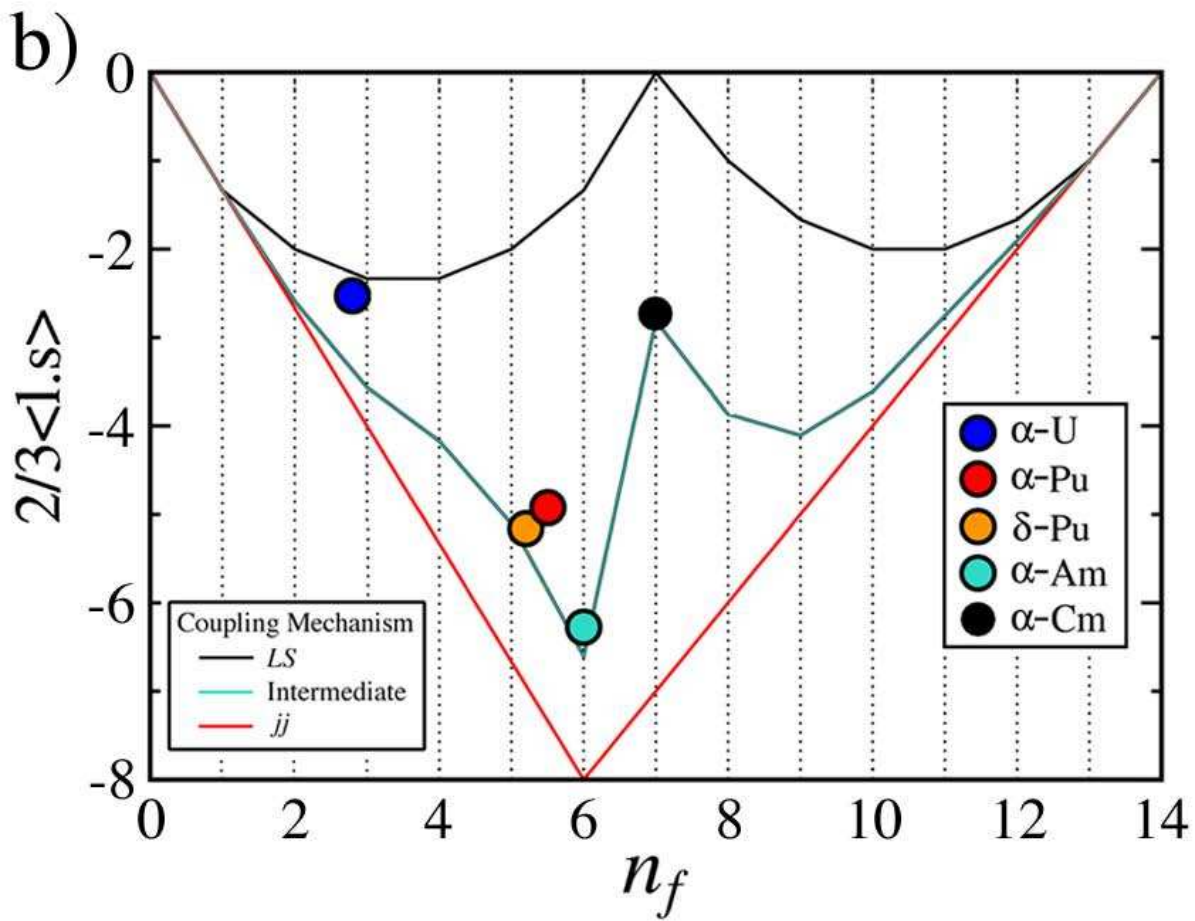

Figure 27   RG10015   06MAY2008

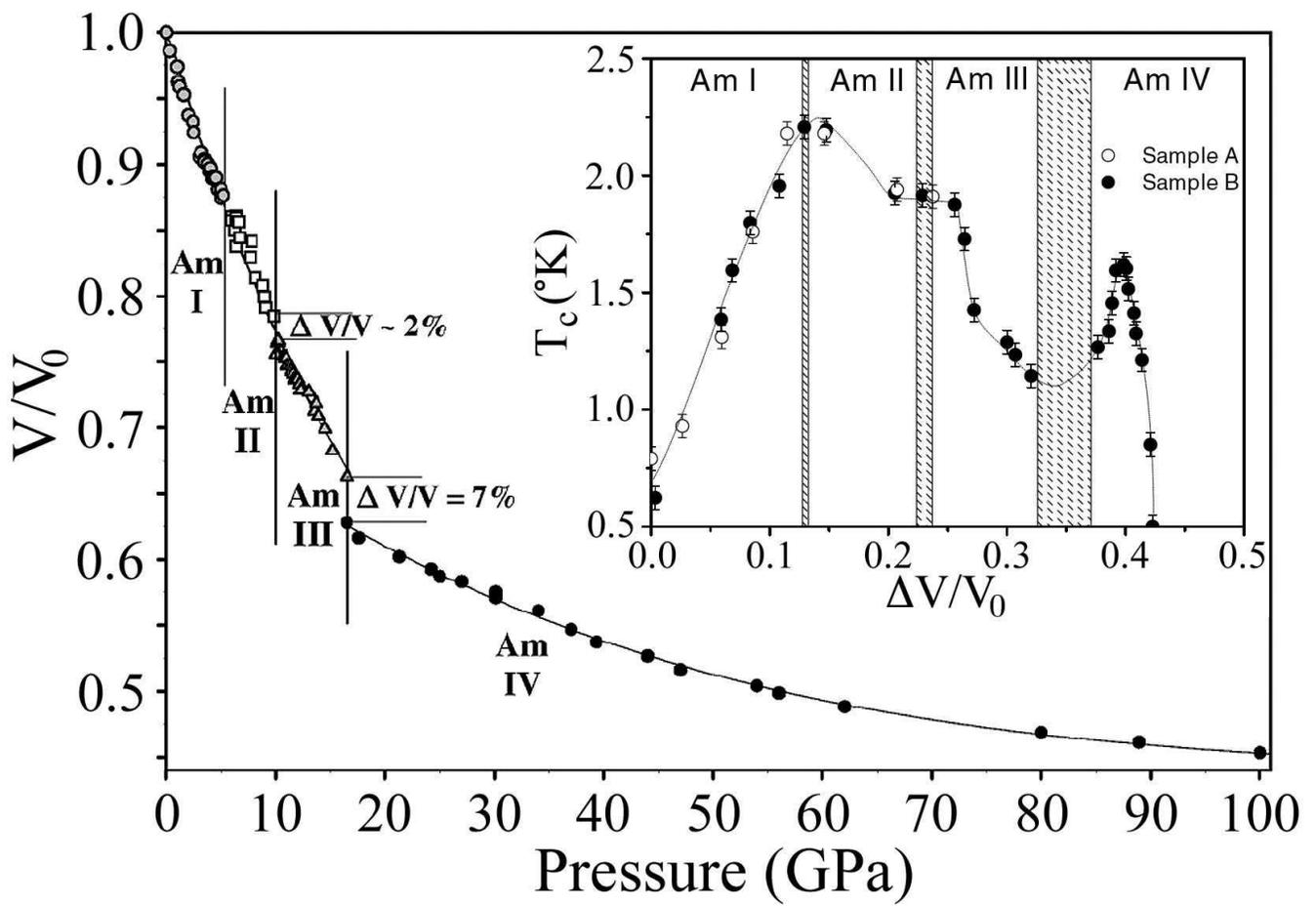

Figure 28    RG10015    06MAY2008

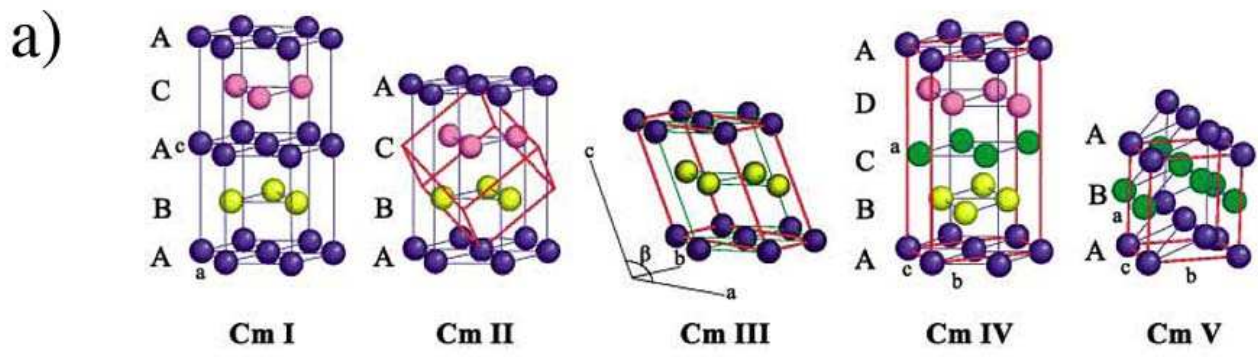
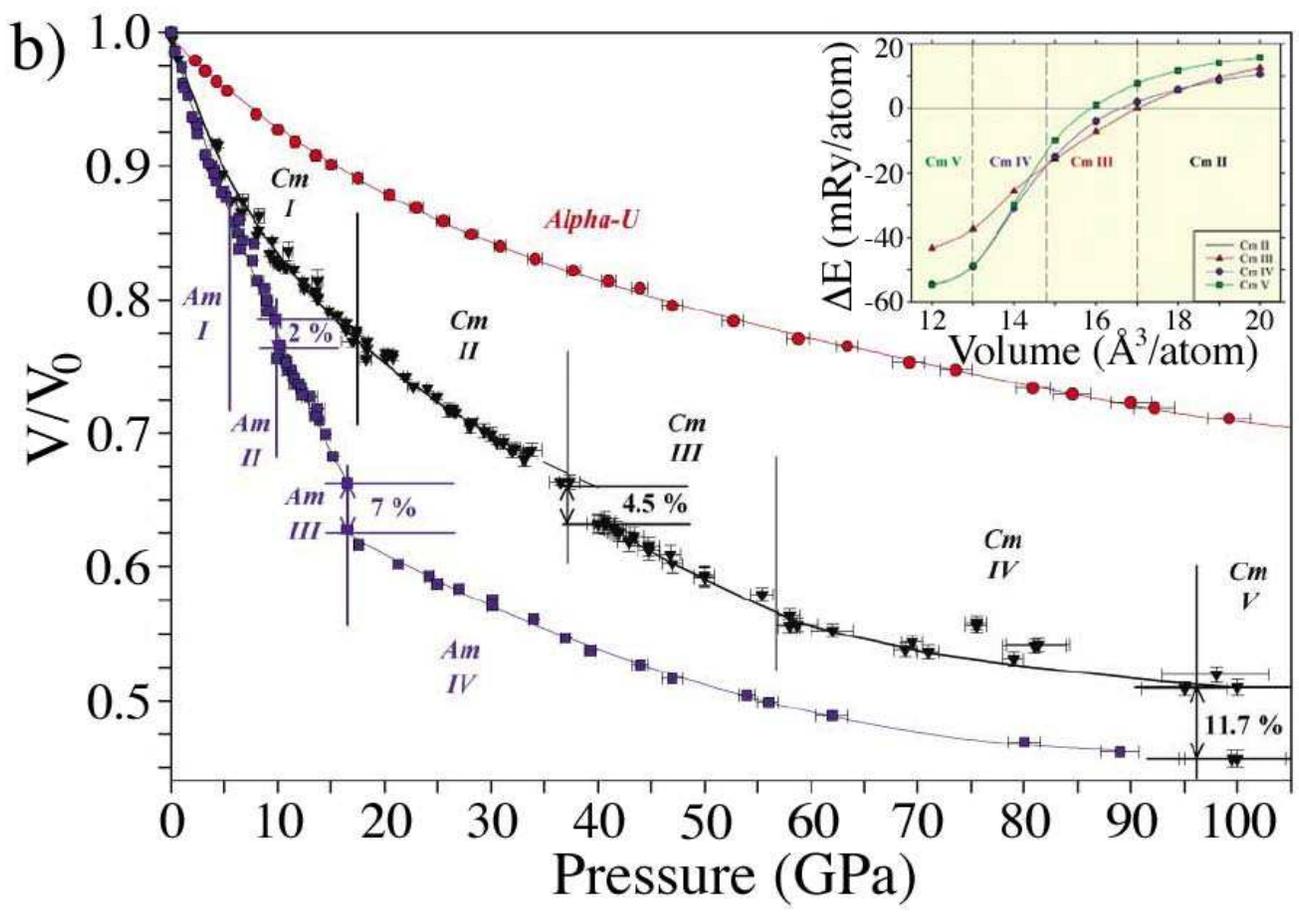

Figure 29    RG10015    06MAY2008